\documentclass[]{sig-alternate}
\usepackage{amsmath, amssymb}
\usepackage{graphicx}
\usepackage{subfig}
\usepackage{comment}
\usepackage{braket}
\usepackage{multirow}

\usepackage{paralist}

\usepackage[colorlinks,urlcolor=blue,citecolor=blue,linkcolor=blue]{hyperref}

\newcommand{\Sec}[1]{\hyperref[sec:#1]{\S\ref*{sec:#1}}} 
\newcommand{\Eqn}[1]{\hyperref[eq:#1]{(\ref*{eq:#1})}} 
\newcommand{\Fig}[1]{\hyperref[fig:#1]{Figure\,\ref*{fig:#1}}} 
\newcommand{\Tab}[1]{\hyperref[tab:#1]{Table\,\ref*{tab:#1}}} 
\newcommand{\Thm}[1]{\hyperref[thm:#1]{Thm.\,\ref*{thm:#1}}} 
\newcommand{\Lem}[1]{\hyperref[lem:#1]{Lem.\,\ref*{lem:#1}}} 
\newcommand{\Prop}[1]{\hyperref[prop:#1]{Prop.~\ref*{prop:#1}}} 
\newcommand{\Cor}[1]{\hyperref[cor:#1]{Cor.~\ref*{cor:#1}}} 
\newcommand{\Def}[1]{\hyperref[def:#1]{Defn.~\ref*{def:#1}}} 
\newcommand{\Alg}[1]{\hyperref[alg:#1]{Alg.~\ref*{alg:#1}}} 
\newcommand{\Ex}[1]{\hyperref[ex:#1]{Ex.~\ref*{ex:#1}}} 
\newcommand{\Clm}[1]{\hyperref[clm:#1]{Claim~\ref*{clm:#1}}} 

\newcommand{\CC}{\overline{C}}

\newcommand{\dmin}[1]{d_{\rm min}{(#1)}}
\newcommand{\dmid}[1]{d_{\rm mid}{(#1)}}
\newcommand{\dmax}[1]{d_{\rm max}{(#1)}}

\newcommand{\B}[1]{\mathcal{B}({#1})}

\newcommand{\da}[1]{d_{1}{(#1)}}
\newcommand{\db}[1]{d_{2}{(#1)}}
\newcommand{\dc}[1]{d_{3}{(#1)}}

\newcommand{\rba}[1]{r_{21}{(#1)}}
\newcommand{\rca}[1]{r_{31}{(#1)}}
\newcommand{\rcb}[1]{r_{32}{(#1)}}

\begin{document}

\title{Degree Relations of Triangles in Real-world Networks and Models }
\author{
%
%
\alignauthor
Nurcan Durak, Ali Pinar, Tamara G. Kolda, and C. Seshadhri\\
       \affaddr{Sandia National Laboratories}\\
       \affaddr{Livermore, CA}\\
       \email{\{ndurak, apinar, tgkolda, scomand\}@sandia.gov}
}

\maketitle

\begin{abstract}
Triangles are an important building block and distinguishing feature
of real-world networks, but their structure is still poorly
understood. Despite numerous reports on the abundance of triangles,
there is very little information on what these triangles look like.
We initiate the study of \emph{degree-labeled triangles} ---
specifically, degree homogeneity versus heterogeneity in triangles.
This yields new insight into the structure of real-world graphs.
We observe that networks coming from social and collaborative situations are dominated
by homogeneous triangles, i.e., degrees of vertices in a triangle are quite similar to each other.
On the other hand, information networks (e.g., web graphs) are dominated by heterogeneous triangles,
i.e., the degrees in triangles are quite disparate. Surprisingly,
nodes within the top 1\% of degrees participate in the \emph{vast majority}
of triangles in heterogeneous graphs.
We also ask the question of whether or not current graph models
reproduce the types of triangles that are observed in real data and
showed that most models fail to accurately capture these salient features.

\end{abstract}


\section{Introduction}
\label{sec:intro}

There is a growing interest in understanding the structure, dynamics,
and evolution of large scale networks. Observing the commonalities and
differences among real-world networks improves graph mining in many
aspects ranging from community detection to generation of more
realistic random graphs.

A triangle is a set of three vertices that are pairwise connected and
is arguably one of the most important patterns in terms of
understanding the inter-connectivity of nodes in real graphs
\cite{EcMo02, WaSt98}. Note that the community structure is closely
tied to triangles, and the degree behavior of triangles is an integral
part of this structure \cite{SeKoPi11}. Whether these graphs come from
communication networks, social interaction, or the Internet, the
presence of triangles is indication of community behavior.  In social
networks, it is considered highly probable that friends of friends
will themselves be friends, thus forming many triangles. Social
scientists have long observed the significance of triangles, as they
are manifestations of specific interaction patterns
\cite{Co88,Po98,FoDeCo10,Burt07,La06}.  For  example, in friendship networks, a
significant difference among degrees in triangle vertices might
indicate an anomaly such as existence of a spam bot \cite{BeBoCaGi08}.

\begin{figure}[thb]
\centerline{
\includegraphics[width=0.3\textwidth]{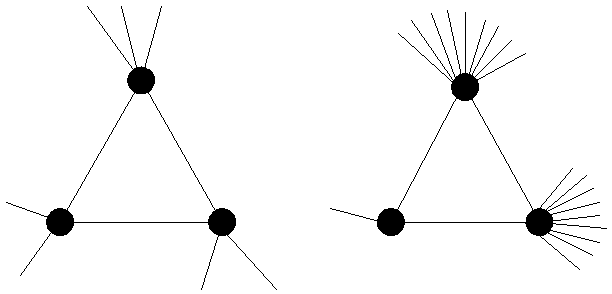} }
\caption{ \label{fig:deg-tri}  Homogeneity versus heterogeneity in triangles.  The triangle on the left is homogeneous since degrees of all its vertices are close, and the one on the right is heterogeneous  due to a mixture of high and low degree vertices. }
\end{figure}

In this paper, we take a closer look at the structure of triangles,
specifically, the degrees of the triangle vertices.  Consider the
two triangles in \Fig{deg-tri}. How are the degrees of the three
vertices related? Do these represent fundamentally
different types of relationships and so appear in different sorts of
networks? When we look at real-world networks, we may ask if there
is a high incidence of \emph{degree homogeneity}, wherein vertices of
similar degree come together to form triangles? Or do triangles tend
to show \emph{degree heterogeneity}, i.e., connecting vertices of
disparate degree?

\subsection{Background and Previous Work}
\label{sec:rel}
The study of triangles is quite prevalent in the social sciences
community.  Coleman~\cite{Co88} and Portes~\cite{Po98} used the
clustering coefficient to predict the likelihood of going against
social norms.  Welles et al.
  studied the variance of clustering
coefficients for different demographics groups and found that
adolescents are more likely to have connected friends than adults and
are even more likely to terminate connections with friends that are
not connected to their other friends~\cite{FoDeCo10}.  Burt underlined
the importance of nodes that could serve as a bridge between various
communities~\cite{Burt04} and tied this to the number of open
triangles in a vertex~\cite{Burt07}.  Lawrence~\cite{La06} observed
that there are powerful homophily effects in who connects to whom in
organizational environments, who people are aware of, or whose
opinions people attend to.
Bearman, Moody, and Stovel  studied homogeneity of partners in a
romantic network of adolescents~\cite{BeMoSt04}. One noteworthy
observation in this study was the similarity  on partnership
experience, which corresponds to similarity in vertex degrees.

The notion of describing graph structure based on the frequency of
small patterns such as triangles has been proposed under
different names such as  motifs~\cite{BoLaMoCh06, Milo2002},
graphlets~\cite{Pr07}, and structural signatures \cite{CoWaFa06}.
Triangle counts form the basis for community detection algorithms in
\cite{BeHe11,GlSe11}.
They have also served as the driving force for generative
models~\cite{SeKoPi11,WaSt98}.
Eckman and Moses~\cite{EcMo02}
interpreted the clustering coefficients as a curvature and  showed
that connected regions of high curvature on the WWW characterized
common topics.
Directed triangles are important motifs for comparing and
characterizing
graphs~\cite{BoLaMoCh06, CoWaFa06,MaHuKrHu07,MaKrFaVa06,Pr07}.  For
graph databases, exploiting frequent patterns have also been proposed
for efficient query processing~\cite{ShWaGi02,YaYuHa04}.

The frequency of triangles is often measured using the \emph{clustering coefficient},
as defined by Watts and Strogatz~\cite{WaSt98}.
We first establish some notation.
Consider an undirected graph $G$ with $n$ vertices.
Let $d_j$ denote the
degree of node $j$ and $t_j$ denote the number of triangles containing
node $j$. If
we define a wedge to be a path of length 2, then the number of wedges
centered at node $j$ is ${d_j \choose 2}$.
Now we can define various clustering coefficients.
The clustering coefficient of vertex $j$, $C_j$, is defined as the
number of triangles incident to $j$ divided by the number of wedges
centered at $j$, i.e., $C_j = t_j / {d_j\choose 2}$. The average of  clustering coefficients across all vertices (called the local clustering coefficient) is defined as $\CC = \frac{1}{n}\sum_j C_j$.
Let $V_d = \{ j : d_j = d \}$ be the set of vertices of degree $d$.  We define the
clustering coefficient of degree $d$ to be $C_d = \frac{1}{|V_d|} \sum_{j \in V_d} C_j$.

The (global) clustering coefficient, also known as the
{\em transitivity}, is
\begin{displaymath}
 \label{eq:GCC}
C = \frac{3 \times \text{total number of triangles}}{\text{total
    number of wedges}}
= \frac{\sum_j t_j}{\sum_j {d_j \choose 2}}.
\end{displaymath}
For  \emph{random}
graphs with no structure, $C$ and $\CC$ values are extremely
small~\cite{PiSeKo12}.

Most of the studies on degree-based similarity is based on
\emph{assortativity}, which was introduced by Newman~\cite{Ne02}.
Various studies have been
conducted on the assortativity (or lack thereof) of real
graphs~\cite{HoZh07,LiHo12,WhAl06}.
However, Newman's assortativity measure is misleading to classify networks with
heavy-tailed degree distributions because it produces either neutral or
negative assortativity (disassortativity) values for most of the large
scale networks as shown in \Tab{graphs} with $r$ values.

Most relevant to our work is that of Tsourakakis~\cite{Ts08}, which observed various power laws
in triangle behavior.
 This work focused on triangle counts for nodes
(how many triangles a node participates). He finds that the average number of  triangles per vertices of a given degree
  follows a
power-law distribution and the slope of the degree-triangle plot has
the negative slope of the degree distribution plot of the
corresponding graph.
It is argued that low degree nodes form fewer triangles than higher degree nodes.
Our analysis shows that while this is certainly true for social networks, it does not
hold for information networks, such as the autonomous  systems networks.

\subsection{Contributions} \label{sec:cont}

Our contributions fall into two categories. Our first set of results comes
from empirical studies of degree relations in triangles of real graphs. Then,
we perform experiments on a variety of graph models to show their (in)ability
to reproduce the behavior of real graphs.

{\bf Triangle homogeneity vs heterogeneity:}
We take a collection of graphs
from diverse scenarios (collaboration, social networking, web, infrastructure)
and measure triangle degree relationships.
We compute various correlations between degrees of vertices in triangles to understand the homogeneity nature of these triangles.

Our experiments showed that graphs coming from social or collaborative scenarios are completely
dominated by homogeneous triangles. There are a few heterogeneous triangles. This may not be surprising from a sociological viewpoint, since like should attract like. But graphs
coming from web, routing, or communication are dominated by heterogeneous
triangles. It is interesting that in communication or routing networks, majority of triangles are formed by the vertices within top 1\% degrees.

We observed that there is a high correlation between global clustering coefficient $C$ and the homogeneity tendency of triangles. The higher $C$ a network has, the stronger homogeneity tendency the network has.

We also showed that the triangles in networks have diverse degrees of vertices and there is a varying distance among triangle degrees.

{\bf The triangle behavior of graph models:} Our result
can be stated quite succinctly. No existing graph model reflects the homogeneous and heterogeneous triangle behavior together.
Many standard graph models like Preferential Attachment, Edge Copying, Stochastic Kronecker, etc. do not generate enough triangles and they cannot approximate the clustering coefficients of the real graphs \cite{SaCaWiZa10}. The Chung-Lu \cite{ChLu02} model cannot generate homogeneous graphs and cannot approximate the clustering coefficient of the high clustering coefficient networks. However, Chung-Lu is the only model imitates the networks with low clustering coefficient and it is enable to generate heterogeneous triangles.

The Forest Fire \cite{LeKlFa05} and BTER~\cite{SeKoPi11} models  generate a reasonable number of triangles (especially incident
to low degree vertices) but these triangles are extremely homogeneous. Low degree vertices, when they participate in triangles, exclusively form
triangles with other low degree vertices. This happens regardless of  parameter
choices, and is a fundamental property of these models. This shows that
while they can qualitatively look like social and interaction networks, the
behavior of, say, heterogonous networks cannot be reproduced by these models.

\begin{table*}[bt]
\caption{Properties of networks we analyzed. }
\begin{center}
\begin{tabular}{ |c | c | c |c|c|c| c|c| c|c|c|c | c|}
\cline{2-12}
        \hline
	&Graph Name	&$N$	 &$E$ &$\rho$ & $C$ & $\CC$ & $T$  &$\alpha$  &$\kappa_{90}$ &$\kappa_{99}$ &$d_{\rm{max}}$ &r\\ \hline\hline

\multirow{4}{*}{\rotatebox{90}{high-$C$}} &
amazon0312     & 400K	&2,349K  & 5.9& 0.260    &0.41 & 3,686K   & 3.1 &19& 55& 2747 &-0.02\\ 
&ca-AstroPh	   & 18K	&198K  & 11 &0.318  &0.63  &1,351K  &1.52 &56 &145 & 504 &0.2\\ 
&cit-HepPh		&34K	&420K &12 	&0.146  &0.30 &1,276K &1.53 &56 &147 &846 &0\\ 
&soc-Epinions1	&76K	&405K &  5.3 &0.066   &0.228 &1,624K  &1.68 &65 &307 &3044 &-0.04\\ 

\hline \hline
\multirow{4}{*}{\rotatebox{90}{low-$C$}} &
as-caida20071105	&26K	&53K &2 	&0.007     &0.21 &36K 	&1.52 &12 &99 &2628 &-0.19\\ 
& oregon1\_010331	&10K	&22K &2.1	&0.009	&0.45	&17K  &1.5 &10 &839 &2312 &-0.18\\ 
&web-Stanford  &281K	&1,992K  &7.1  &0.009 &0.61 &11,329K  &1.51 &30 &92 &38625 &-0.11\\ 
&wiki-Talk	&2,394K		&4,659K &1.9		&0.002  &0.20	 &9,203K  &1.67 &21 &401 & 100029 &-0.06 \\ 
\hline\hline
\end{tabular}
\end{center}
\label{tab:graphs}
\end{table*}

\section{Real-World Triangle Behavior}
\label{sec:tri}

\subsection{Data}

We analyze the degree relations among vertices of  triangles  on  a diverse
set of real-world graphs: collaboration network (ca-AstroPh), citation
network (cit-HepPh), trust network
(soc-Epinions1), co-purchasing network (Amazon0312), autonomous
systems (as-caida20071105), routing network (oregon1\_010331), web
network (web-Stanford), and communication network (wiki-Talk). All
these graphs were obtained from the SNAP database \cite{Snap}.
In our studies, we have symmetrized the graphs  by
treating all edges as undirected; made each graph simple, by removing self loops  and parallel edges; and did  not use edge weights.
Cohen's algorithm~\cite{Co09} was used to enumerate all triangles.

In \Tab{graphs}, we provide the following properties of the graphs we
analyzed: $N$ = number of nodes; $E$ = number of edges; $\rho = E/N$ (density); $C$ = global clustering coefficient; $\CC$ = local
clustering; $T$ = number of triangles; $\alpha$ = the power-law exponent, which is computed
by fitting power-law distribution to degree distribution plots of the
graphs \cite{ClShNe09}; $\kappa_{90}$ and $\kappa_{99}$, respectively, are the 90th and 99th percentiles of
degree of all nodes participating in triangles (i.e., we obtain all
nodes participating in any triangle (each node is only counted once),
put their degrees in a list, and then pick the 99th percentile of the
degree list); $d_{\rm{max}}$ = maximum degree; and $r$ = assortativity value.

In this paper, a network whose global clustering coefficient, $C$ is greater than 0.01,  is referred to as a \emph{high-$C$} network, and as a \emph{low-$C$} network otherwise. In \Tab{graphs}, the first 4 graphs are \emph{high-$C$} graphs, whereas the last 4 are \emph{low-$C$} graphs.

\subsection{Analysis}

We analyze the degree
similarity of triangle vertices by grouping the triangles according to
their minimum degree vertex. We first present the notation. For $t=1,\dots,T$, let $\dmin{t}$, $\dmid{t}$, and $\dmax{t}$ denote
the minimum, middle, and maximum degree of the $t$-th triangle. For instance, if the $t$-th triangle has vertices of degrees $5, 10$, and $4$, then $\dmin{t} = 4$, $\dmid{t} = 5$, and $\dmax{t} = 10$.
Define the set $\B{i}$ to be the set of all triangles whose minimum degree is $i$, i.e.,
\begin{displaymath}
  \B{i} = \Set{ t \in T| \dmin{t} = i }.
\end{displaymath}

We may then define some average statistics for each set
$\B{i}$. Define $\da{i}$, $\db{i}$, and $\dc{i}$ to be the median of
minimum, middle, and maximum degree, respectively, of triangles in
$\B{i}$. In other words,
\begin{align*}
  \da{i} &= {\rm median} \Set{ \dmin{t} | t \in \B{i} } = i, \\
  \db{i} &= {\rm median} \Set{ \dmid{t} | t \in \B{i} }, \\
  \dc{i} &= {\rm median} \Set{ \dmax{t} | t \in \B{i} }.
\end{align*}

For instance, if the triangles in $\B{2}$ were given by $[2 \ 2 \ 3]$, $[2 \ 4 \ 5]$,
and $[2 \ 3 \ 3]$. Then, the $\da{2}=2$, $\db{2}=3$ and $\dc{2}=3$.

\begin{figure*}[tb]
  \centering
  \subfloat[amazon0312]{\label{fig:degs-Amazon}
    \includegraphics[width=.25\textwidth,trim=0 0 0 0]{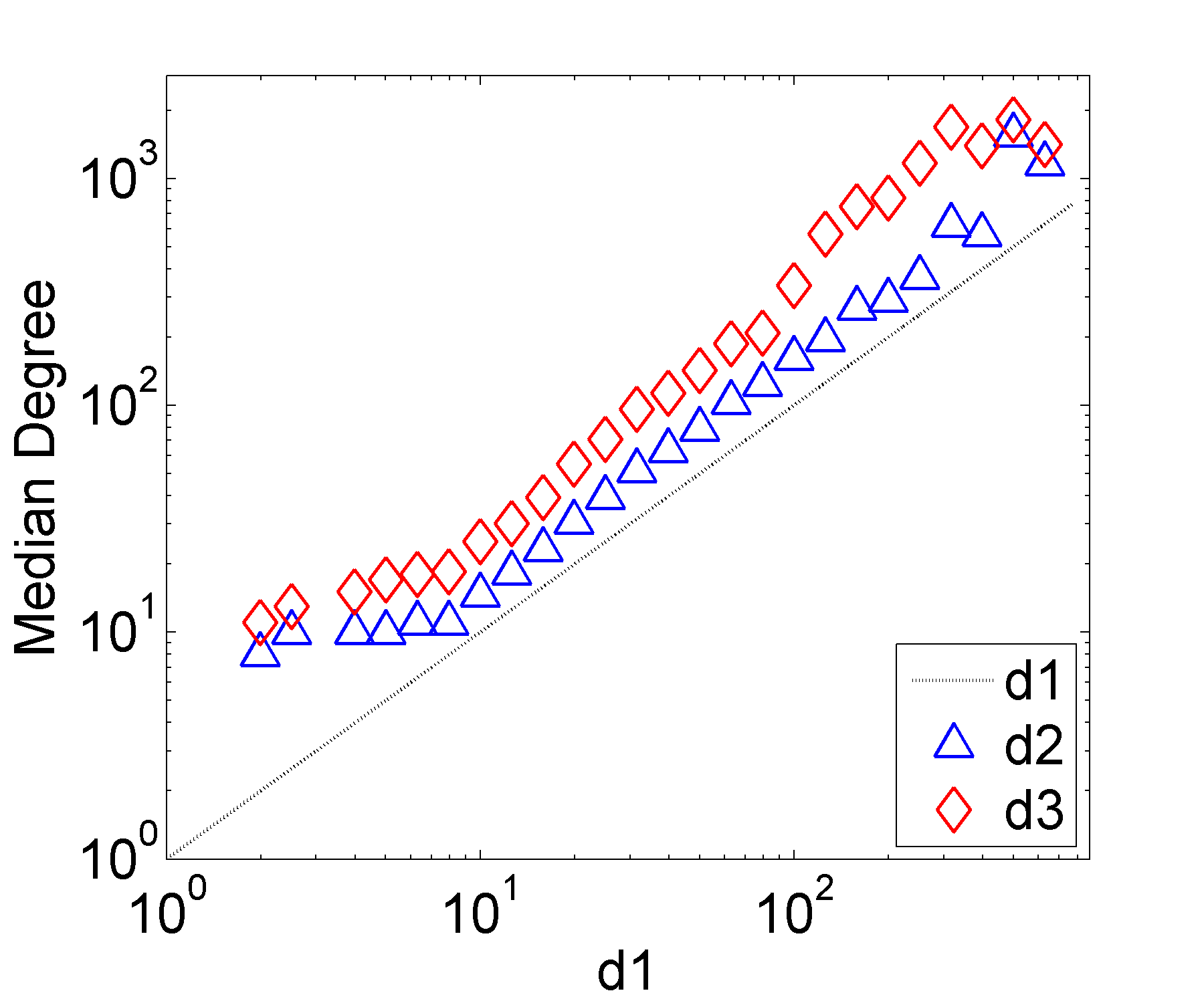}
  }
  \subfloat[ca-AstroPh]{\label{fig:degs-ca-Astro}
    \includegraphics[width=.25\textwidth,trim=0 0 0 0]{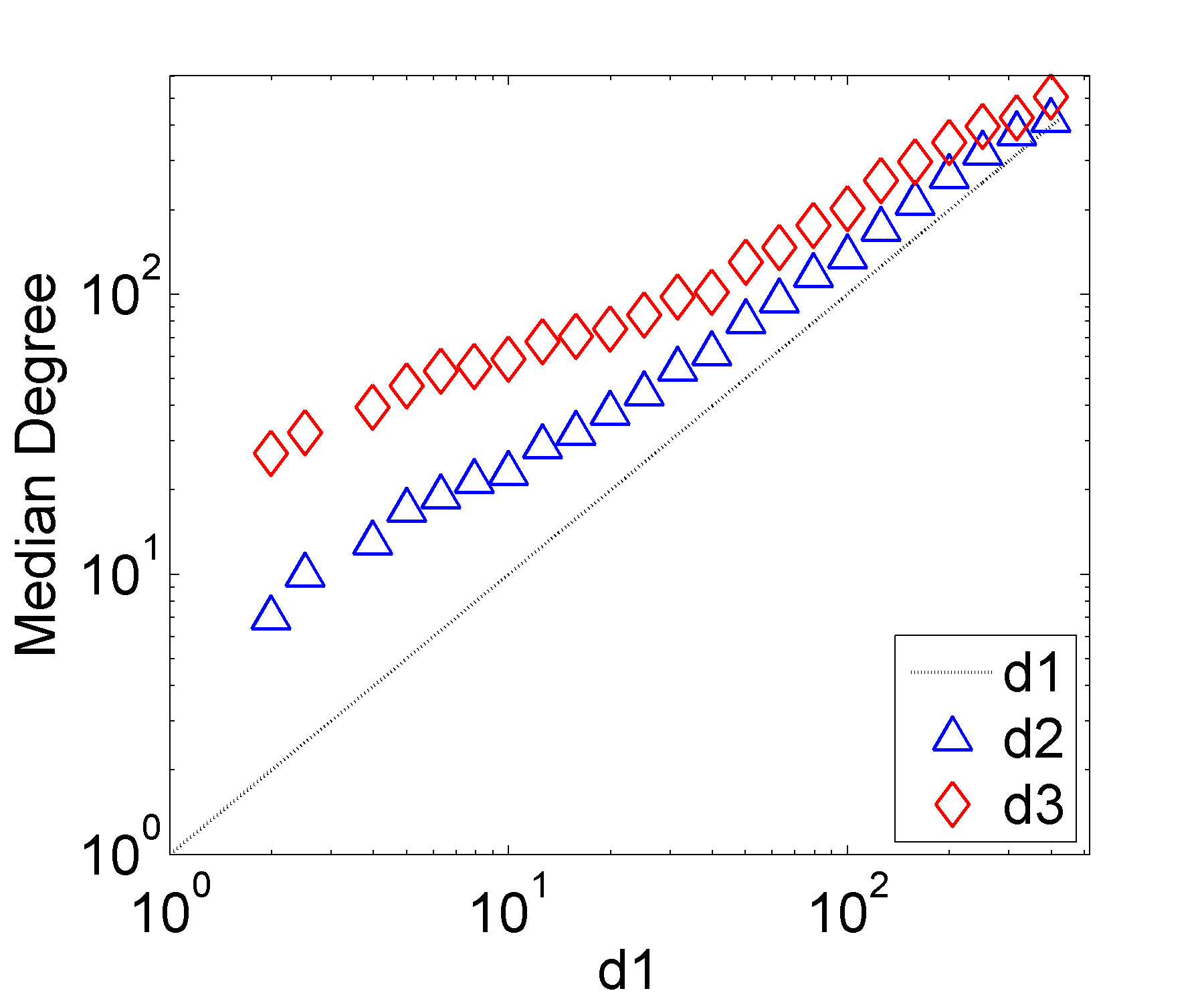}
  }
  \subfloat[cit-HepPh]{\label{fig:degs-cit-HepPh}
    \includegraphics[width=.25\textwidth,trim=0 0 0 0]{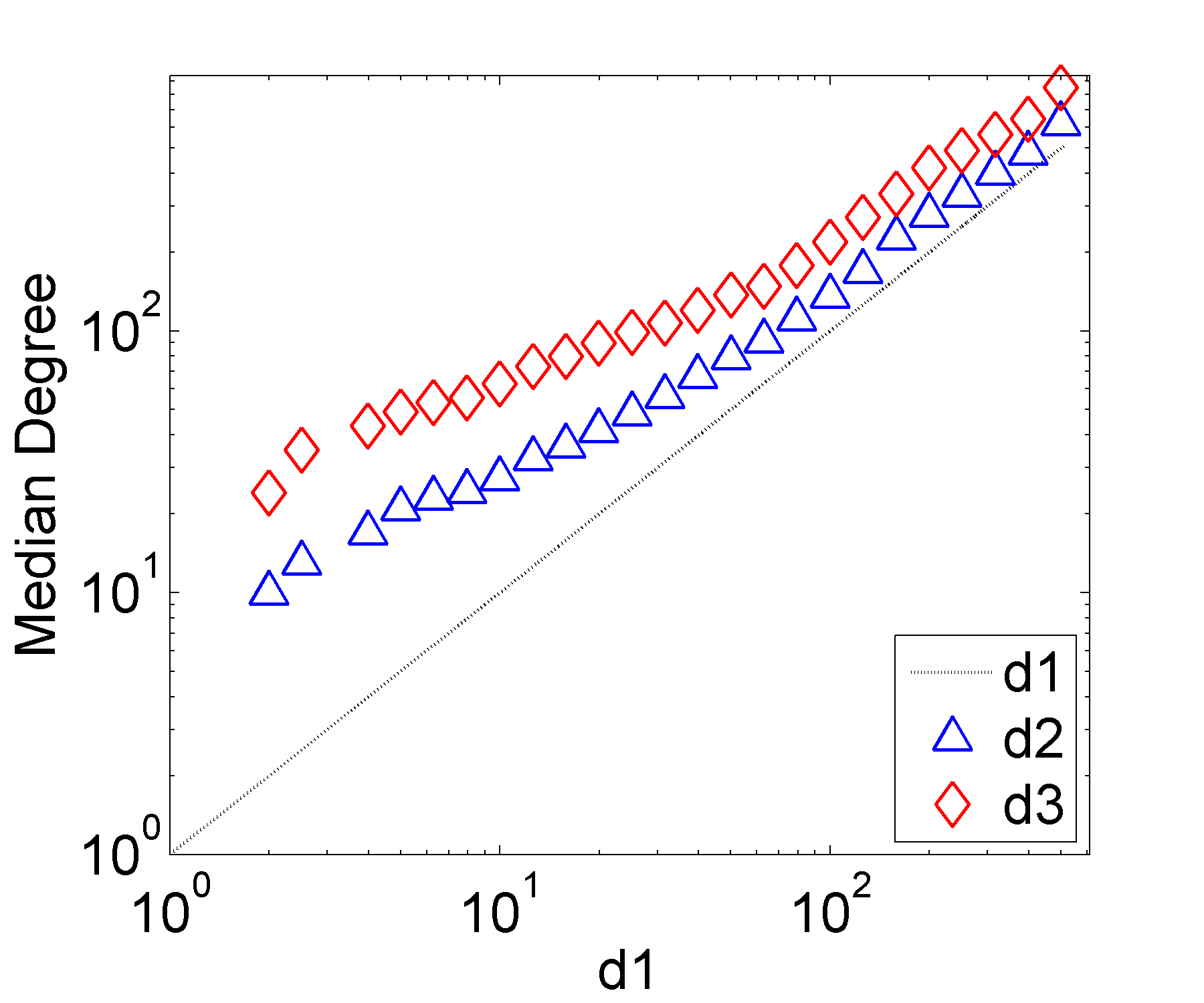}
  }
  \subfloat[Soc-Epinions]{\label{fig:degs-Soc_Epinions}
    \includegraphics[width=.25\textwidth,trim=0 0 0 0]{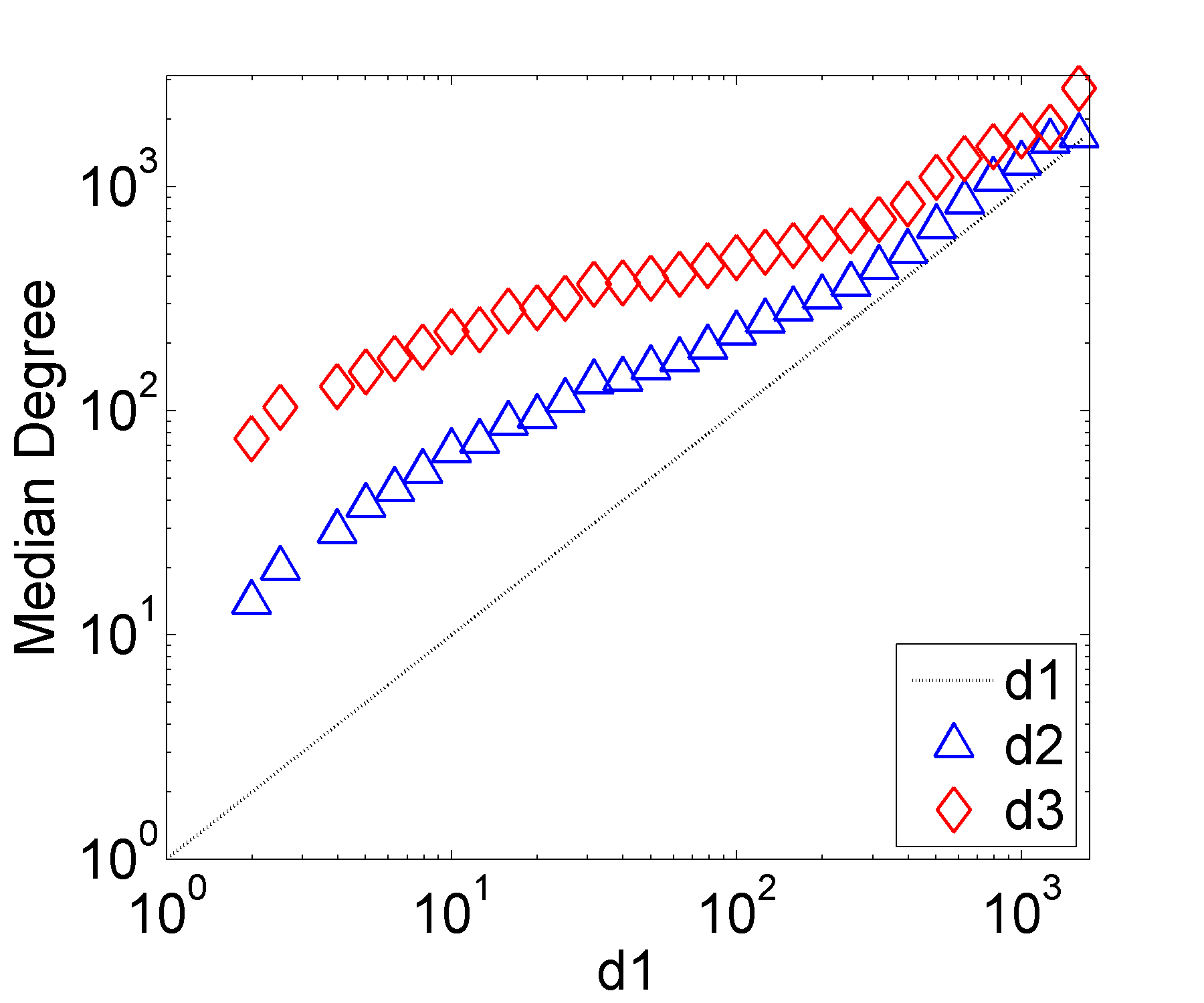}
  }
  \\
  \subfloat[as-caida20071105]{\label{fig:degs-as-caida}
    \includegraphics[width=.25\textwidth,trim=0 0 0 0]{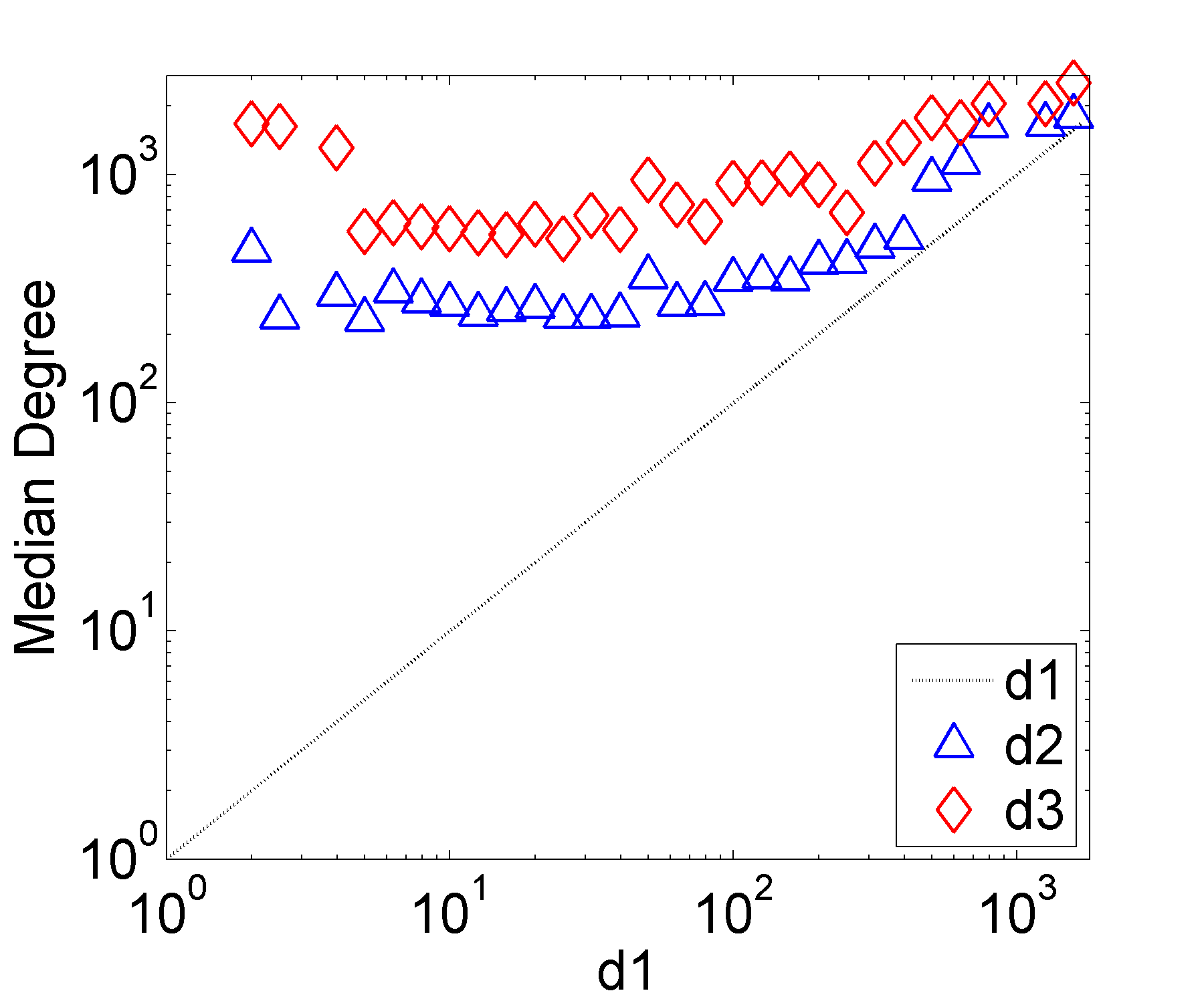}
  }
   \subfloat[oregon1\_010331]{\label{fig:degs-oregon}
    \includegraphics[width=.25\textwidth,trim=0 0 0 0]{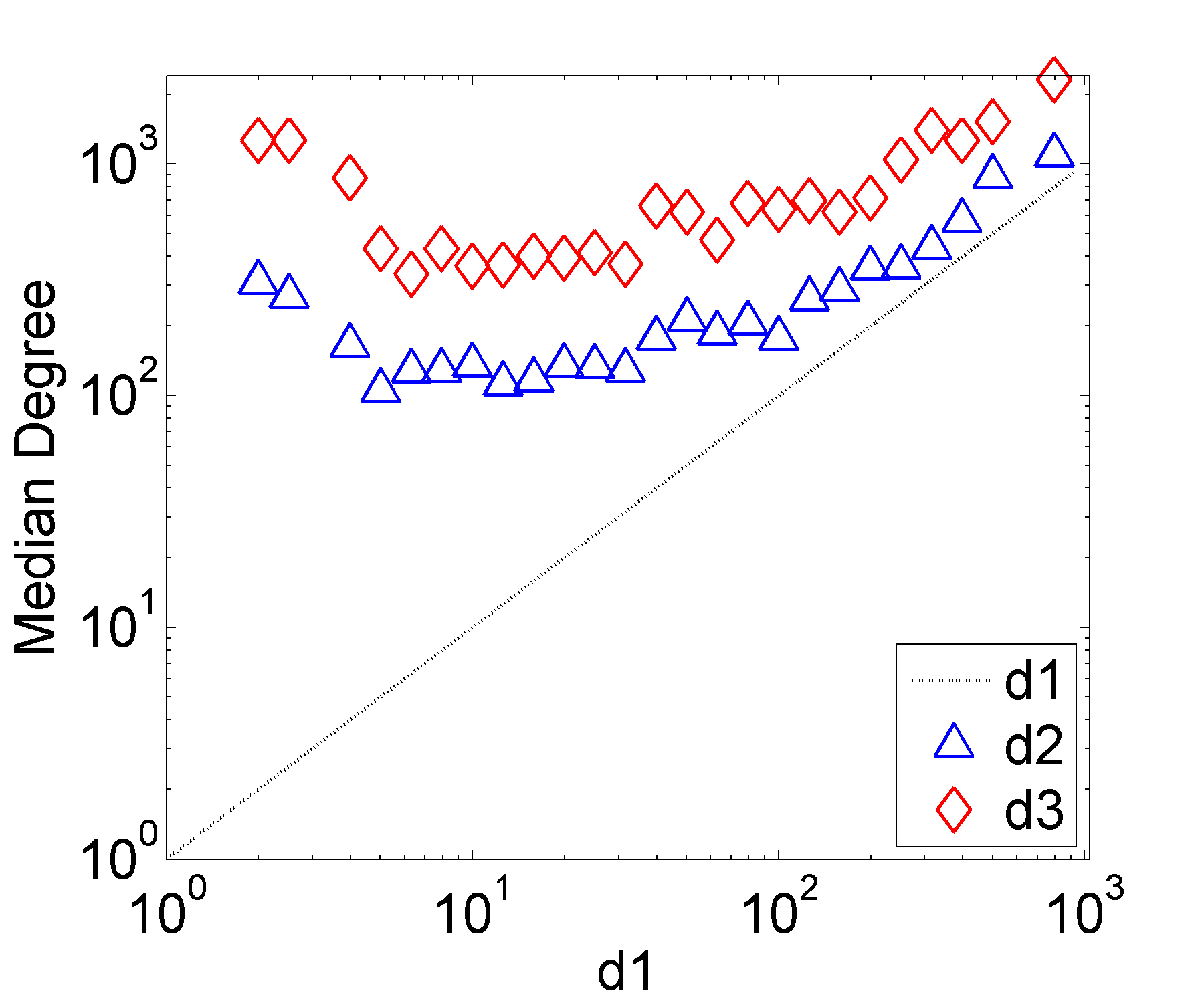}
  }
  \subfloat[web-Stanford]{\label{fig:degs-web-Stanford}
    \includegraphics[width=.25\textwidth,trim=0 0 0 0]{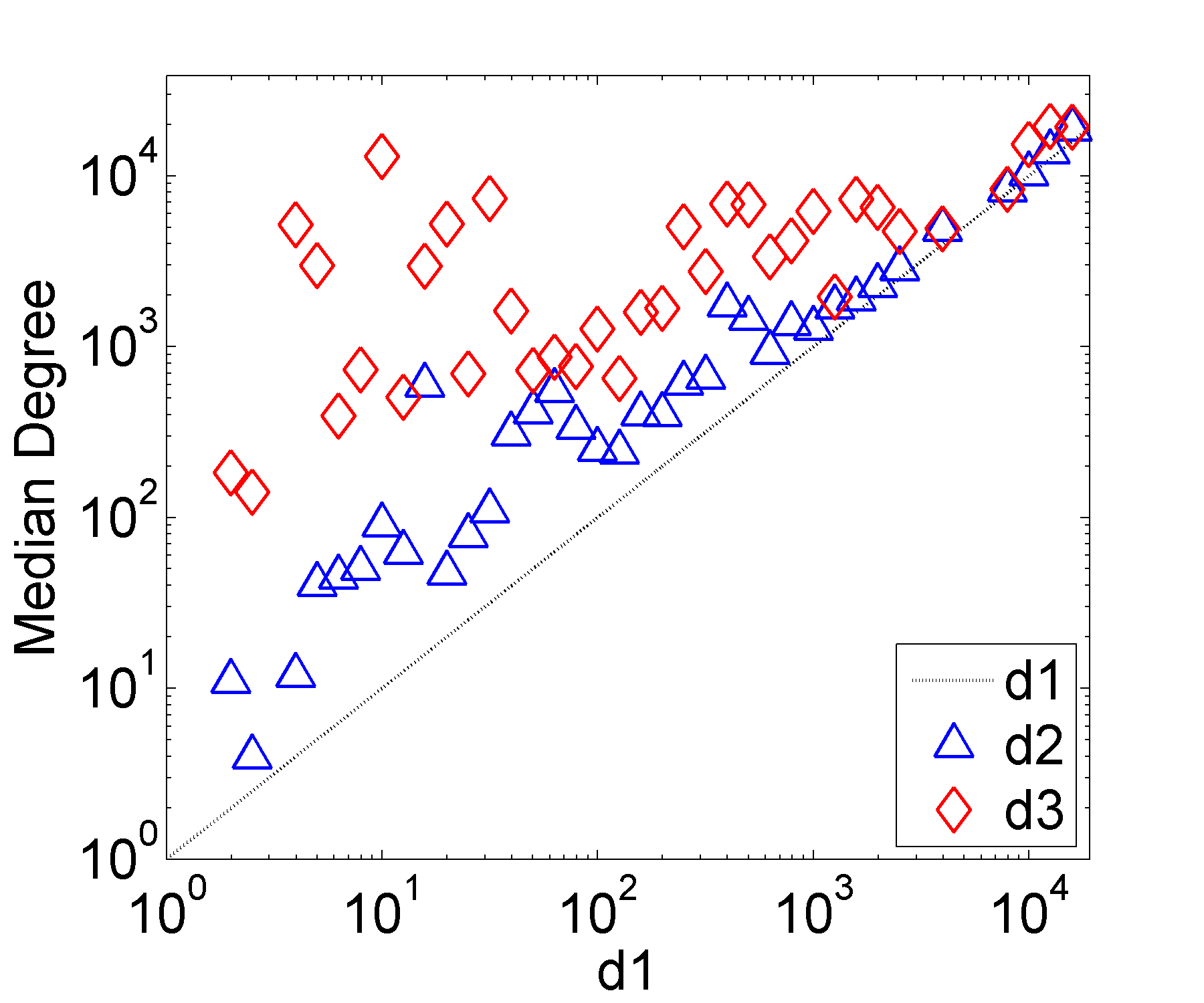}
  }
   \subfloat[wiki-Talk]{\label{fig:degs-wiki-talk}
    \includegraphics[width=.25\textwidth,trim=0 0 0 0]{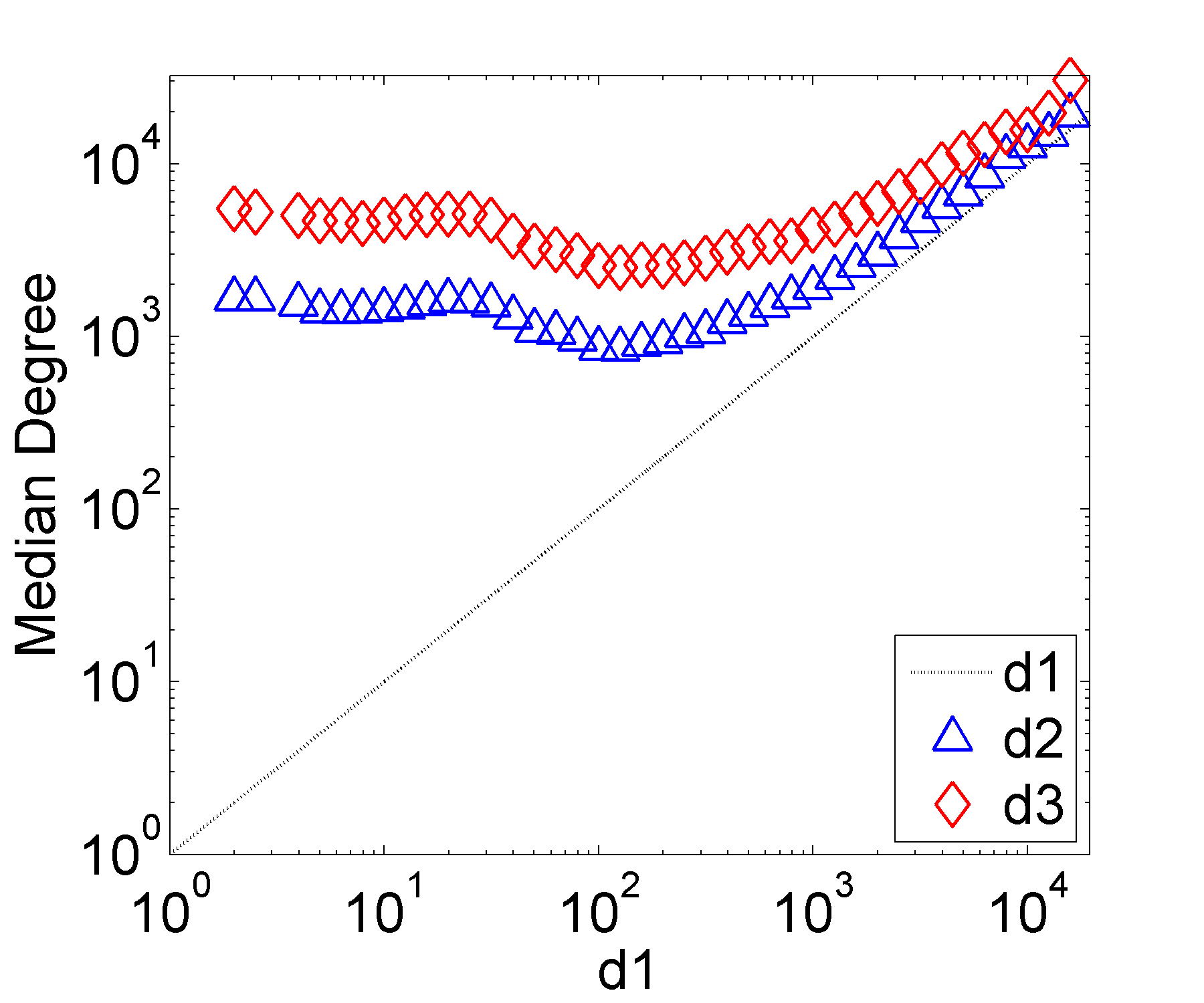}
  }
  \caption{Triangle degree-comparison plots which compare the medians of minimum degree, $\da{i}$, the middle degree, $\db{i}$, and the maximum degree $\dc{i}$ }
  \label{fig:d1-vs-other1}
\end{figure*}

To compare the relations among triangle degrees, we plot the $\db{i}$ and $\dc{i}$ versus $\da{i}$ as in \Fig{d1-vs-other1} and call them degree-comparison plots. Note that in these and all other log-log and semi-log plots, we use the exponential binning which is a standard procedure to de-noise the data when plotting on logarithmic scale.

\subsection{Observations}
\label{subsec:obs}

By considering the degree relations of the triangle vertices, we make the following observations.

\smallskip
\textbf{Observation 1}: \textit{The global clustering coefficient is an indicator for the triangle degree relations.}

\smallskip When we analyze \Fig{d1-vs-other1}, we can see a clear
relation between global clustering coefficient $C$ and the type of
triangles. In \emph{high-$C$} networks, minimum, middle, and maximum degrees
of triangle vertices are close in value. While, in \emph{low-$C$} networks,
triangles are highly heterogeneous. Observe how very small values of
$\da{i}$ connect to quite large $\db{i}$ or $\dc{i}$. In \Fig{degs-as-caida}, \Fig{degs-oregon} and \Fig{degs-wiki-talk} low degree vertices are connecting to two high vertices ($\db{i}$ and $\dc{i}$). Web-Stanford (\Fig{degs-web-Stanford}) is less structured and has a different tendency from the rest but it is still visible that low degree vertices are connecting to high $\dc{i}$ value.

The average clustering coefficient $\CC$ is not a very distinguishing metric for our study. The global clustering coefficient $C$ shows wide variance and is a better indicative of the triangle behavior.

\smallskip
\textbf{Observation 2}: \textit{There is a non-trivial gap between the minimum, medium, and maximum degrees of triangles.}

\smallskip
Even though the degrees of triangle vertices are similar to each other in \emph{high-$C$} networks, still there is a non-trivial between $\da{i}$ and $\db{i}$ and between $\db{i}$ and $\dc{i}$ in \Fig{d1-vs-other1}. This observation tells us that inside the communities or clusters in real networks, there exist triangles with varying degrees of vertices.

\smallskip
\textbf{Observation 3}:  \textit{The ratios among degrees of triangle vertices are small in \emph{high-$C$} networks and large in \emph{low-$C$} networks.}

\smallskip
The ratios of triangle degrees provide valuable information to see the distinction between networks. For the $t$-th triangle, three degree ratios are defined as follow: $\rba{t} = \frac{\dmid{t}}{\dmin{t}}$, $\rca{t} = \frac{\dmax{t}}{\dmin{t}}$, and $\rcb{t} = \frac{\dmax{t}}{\dmid{t}}$. These ratios are computed for all the triangles separately and their average is computed as $\bar{r}_{21}$, $\bar{r}_{31}$, and $\bar{r}_{32}$, respectively. \Tab{degree-ratio} lists the average ratios for all the networks.

\begin{table}[htb]
\caption{The average of triangle degree ratios}
\begin{center}
\begin{tabular}{|c | c | c |c| c|}
\hline
& Graph Name &$\bar{r}_{21}$ & $\bar{r}_{31}$ &$\bar{r}_{32}$ \\ \hline \hline
\multirow{4}{*}{\rotatebox{90}{high-$C$}} &
amazon0312 & 1.98& 4.95& 2.53 \\
& ca-AstroPh & 1.88& 3.46& 1.89 \\
& cit-HepPh & 2.20& 4.96& 2.38 \\
&  soc-Epinions1 & 3.34& 9.41& 2.95 \\ \hline\hline
\multirow{4}{*}{\rotatebox{90}{low-$C$}} &
as-caida20071105 & 70.99& 164.35& 8.14 \\
&  oregon1\_010331 & 54.80& 175.69& 9.09 \\
&  web-Stanford & 87.91& 300.37& 114.43 \\
&  wiki-Talk & 42.64& 138.01& 4.75 \\ \hline \hline
\end{tabular}
\end{center}
\label{tab:degree-ratio}
\end{table}

There is a clear distinction between \emph{high-$C$} and \emph{low-$C$} networks. The average ratios are very small in \emph{high-$C$} networks, which also supports the triangle homogeneity in \emph{high-$C$} networks. The average of $\bar{r}_{21}$ and $\bar{r}_{32}$ are close to each other, in other words, the middle degree is both close to the minimum degree and the maximum degree.

The average degree ratios are significantly large in \emph{low-$C$} networks. Particularly, the ratio between the maximum and the minimum degree is  very high.
We can see that in as-caida20071105, oregon1\_010331, wiki-Talk, middle and maximum degrees are close to each other but these degrees differ from the minimum degree significantly. Web-Stanford acts little different. In this graph, minimum and middle degrees are close to each other but the maximum degree is distant to both the minimum and the medium degrees.

\begin{figure}[th]
  \centering
   \includegraphics[width=.45\textwidth, height = 45mm, trim=0 0 0 0]%
    {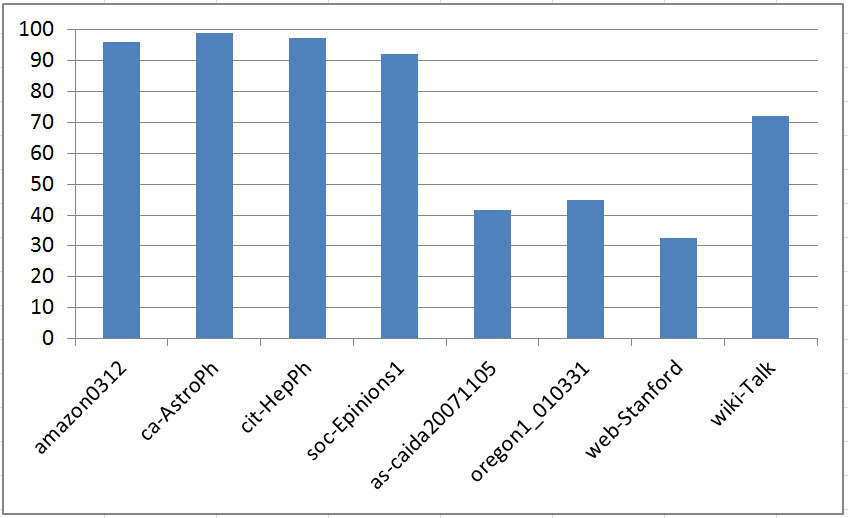}
    \caption{Percentage of triangles such that $d_{\max}/d_{\min} \leq 10$ }
  \label{fig:homo-stat}
\end{figure}

We have also looked at the percentage of homogeneous triangles, which
we define triangles where $\bar{r}_{31} < 10$.  \Fig{homo-stat} shows
that more than 90\% of the triangles in \emph{high-$C$} networks are
homogeneous.

\smallskip
\textbf{Observation 4}: \textit{In \emph{low-$C$} networks, high degree vertices within the top 1\% participate in the vast majority of the triangles.}

\smallskip
In \emph{high-$C$} networks, the triangles incident to low degree vertices are mostly connecting to two low degree vertices. On the other hand, in \emph{low-$C$} networks (particularly in the low density networks), a significant portion of the triangles contain at least one high degree vertex.

\begin{figure}[h]
  \centering
   \includegraphics[width=.45\textwidth, height = 45mm, trim=0 0 0 0]%
    {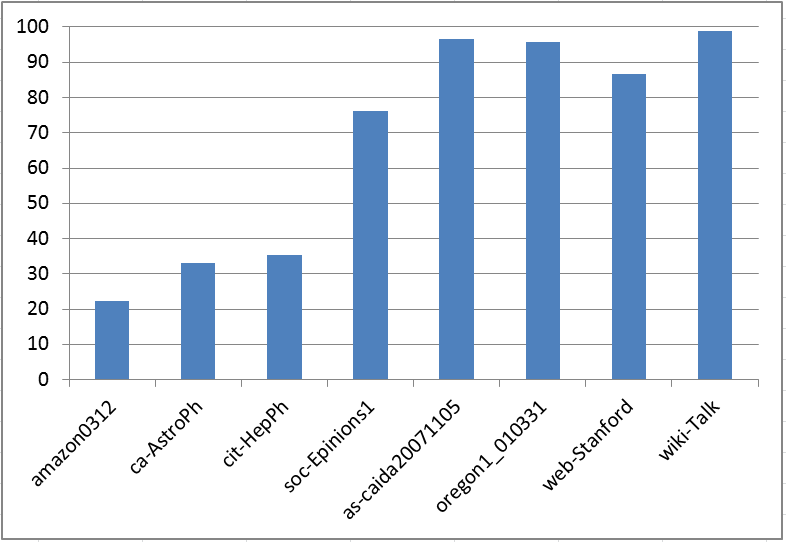}
    \caption{The percentage of triangles produced by vertices in the Top 1\% degrees}
  \label{fig:top1-stat}
\end{figure}

To set a threshold between low degrees and high degrees, we have experimented different percentiles of vertices that  participate in at least one triangle. In \Tab{graphs}, $\kappa_{90}$, $\kappa_{99}$, and the maximum degree of each network are presented. It is interesting that the gap between $\kappa_{90}$ and $\kappa_{99}$ is quite large in \emph{low-$C$} networks. We pick $\kappa_{99}$ as a threshold, since $\kappa_{90}$ is still relatively low compared to the maximum degree in most networks. A degree of a triangle vertex is considered \emph{low}, if the degree is no greater than $\kappa_{99}$, \emph {high} otherwise it is considered.

We look at the percentages of the triangles having at least one \emph{high} degree node in \Fig{top1-stat}. In \emph{low-$C$} networks, we can say that high degree vertices within the top 1\% participate in most of the triangles. In \emph{high-$C$}, \emph{high} degree nodes are participating in less triangles except soc-Epinions network. Soc-Epinions has a very low clustering coefficient value (i.e., less than $0.1$) and it is a border case. Hence, its statistics are higher than the other \emph{high-$C$} networks.

\section{Graph-Model Triangle Behavior }
\label{sec:gm}
In this section, we investigate how well random graph generators match the real graphs in terms of triangle degree similarity. We concentrate on the graph models generating heavy-tailed degree distributions.

\begin{table*}[t]
\caption{The number of triangles generated by graph models }
\begin{center}
\begin{tabular}{|c| c | c | c | c | c | c | c| c| }
        \hline
& Graph Name &Original &BTER &CL &FF &EC &PA &SKG \\ \hline\hline
\multirow{4}{*}{\rotatebox{90}{high-$C$}} &
amazon0312  & 3,686K  &3,704K  &  5K & 4,420K& 12K & 10K & 12K \\
& ca-AstroPh &1,351K & 1,315K &49K &2,937K &43K & 20K &4K\\
& cit-HepPh &1,276K &1,315K &48K &8,502K  &180K &40K &34K\\
& soc-Epinions1 &1,624K &2,128K &641K &1,199K & 24K& 4K &44K\\
\hline \hline
\multirow{4}{*}{\rotatebox{90}{low-$C$}} &
as-caida20071105 &36K &74K &43K &38K  &3K & $\textless$ 1K &3K\\
& oregon1\_010331 &17K &26K &15K &17K & 1K& $\textless$ 1K  & $\textless$ 1K  \\
& web-Stanford &11,329K &14,185K &3,783K & 11,651K& 158K& 16K &114K\\
& wiki-Talk   &9,203K  &  66,740K & 41,427K  & 2,936K   & 16K & $\textless$ 1K & $\textless$ 1K  \\ \hline \hline
\end{tabular}
\end{center}
\label{tab:gm-triangle-num}
\end{table*}

\subsection{Graph Models}
\label{subsec:grmod}

Barabasi-Albert \cite{BaAl99}
proposed the Preferential Attachment (PA) model is often  associated with the ``rich get richer" concept.  In the PA model, a new node connects to a pre-specified number of vertices, where the likelihood of choosing a vertex is proportional to its degree. This procedure leads to graphs with power-law degree distributions.
 As Sala et al. \cite{SaCaWiZa10} pointed out, PA model cannot create communities in the graph and cannot generate high clustering coefficients for low degree nodes.

The Stochastic Kronecker Graph (SKG) model~\cite{LeFa07} starts with a basis matrix (typically $2\times 2$), and generates a matrix that specifies edge probabilities  by repeated Kronecker products.
 A noise-added version of SKG has been proven to generate log-normal degree distributions~\cite{CPK11}, but  the clustering coefficients of SKG graphs are very low~\cite{PiSeKo12}.

The Chung-Lu (CL) model \cite{ChLu02}  can be considered as picking a random graph among all graphs with the same degree distribution. In this model, the probability of  an edge is proportional to the product of the degrees of its endpoints, (i.e., $Pr\left(e_{ij}\right) = \frac{d_id_j}{ \left(2m\right)^2}$).  Pinar et al. showed that many properties of graphs generated by SKG and CL models are similar \cite{PiSeKo12}.

The Block Two-Level Erd\H{o}s-R\'{e}nyi (BTER) model~\cite{SeKoPi11}  is built on the observation of high-clustering coefficients and skewed degree distributions.  This model achieves high clustering coefficients by embedding communities with an Erd\H{o}s-R\'{e}nyi structure, which is typically much denser compared to the rest of the graph. Edges are added in a subsequent phase using the CL model, to satisfy the degree distribution requirements. It has been shown that BTER graphs can match many properties of real world graphs~\cite{SeKoPi11}.

The Edge Copying (EC) model ~\cite{KlKu99} imitates the topic-based communities in the Web, and generates an evolving directed graph.
When a new node arrives, it selects a random vertex and copies  a specified number of links.
For the EC model, both the in-degree and out-degree distributions follow power laws. However, the model does not create back links and does not generate many triangles.

The Forest Fire (FF) model~\cite{LeKlFa05} combines the PA model~\cite{BaAl99} to obtain a heavy-tailed degree distribution, the EC model ~\cite{KlKu99} to obtain communities, and community guided attachment for densification. The model has a forward burning parameter $p$. In the FF model, a node $v$ arrives and chooses an ambassador node $w$ randomly and connected to each neighbor of $w$ with probability $p$.  This process  is repeated for each new vertex  $v$ connects to.
Note that EC model only copies the links of a node, it does not hop to the neighbor of the node.

\subsection{Fitting Graph Models to Real Networks}
\label{subsec:fit-gm}

To check whether graph models can reproduce the triangle degree behavior of the real networks, we fit FF, BTER, EC, CL, PA, and SKG models to the real networks listed in \Tab{graphs}.

The inputs to the The PA model are the number of nodes $n$ and the number of edges $k$ created by each node. We pick $k$  by rounding the density of the graph in order to match the number of edges in the real networks.

The BTER model takes the degree distribution of the real networks. The connectivity per block is computed by a table look-up on the average clustering coefficient per degree $C_d$ plot.

For the Chung Lu model, we provide the degree distribution of the real networks as an input to the model.

For the Forest Fire model, we provide the number of nodes $n$ and the forward burning probability $p$. We match the generated graph models to the number of edges in the real networks. For each target graph, we search a range of values by incrementing $p$ value by $\delta p= 0.001$ in range [0-1] to find the best model giving the similar number of edges to the original network.

The EC model takes the number of nodes $n$, number of edges each node creates $k$, and the edge copying probability $p$ which is calculated based on power-law degree distribution. Power-law degree slope is $\frac{1}{1-p}$. We pick $k$ by rounding up the density of graph.
\smallskip

To generate the SKG model, we compute the parameters of the initiator matrix using the Kronfit algorithm \cite{LeChKlFa10}. The size of the final adjacency matrix is $2^{\lceil \log(n) \rceil}$ where $n$ is the number of nodes in the real graph.

Before running our experiments, we symmetrize by treating each edge as undirected  and remove self-links for all the generated graphs.

\subsection{Triangle Analysis in Graph Models}
\label{subsec:gm-tri-an}

After fitting different graph models on the real networks, we enumerate triangles in each randomly generated network using Cohen's algorithm~\cite{Co09}. We analyze the triangle behaviors in these random graphs in different aspects.

\smallskip
{\bf The numbers of triangles:} \textit{None of the graph models capture the triangle numbers for \emph{both} \emph{high-$C$} and \emph{low-$C$} networks. }

\smallskip
Some models are good at generating similar number of triangles for \emph{high-$C$} networks, some of them are good at for \emph{low-$C$} networks, but none of them is good at both.
 The number of triangles generated by different graph models for each target graph is listed in \Tab{gm-triangle-num}. Graph models behave differently in \emph{high-$C$} and \emph{low-$C$} networks in terms of generating triangles.

In \emph{high-$C$} networks, BTER matches the number of triangles in the original graph better than the rest of the models. FF creates significantly more triangles than the original number of triangles. CL generates significantly less triangles than the original number of triangles. For networks with the high clustering coefficient $C$ and high density $\rho$, CL is not a good choice.

In \emph{low-$C$} networks, BTER generates more triangles than the original number of triangles. As a matter of fact, when the $C$ value is very small, BTER generates even more triangles than the original number of triangles. Particularly for wiki-Talk, it generates many more triangles. CL is also generating many more triangles for wiki-Talk. CL is generating less triangles for web-Stanford but it generates similar number of triangles for oregon1\_010331 and as-caida20071105. FF generates similar numbers of triangles as in the original networks except wiki-Talk. It seems that none of the models  can  reproduce the structure of the  wiki-Talk graph.

EC, PA, and SKG generate significantly less triangles than the original triangle numbers. These models also cannot reach the average clustering coefficient per degree for any of the network. Therefore, we will not include them for the rest of the plots.

\begin{figure*}[p]
  \centering
  \subfloat[ca-AstroPh]{\label{fig:gm-d1-d2-ca-Astro}
    \includegraphics[width=.25\textwidth,trim=0 0 0 0]{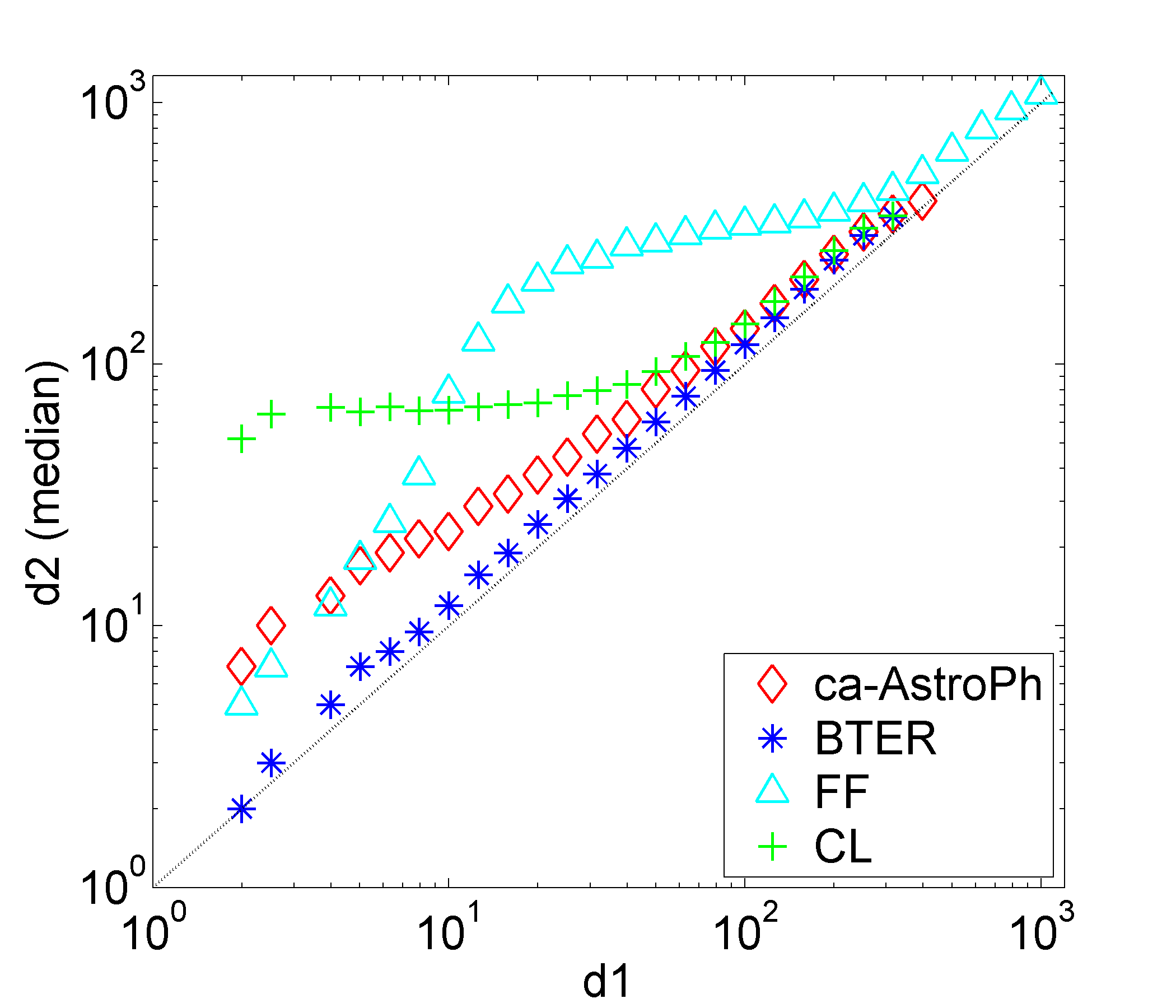}
  }
  \subfloat[cit-HepPh]{\label{fig:gm-d1-d2-cit-HepPh}
    \includegraphics[width=.25\textwidth,trim=0 0 0 0]{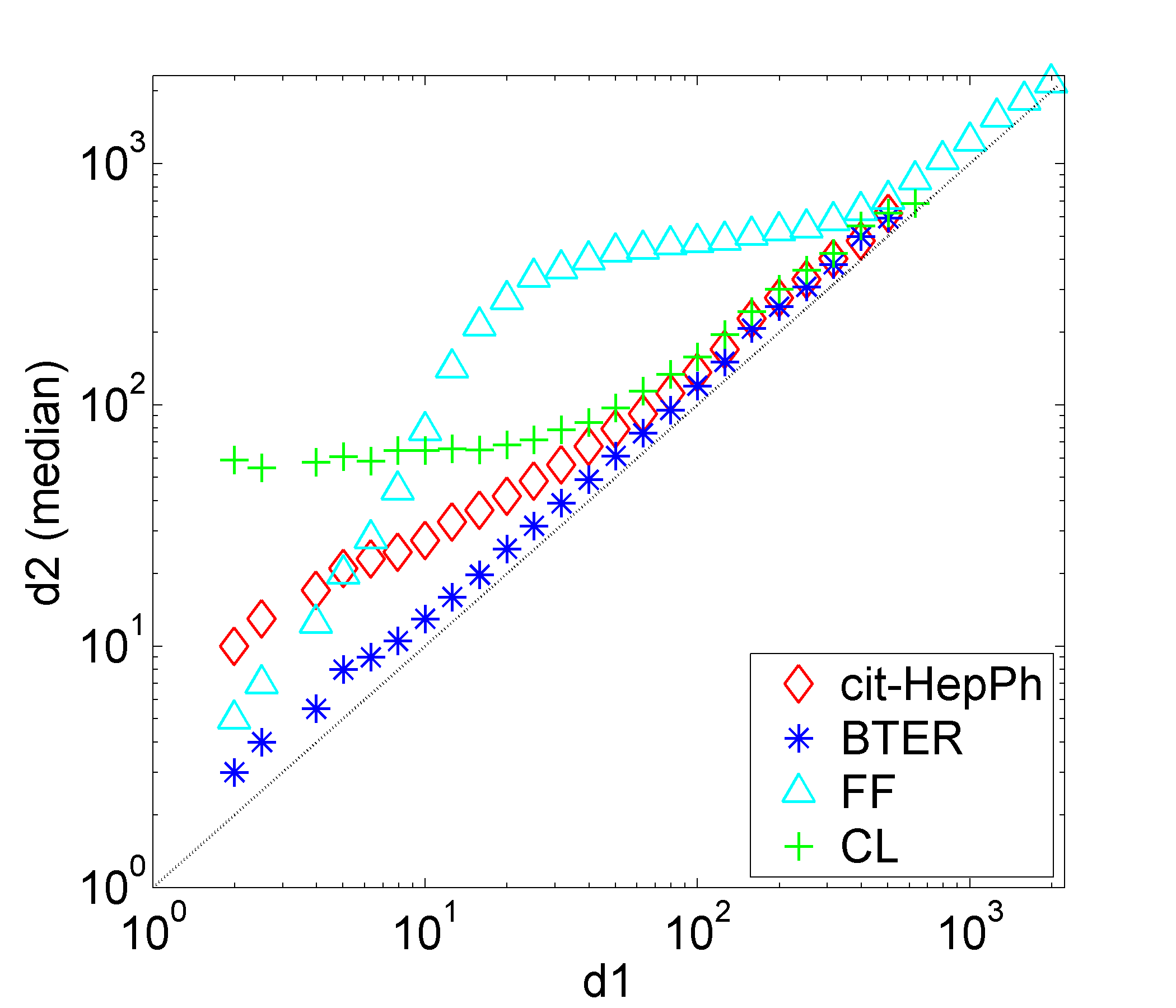}
  }
  \subfloat[Amazon0312]{\label{fig:gm-d1-d2-Amazon0312}
    \includegraphics[width=.25\textwidth,trim=0 0 0 0]{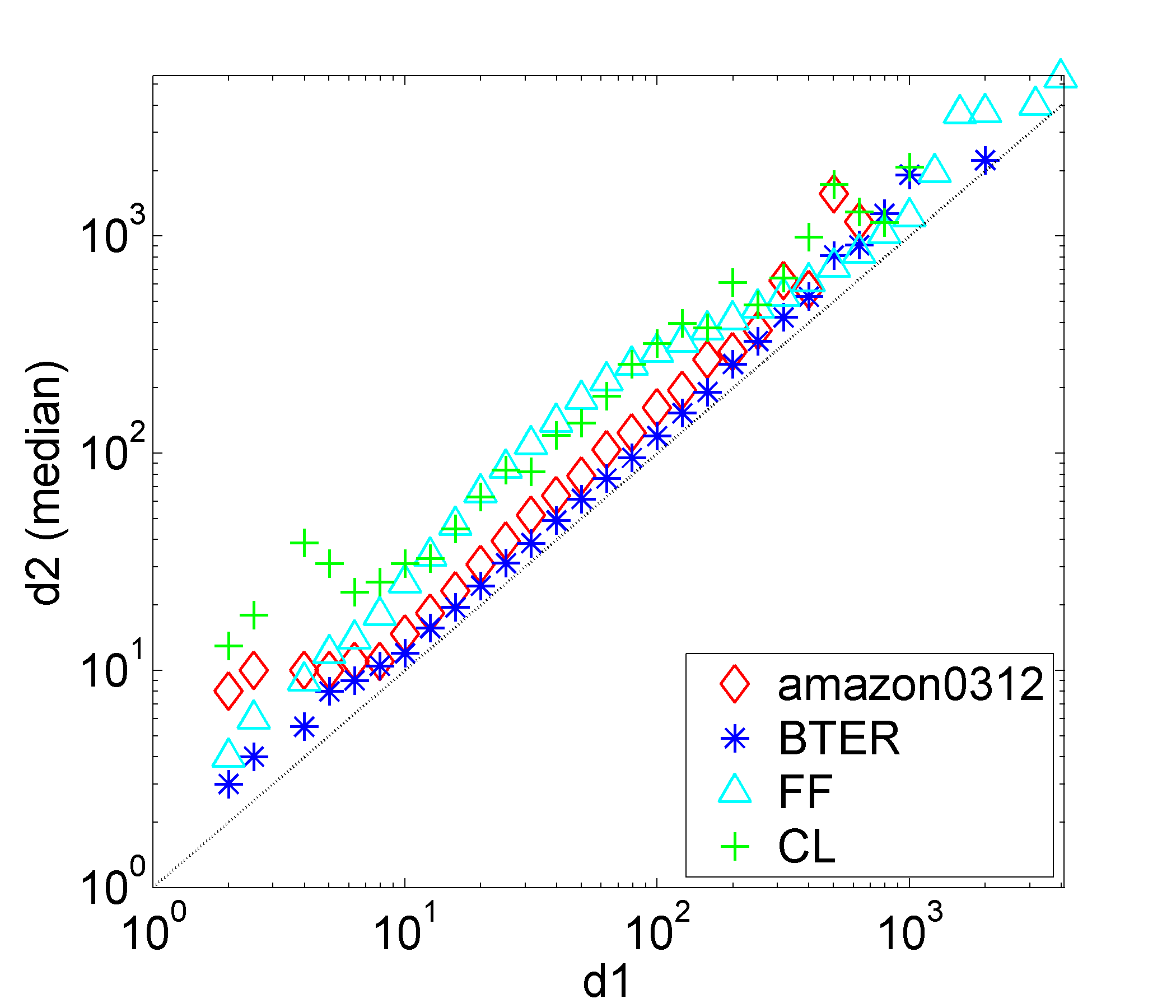}
  }
  \subfloat[Soc-Epinions]{\label{fig:gm-d1-d2-Soc-Epinions}
    \includegraphics[width=.25\textwidth,trim=0 0 0 0]{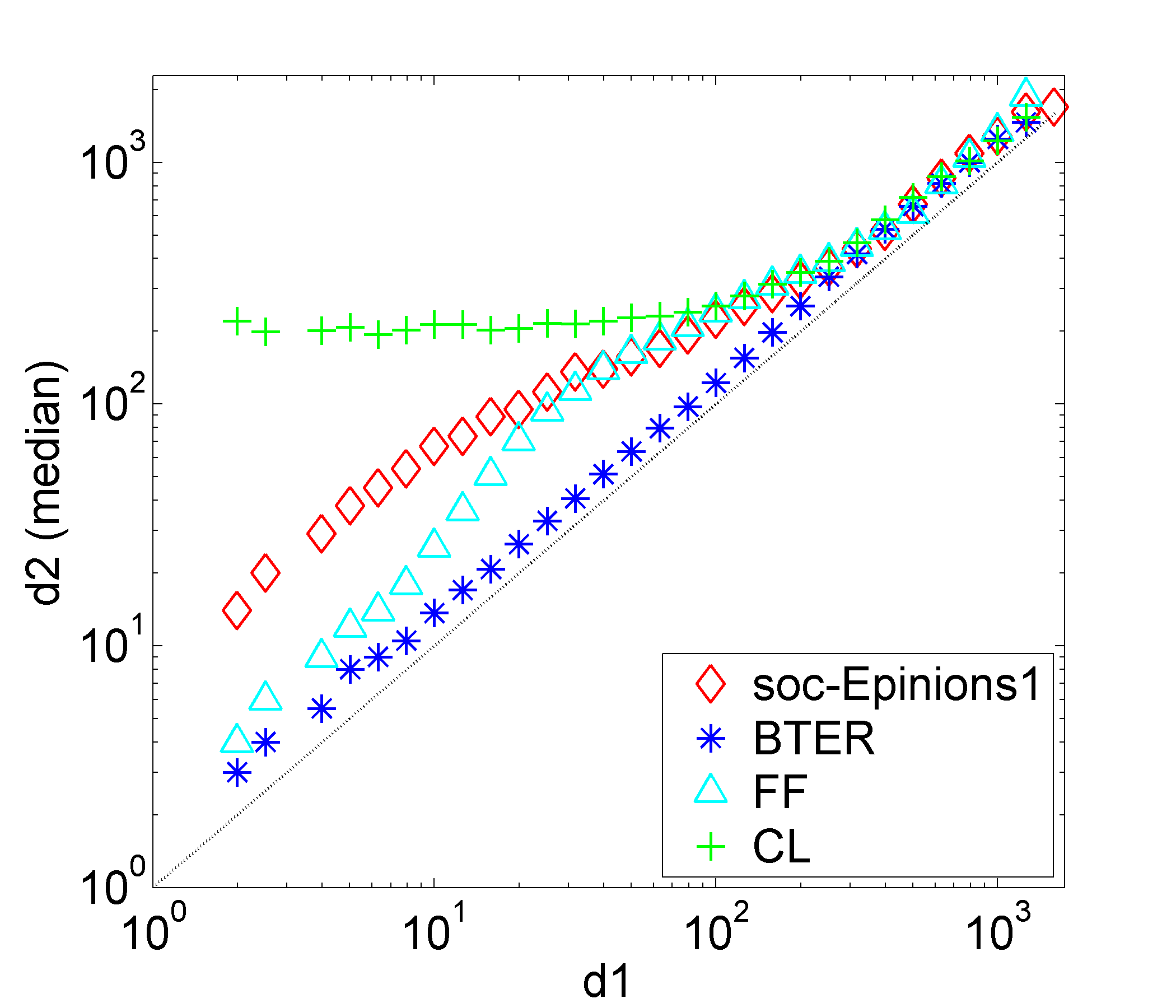}
  }
  \\
  \subfloat[wiki-Talk]{\label{fig:gm-d1-d2-wiki-Talk}
    \includegraphics[width=.25\textwidth,trim=0 0 0 0]{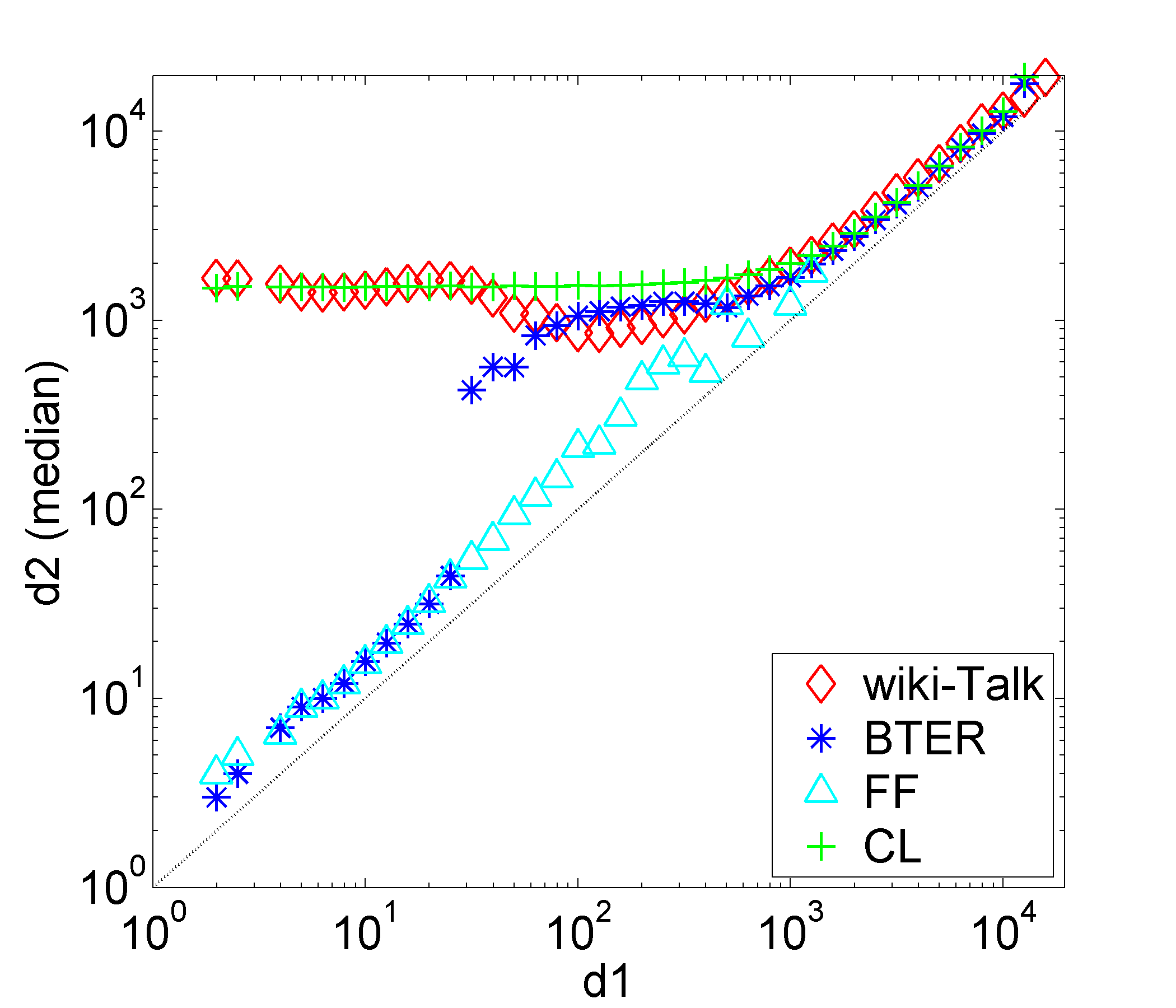}
  }
  \subfloat[as-caida20071105]{\label{fig:gm-d1-d2-as-caida}
    \includegraphics[width=.25\textwidth,trim=0 0 0 0]{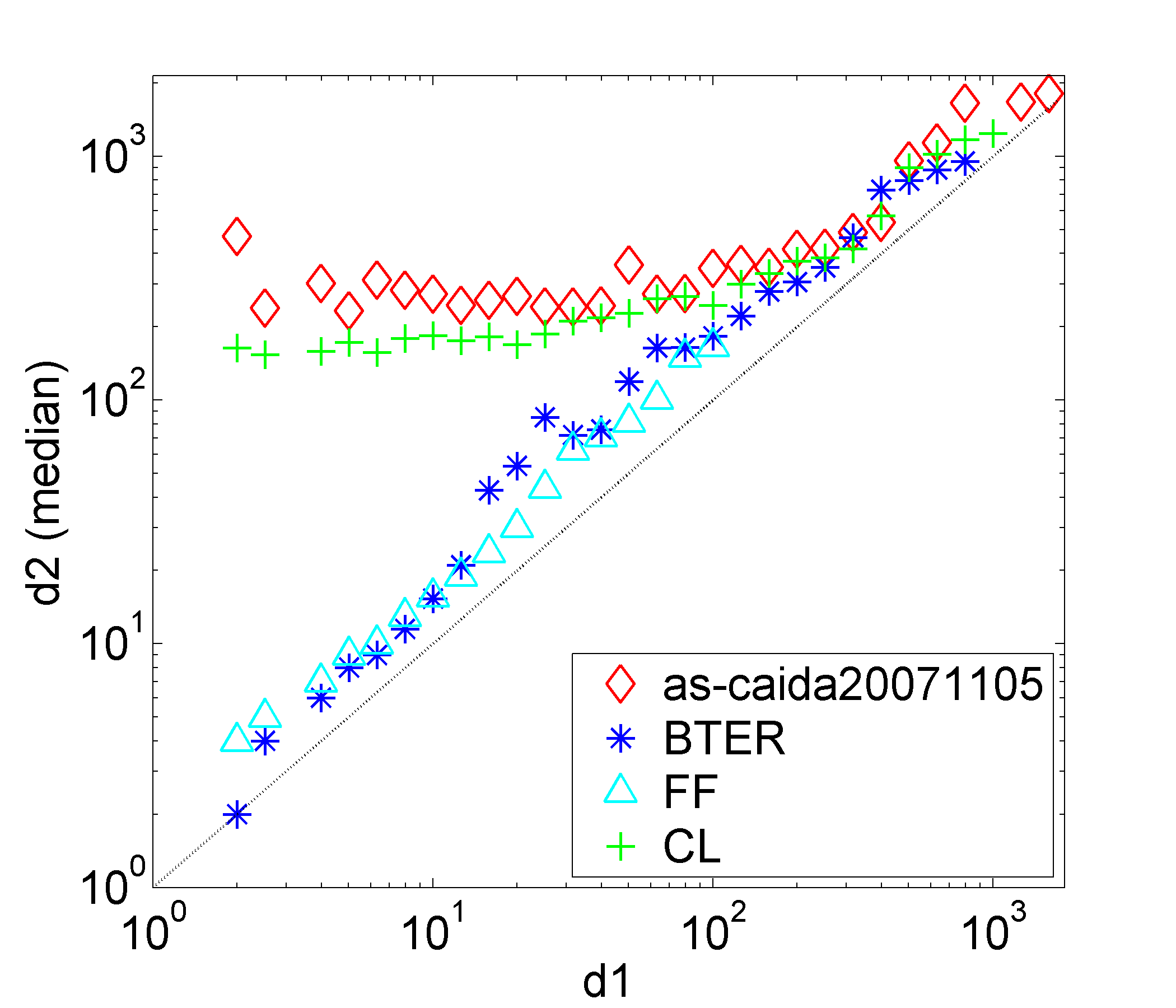}
  }
  \subfloat[oregon1\_010331]{\label{fig:gm-d1-d2-oregon1}
    \includegraphics[width=.25\textwidth,trim=0 0 0 0]{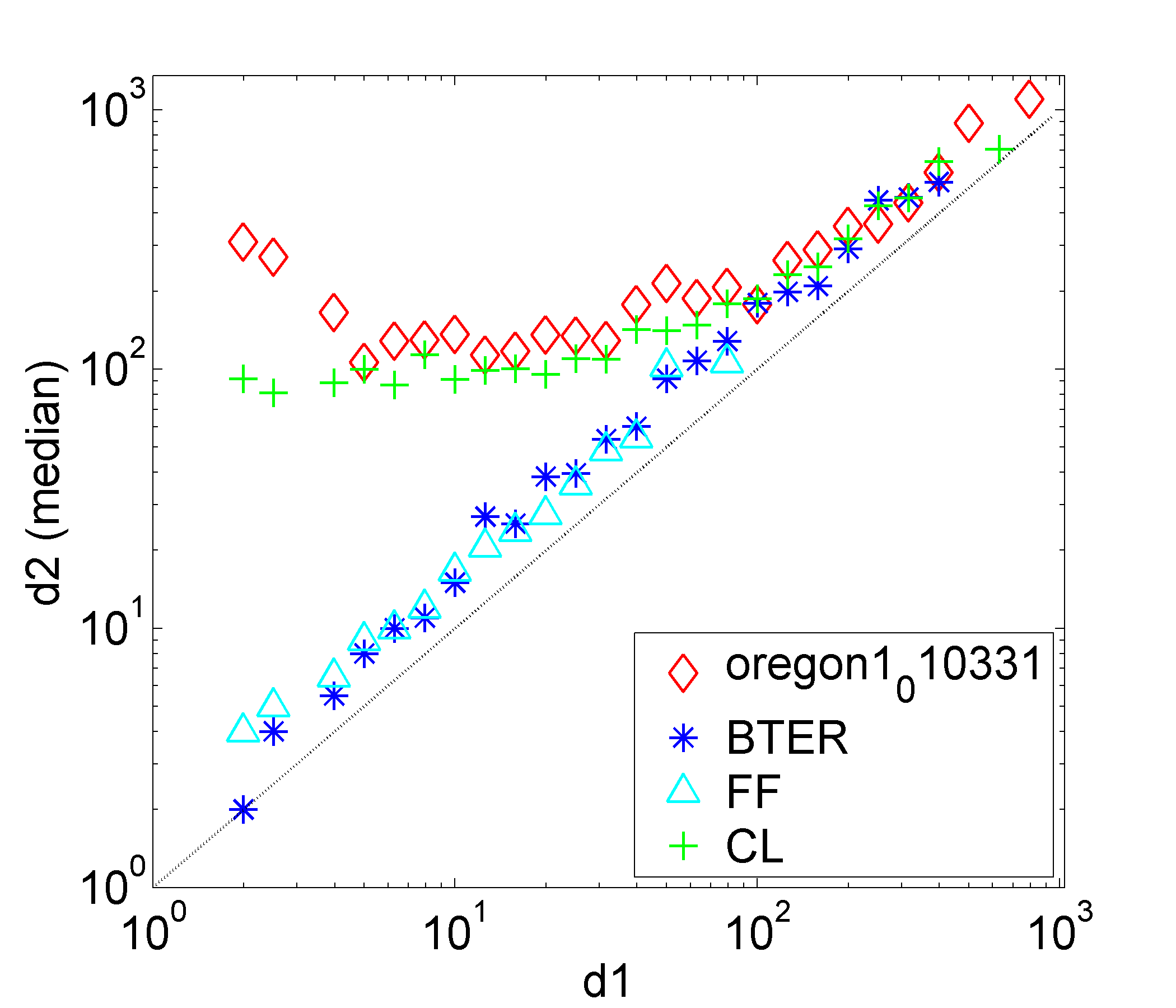}
  }
  \subfloat[web-Stanford]{\label{fig:gm-d1-d2-web-Stanford}
    \includegraphics[width=.25\textwidth,trim=0 0 0 0]{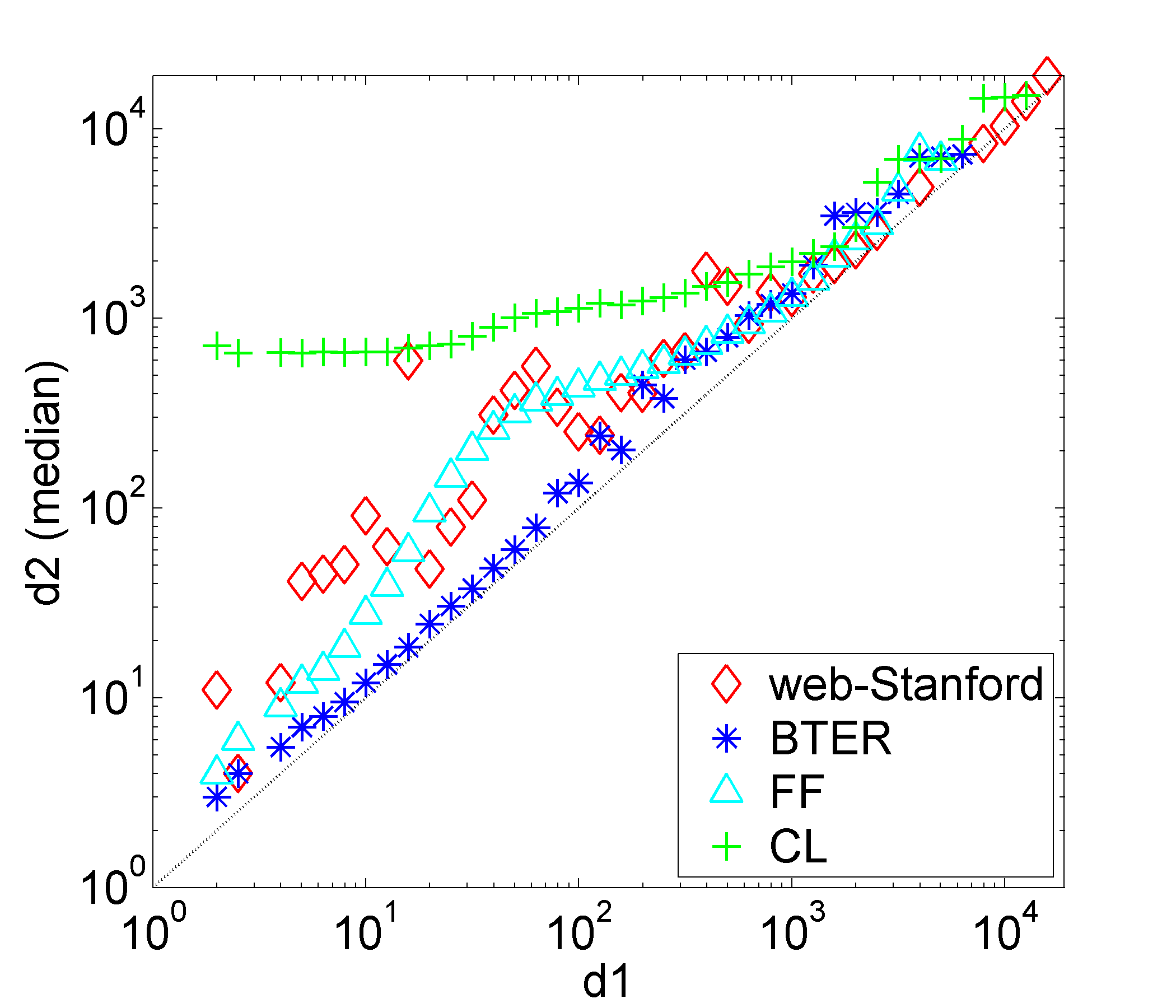}
  }
    \caption{Triangle degree relations between $\da{i}$ and $\db{i}$ in the generated graph models }
    \label{fig:gm-d1-vs-d2}
\end{figure*}

\begin{figure*}[p]
  \centering
 \subfloat[ca-AstroPh]{\label{fig:gm-d1-d3-ca-Astro}
    \includegraphics[width=.25\textwidth,trim=0 0 0 0]{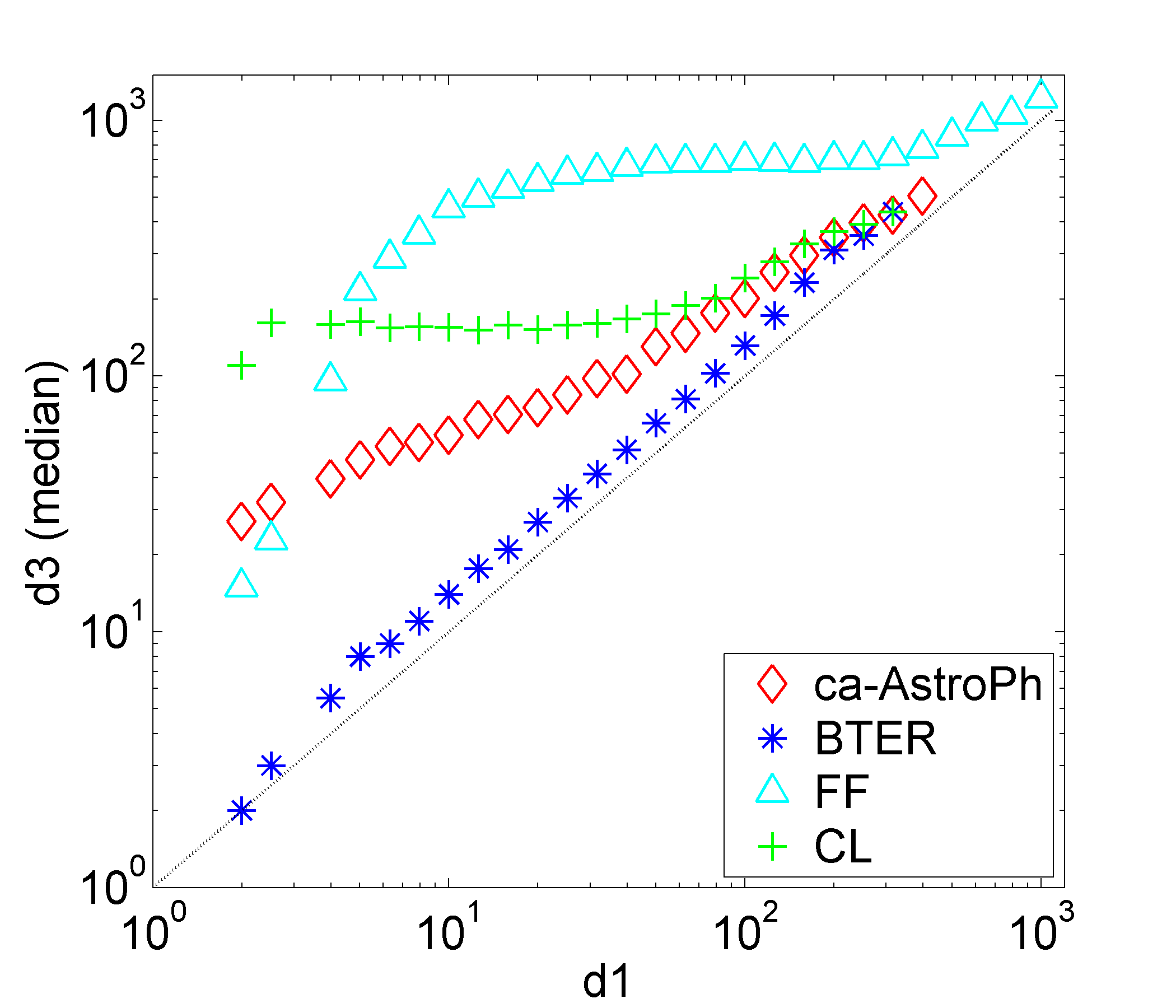}
  }
  \subfloat[cit-HepPh]{\label{fig:gm-d1-d3-cit-HepPh}
    \includegraphics[width=.25\textwidth,trim=0 0 0 0]{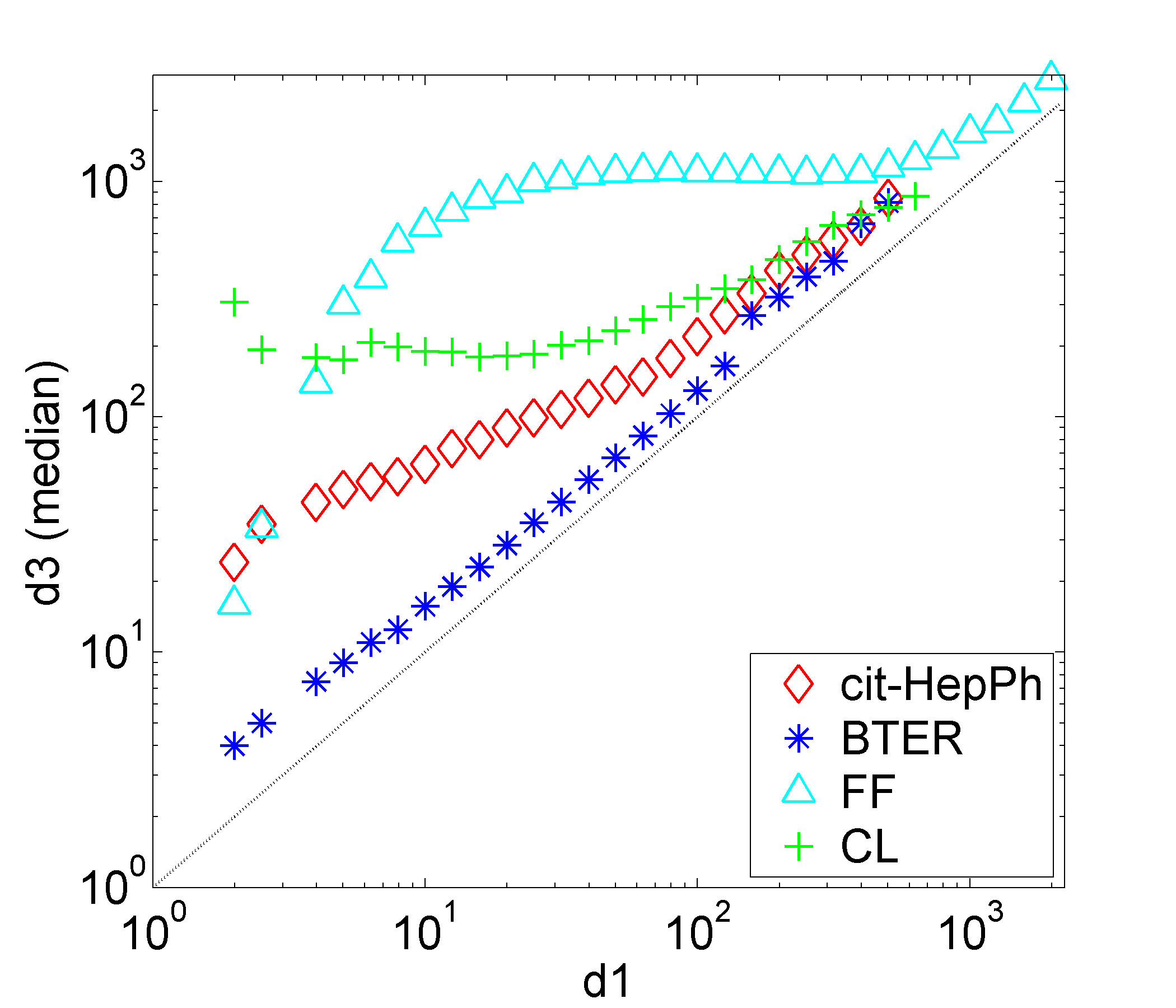}
  }
  \subfloat[Amazon0312]{\label{fig:gm-d1-d3-Amazon0312}
    \includegraphics[width=.25\textwidth,trim=0 0 0 0]{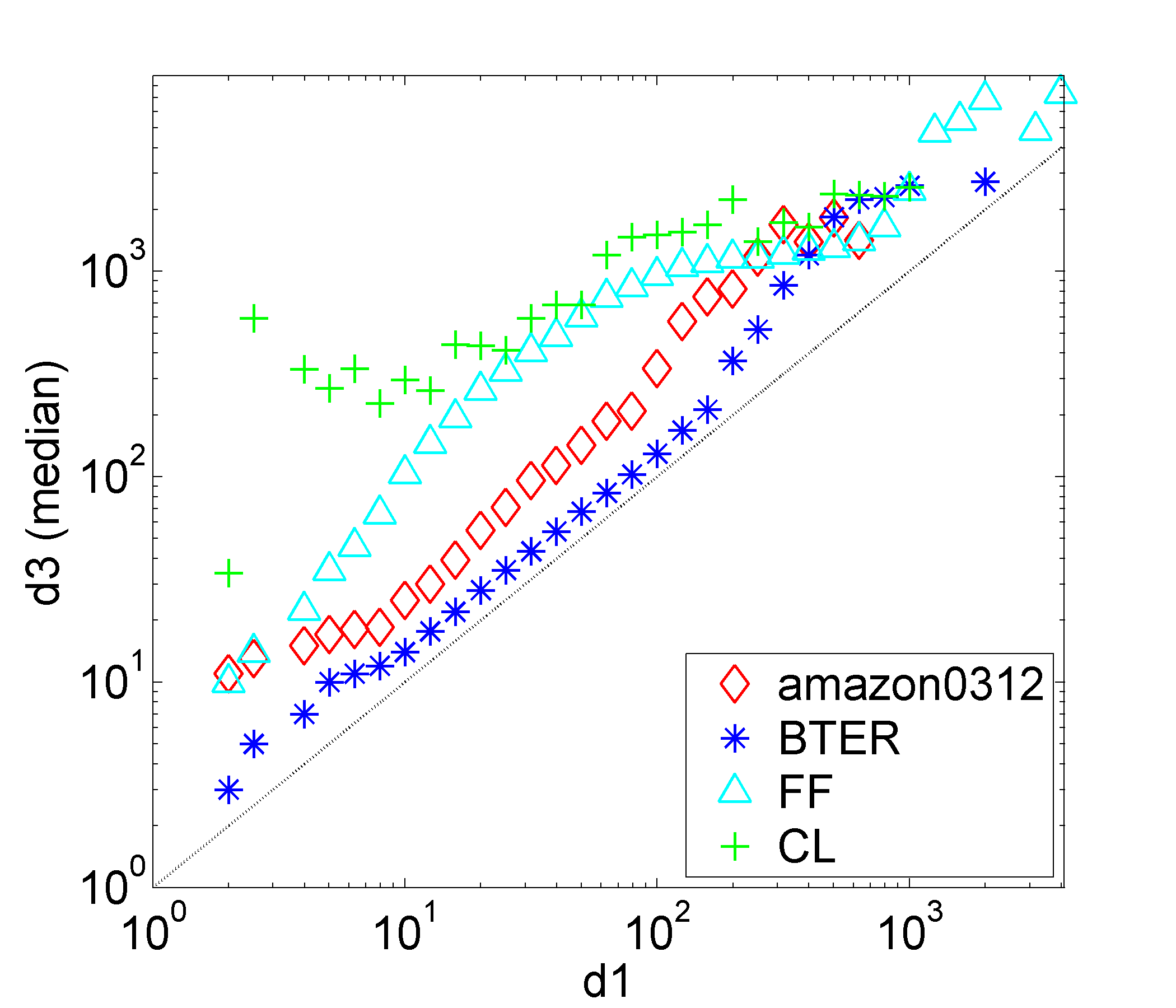}
  }
  \subfloat[Soc-Epinions]{\label{fig:gm-d1-d3-Soc-Epinions}
    \includegraphics[width=.25\textwidth,trim=0 0 0 0]{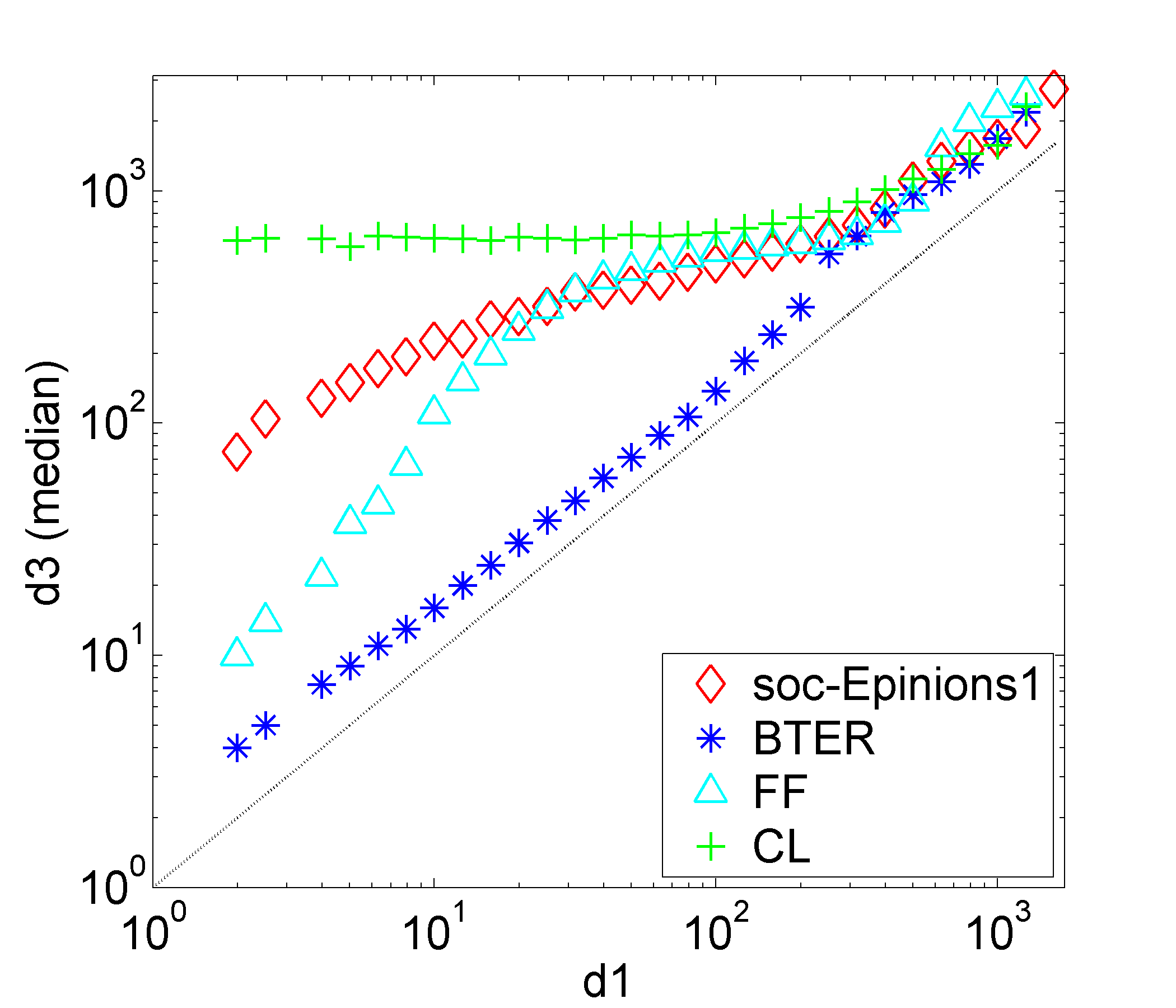}
  }
  \\
  \subfloat[wiki-Talk]{\label{fig:gm-d1-d3-wiki-Talk}
    \includegraphics[width=.25\textwidth,trim=0 0 0 0]{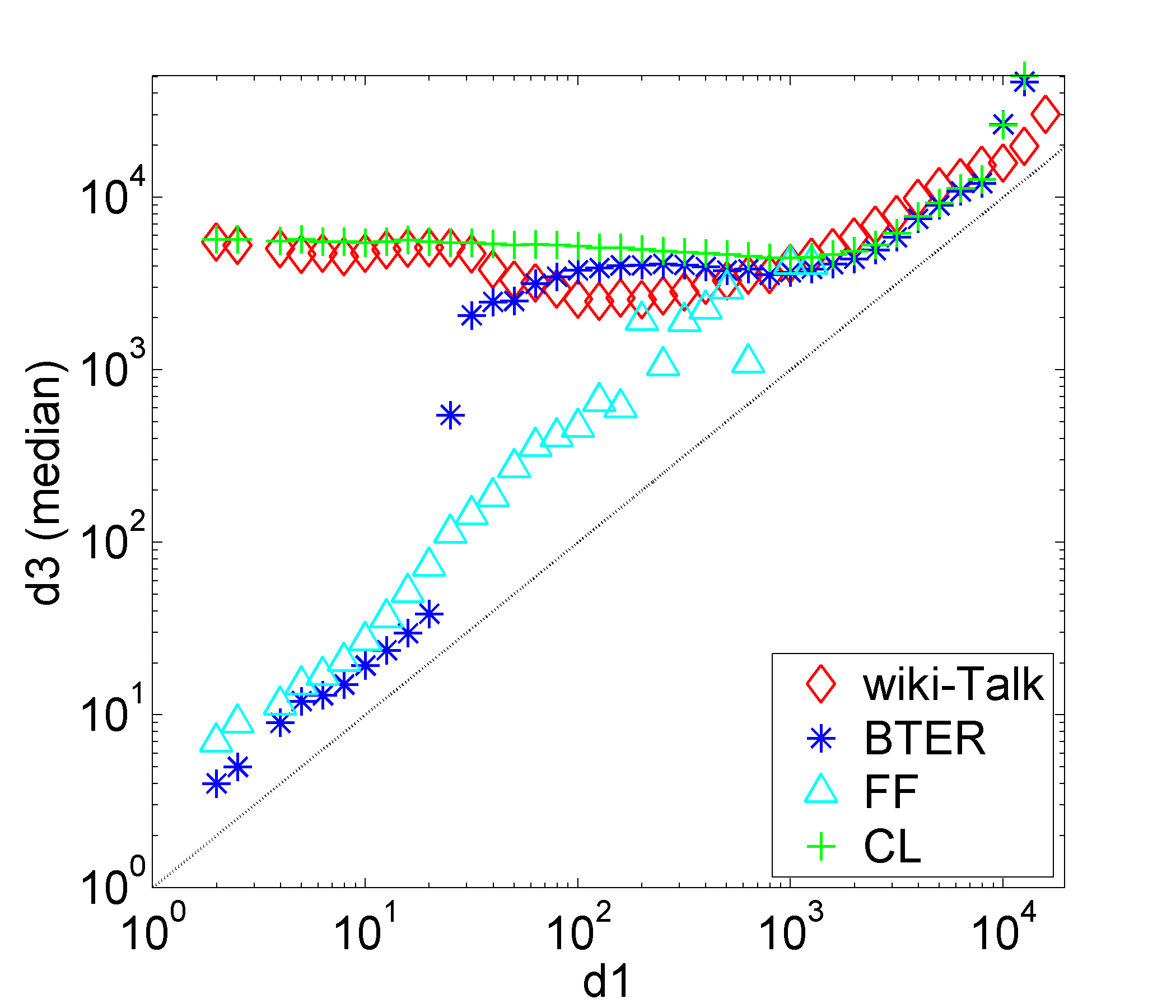}
  }
  \subfloat[as-caida20071105]{\label{fig:gm-d1-d3-as-caida}
    \includegraphics[width=.25\textwidth,trim=0 0 0 0]{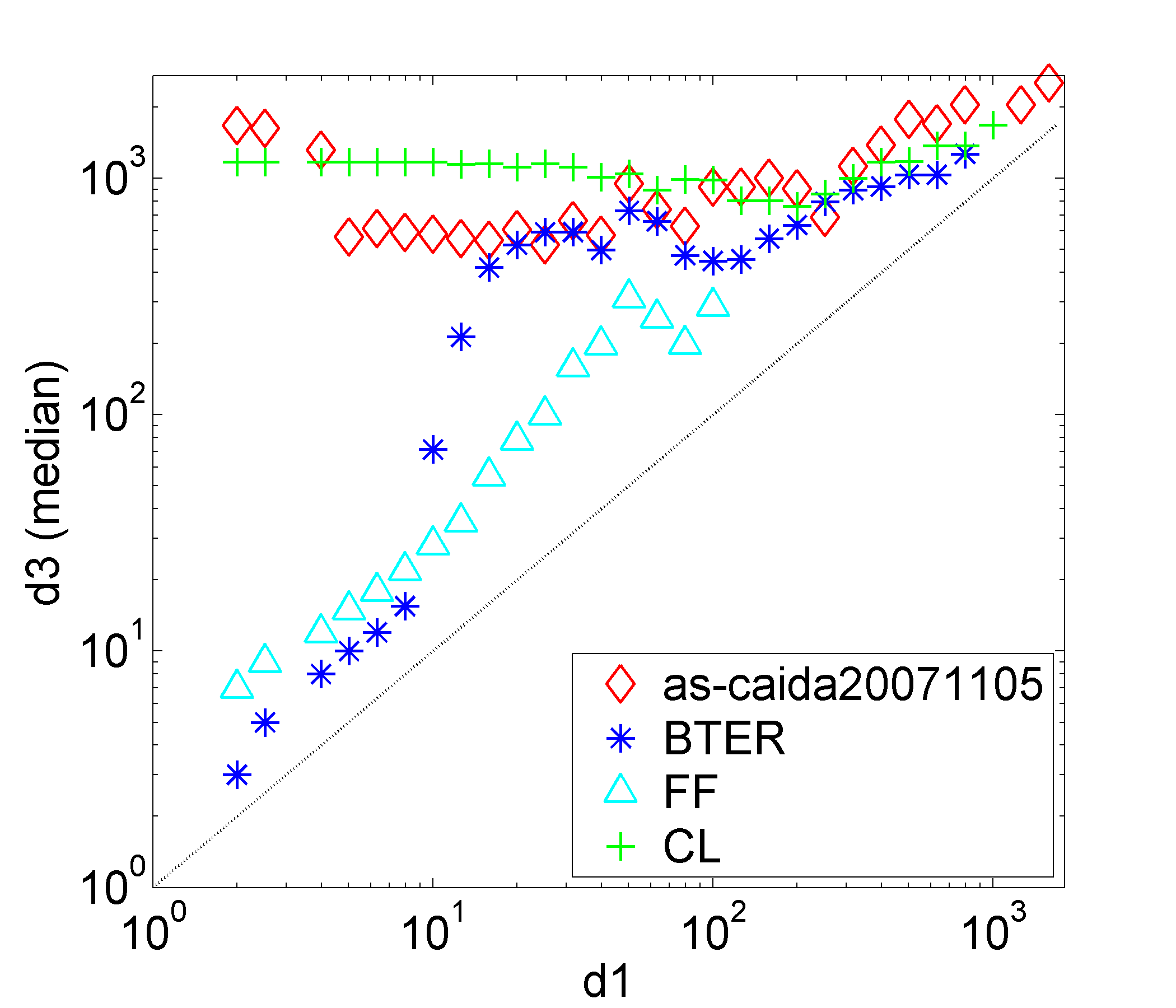}
  }
  \subfloat[oregon1\_010331]{\label{fig:gm-d1-d3-oregon1}
    \includegraphics[width=.25\textwidth,trim=0 0 0 0]{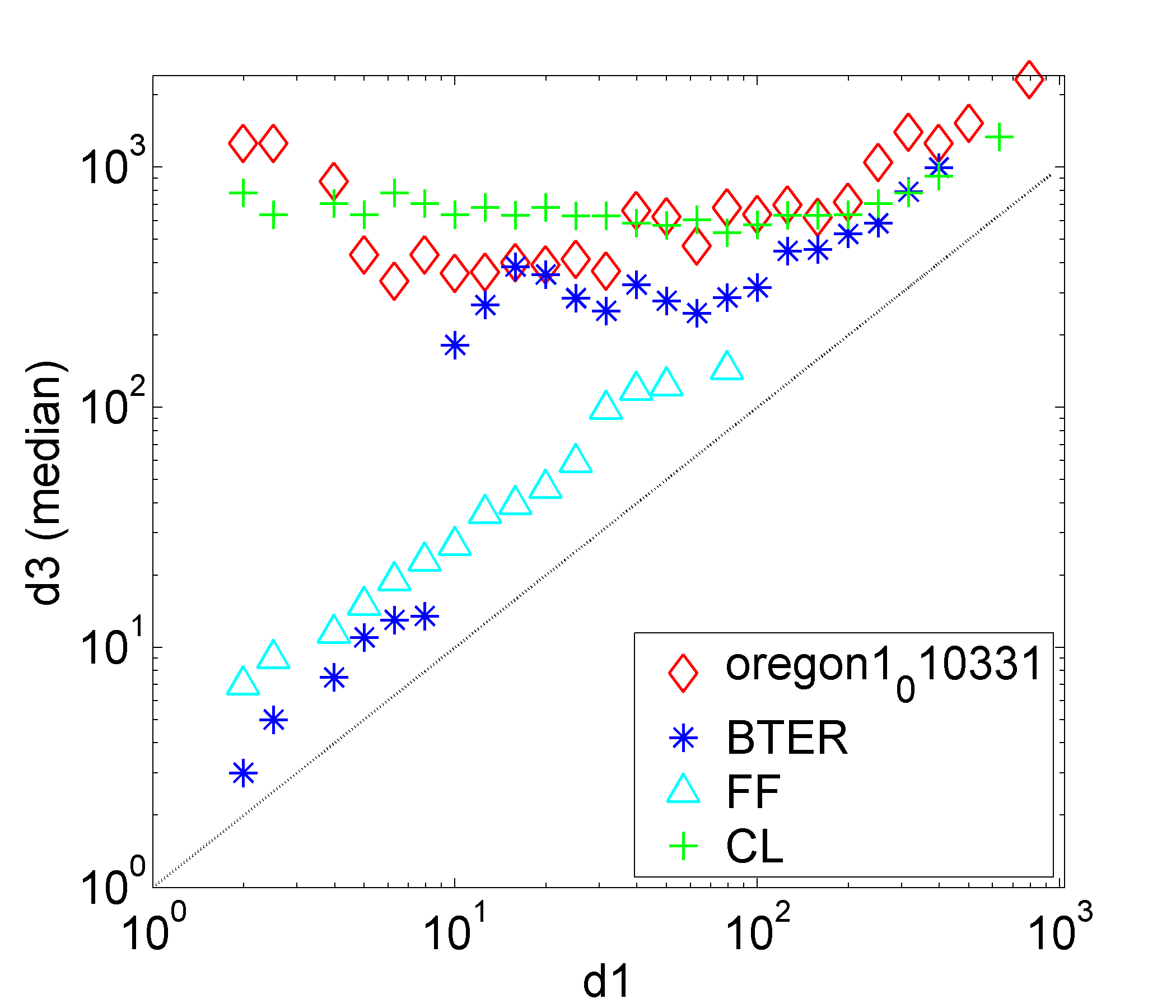}
  }
  \subfloat[web-Stanford]{\label{fig:gm-d1-d3-web-Stanford}
    \includegraphics[width=.25\textwidth,trim=0 0 0 0]{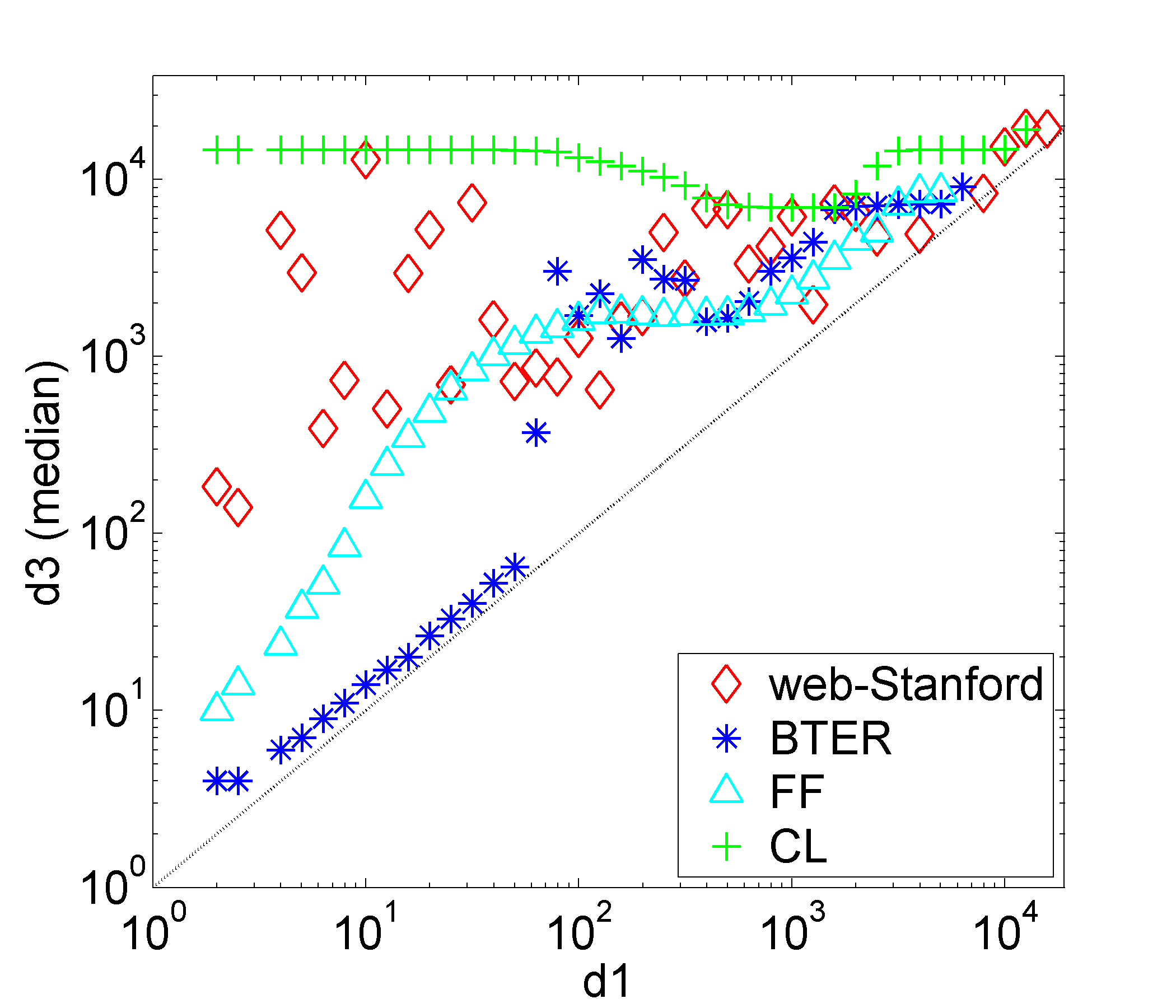}
  }
  \caption{Triangle degree relations between $\da{i}$ and $\dc{i}$ in the generated graph models }
  \label{fig:gm-d1-vs-d3}
\end{figure*}

\smallskip
\textbf{Degree Relations:} \textit{Models generate either homogenous or heterogonous triangles for graphs.}
\smallskip
In \Fig{gm-d1-vs-d2}, we show the relation between $\da{i}$ and  $\db{i}$ and in \Fig{gm-d1-vs-d3}, we show the relation between $\da{i}$ vs $\dc{i}$ for the real graphs as well as their modeled counterparts.

CL produces \emph{heterogeneous} triangles for both \emph{high-$C$} and \emph{low-$C$} networks in both \Fig{gm-d1-vs-d2} and \Fig{gm-d1-vs-d3}. For \emph{low-$C$} networks, it is very intriguing that CL graphs are generating the right \emph{type} of triangles, since the $\da{i}$  vs $\db{i}$ and $\da{i}$  vs $\dc{i}$ plots follow the true graph quite well. But we feel that this indicates that \emph{low-$C$} networks have a CL flavor to them (i.e., triangles are random).

BTER generates homogeneous triangles for both \emph{high-$C$} and \emph{low-$C$} networks.
For \emph{high-$C$} networks, BTER matches the $\da{i}$ vs $\db{i}$ but generates lower $\dc{i}$ values than original $\dc{i}$ values.

FF behaves like BTER for \emph{low-$C$} networks in $\da{i}$ vs $\db{i}$ and $\da{i}$  vs $\dc{i}$ plots. Low degree $\da{i}$ values cannot connect to high degree vertices. For \emph{high-$C$} networks, there is a steep increase after $\da{i}$ reaches $10$ in some of the networks in both $\da{i}$ vs $\db{i}$ and $\da{i}$  vs $\dc{i}$ plots. Distance between FF's $\db{i}$ and original $\db{i}$ and FF's $\dc{i}$ and original $\dc{i}$ is considerable large. FF also reaches higher $\da{i}$ values than the original $\da{i}$ values.

\begin{figure*}[htb]
    \centering
     \subfloat[ca-AstroPh]{\label{fig:degs-gm-ca-Astro}
    \includegraphics[width=.25\textwidth,trim=0 0 0 0]{./d1Vsd2Vsd3ca-AstroPh.png}
  }
    \subfloat[ca-AstroPh BTER]{\label{fig:degs-gm-Ca-AstroPh-bter}%
	\includegraphics[width=.25\textwidth,trim=0 0 0 0]%
        {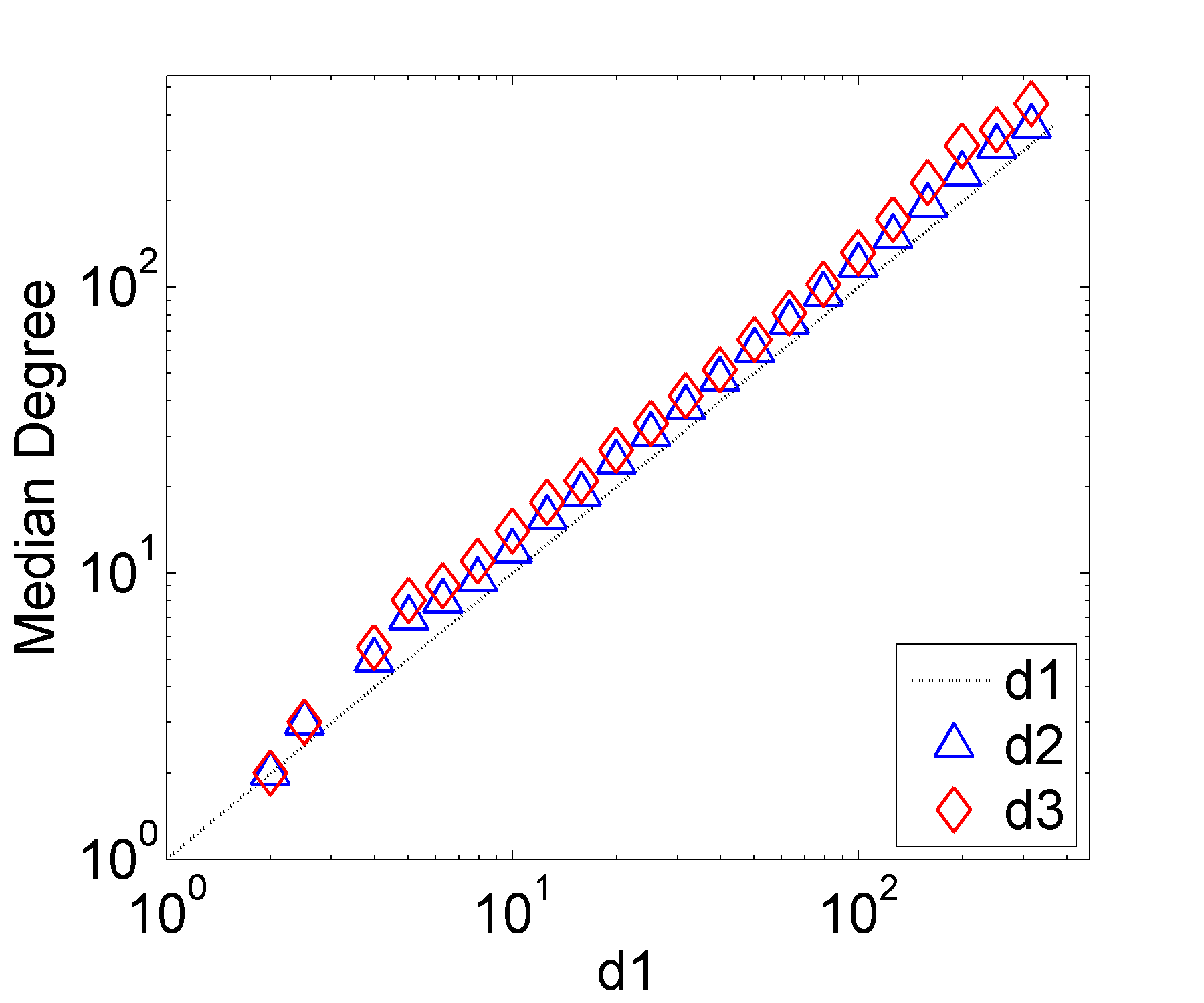}}
    \subfloat[ca-AstroPh FF]{\label{fig:degs-gm-Ca-AstroPh-ff}%
	\includegraphics[width=.25\textwidth,trim=0 0 0 0]%
        {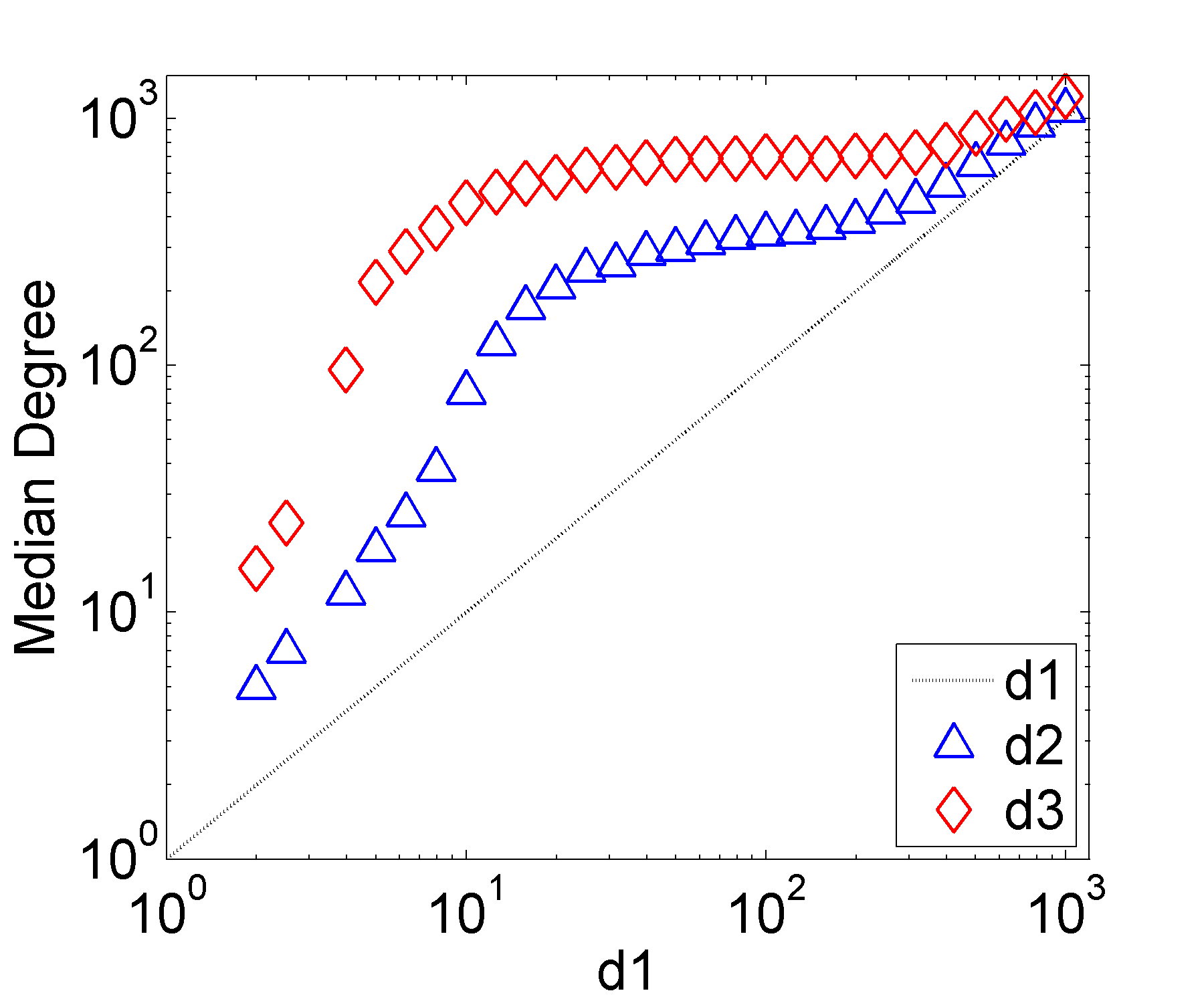}}
    \subfloat[ca-AstroPh CL]{\label{fig:degs-gm-Ca-AstroPh-cl}%
	\includegraphics[width=.25\textwidth,trim=0 0 0 0]%
        {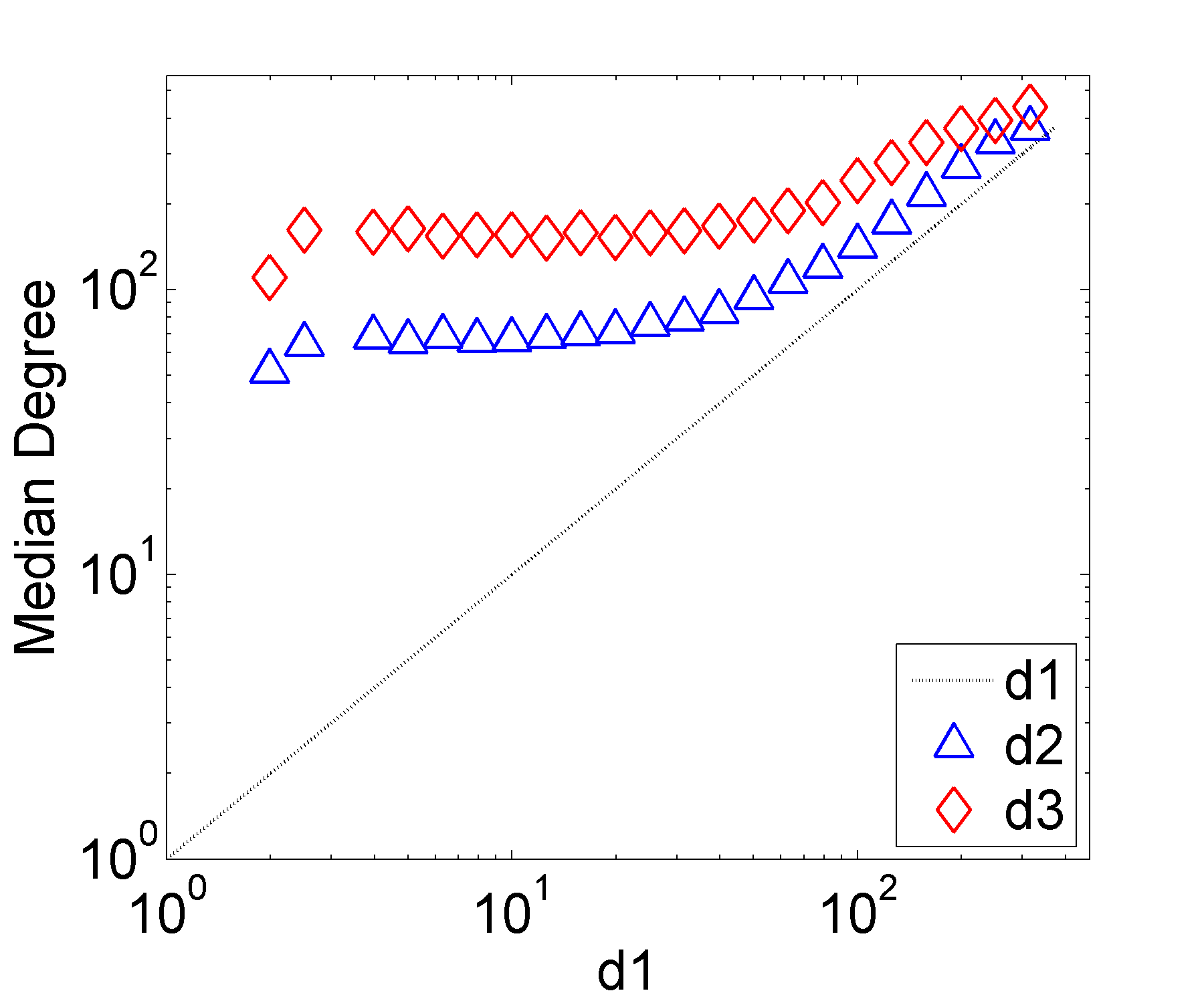}}
    \\
    \subfloat[web-Stanford]{\label{fig:degs-gm-web-Stanford}
    \includegraphics[width=.25\textwidth,trim=0 0 0 0]{./d1Vsd2Vsd3web-Stanford.png}
    }
    \subfloat[web-Stanford BTER]{\label{fig:degs-gm-web-stanford-bter}%
	\includegraphics[width=.25\textwidth,trim=0 0 0 0]%
        {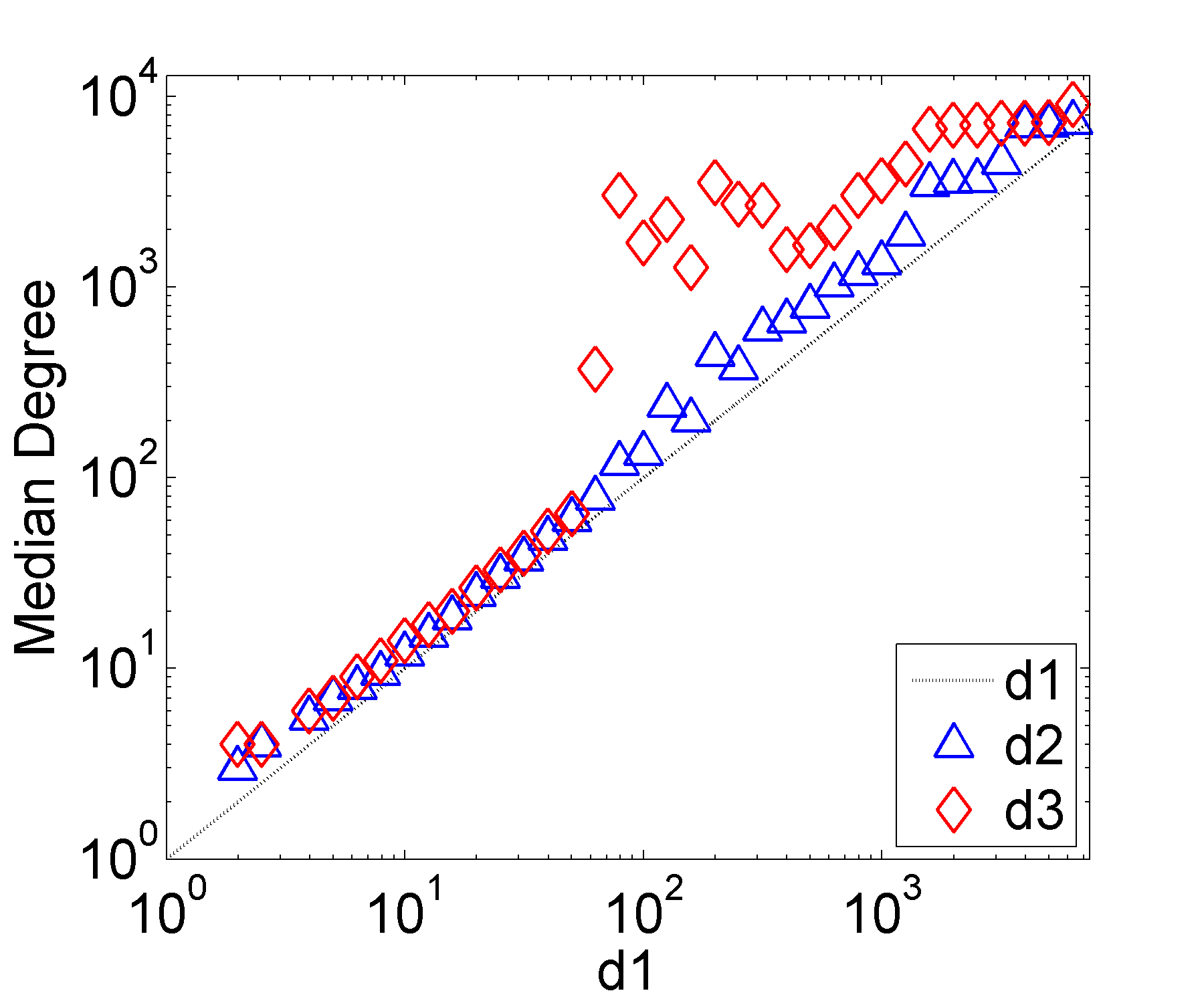}}
    \subfloat[web-Stanford FF]{\label{fig:degs-gm-web-stanford-ff}%
	\includegraphics[width=.25\textwidth,trim=0 0 0 0]%
        {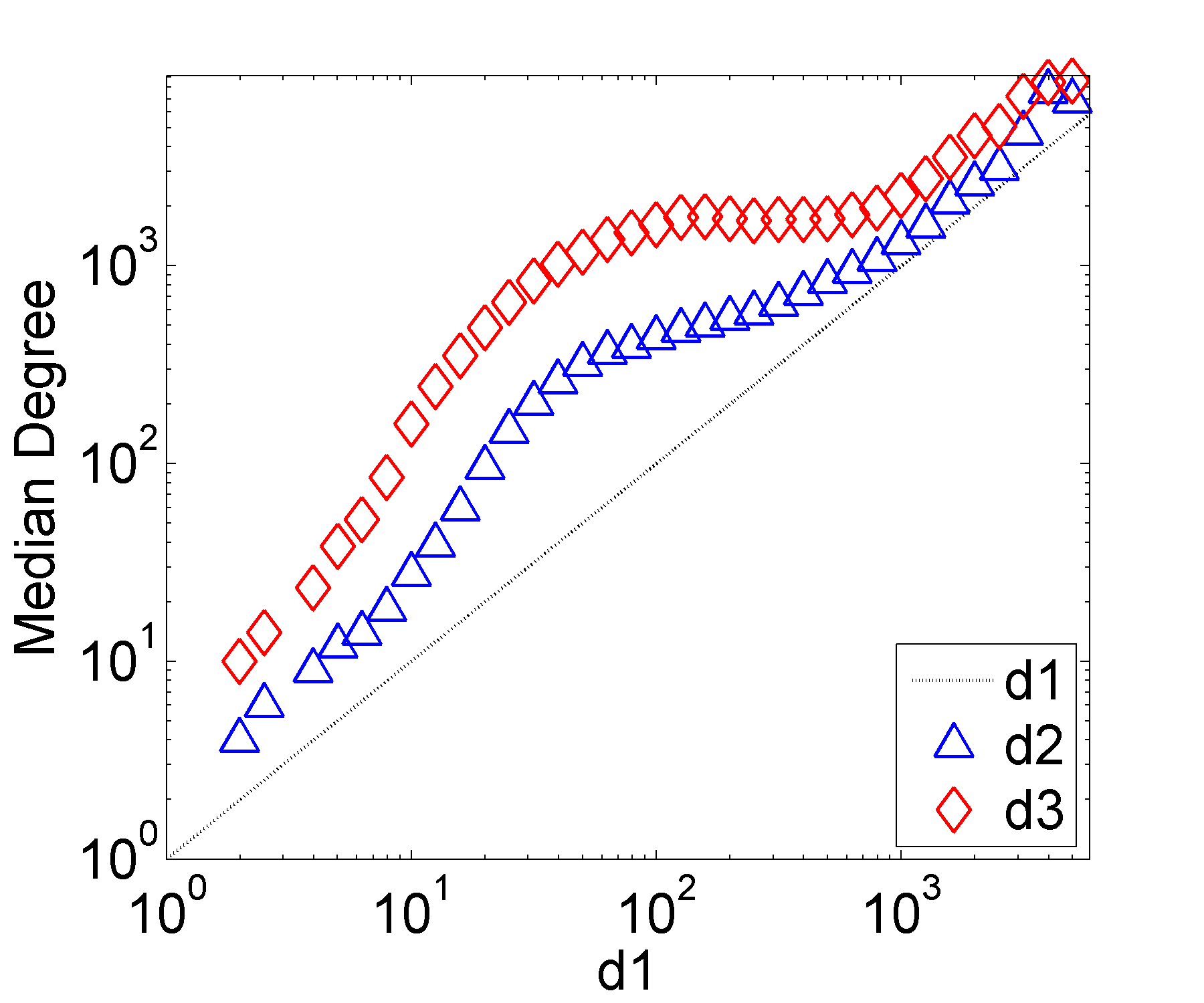}}
    \subfloat[web-Stanford CL]{\label{fig:degs-gm-web-stanford-cl}%
	\includegraphics[width=.25\textwidth,trim=0 0 0 0]%
        {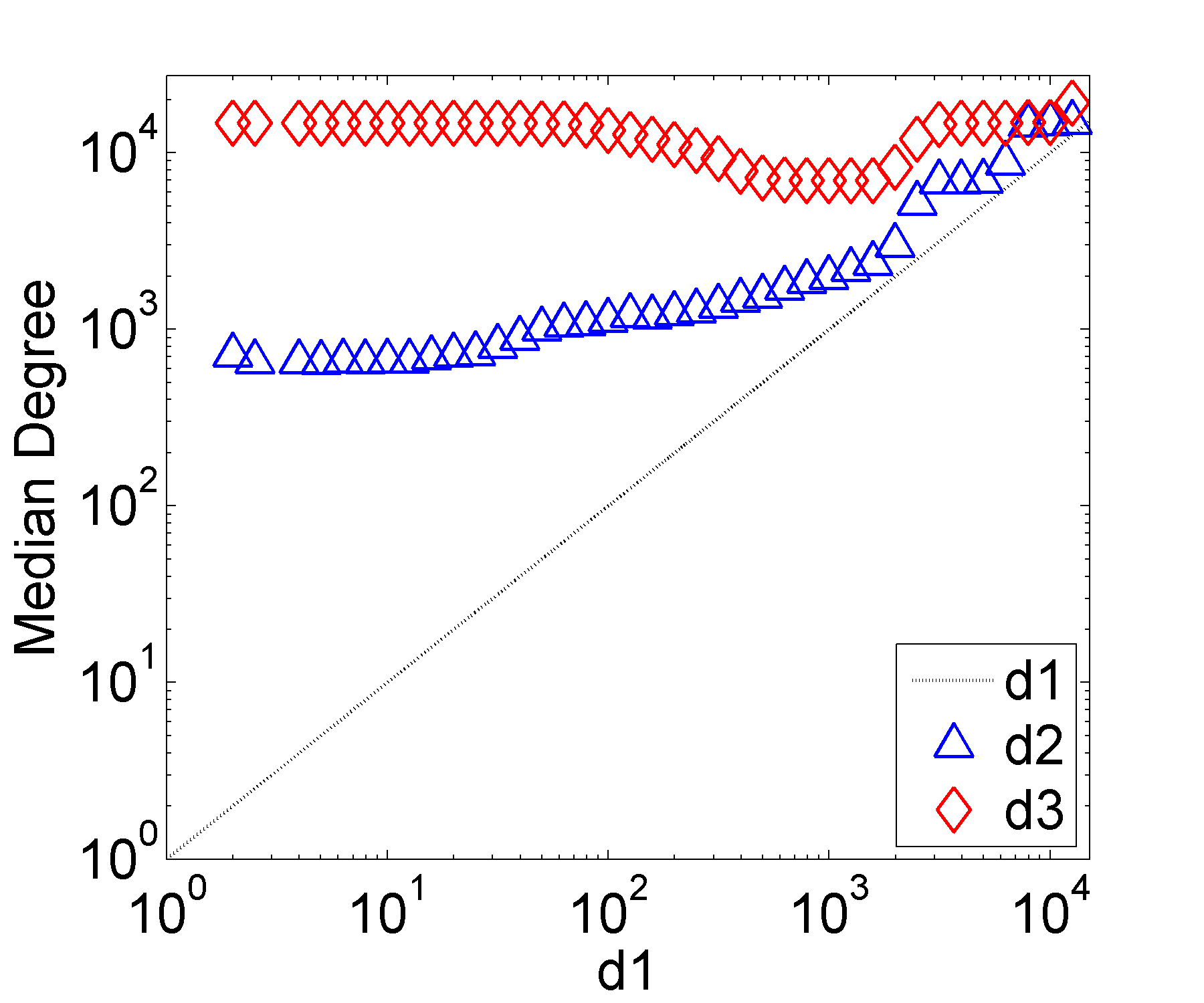}}
    \caption{Triangle degree-comparison plots for the randomly generated graphs}
    \label{fig:gm-degs}
\end{figure*}

\begin{figure*}[htb]
    \centering
    \subfloat[amazon0312]{\label{fig:CDF-gm-Amazon0312}
    \includegraphics[width=.25\textwidth,trim=0 0 0 0]{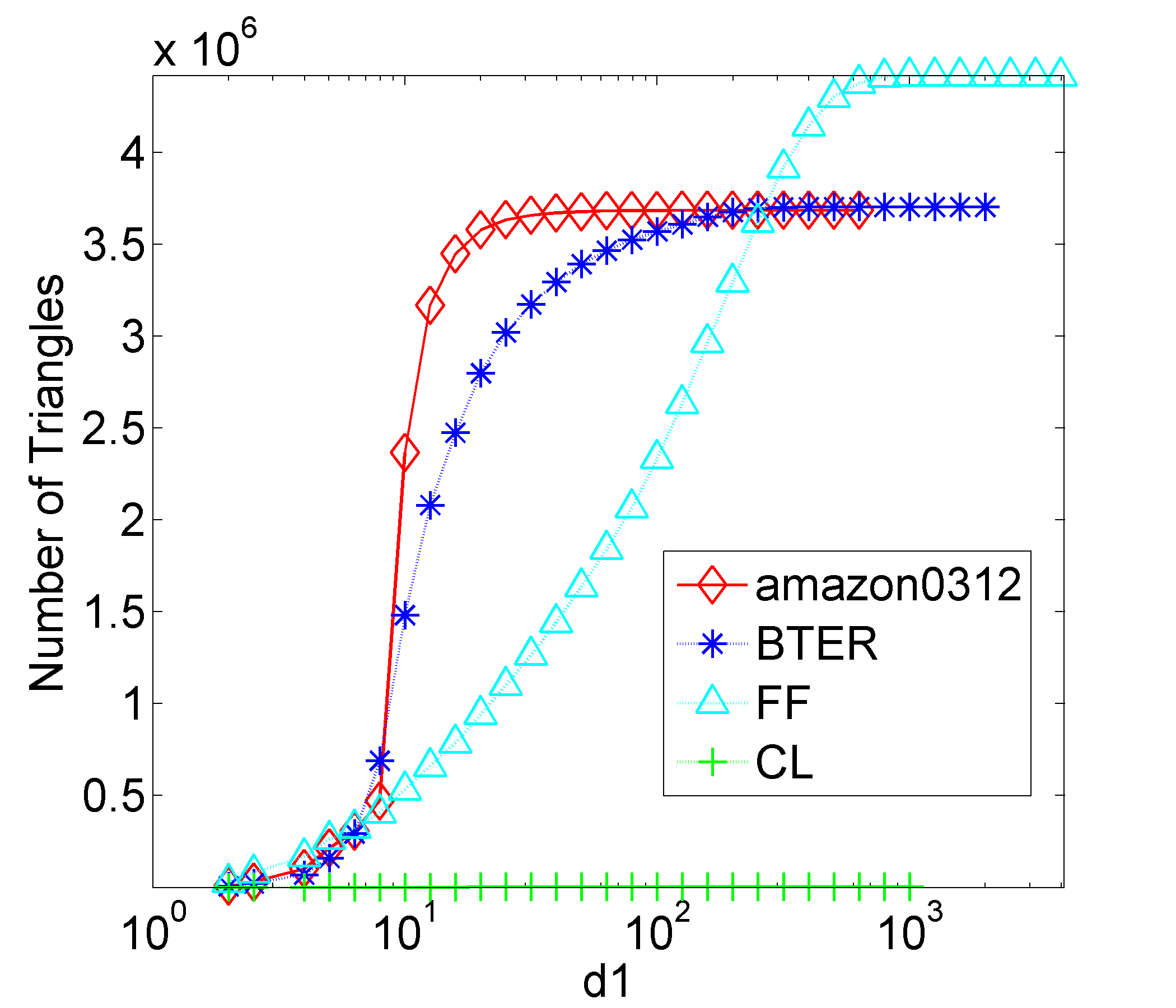}
  }
    \subfloat[ca-AstroPh]{\label{fig:CDF-gm-Ca-AstroPh}%
	\includegraphics[width=.25\textwidth,trim=0 0 0 0]%
        {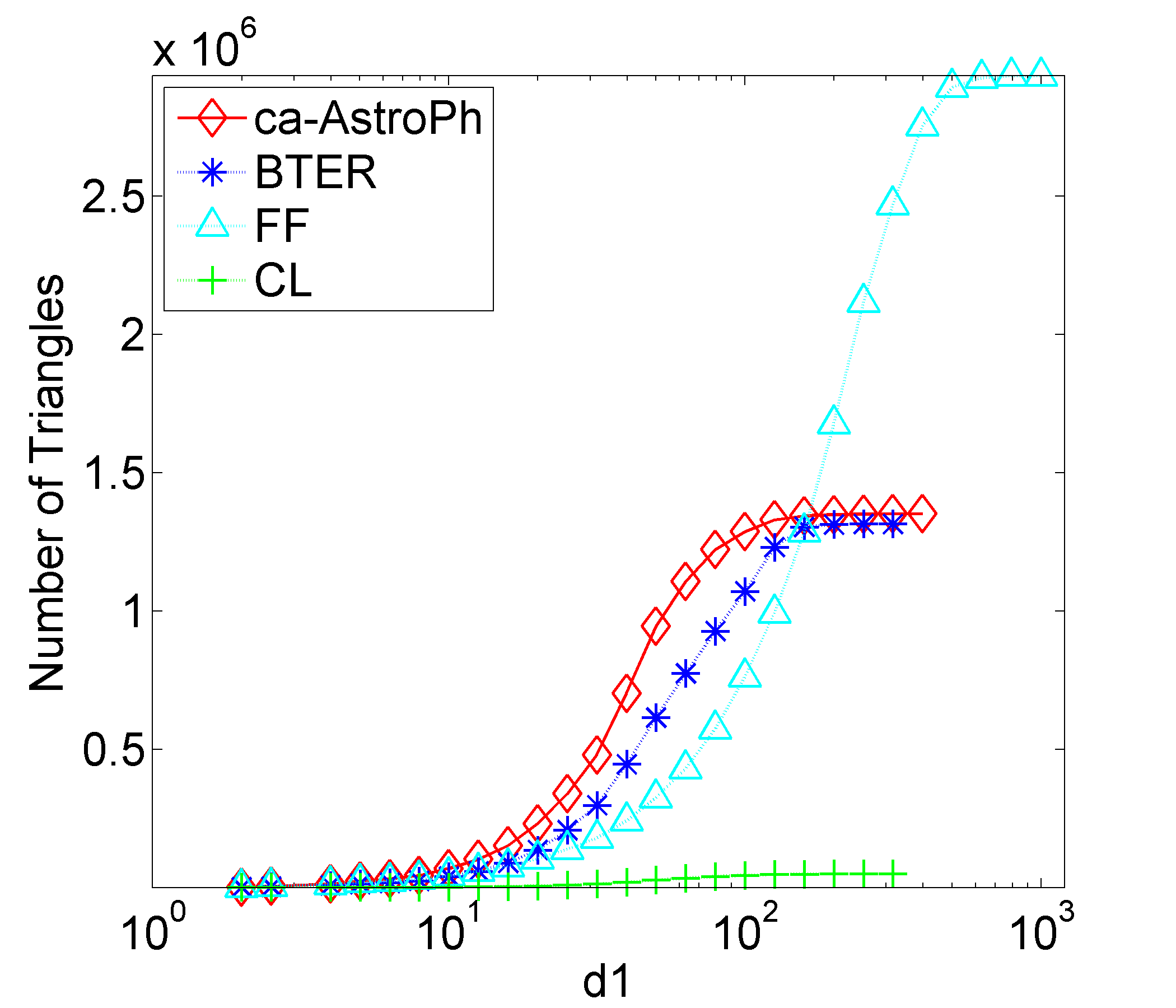}}
  \subfloat[cit-HepPh]{\label{fig:CDF-gm-cit-HepPh}
    \includegraphics[width=.25\textwidth,trim=0 0 0 0]{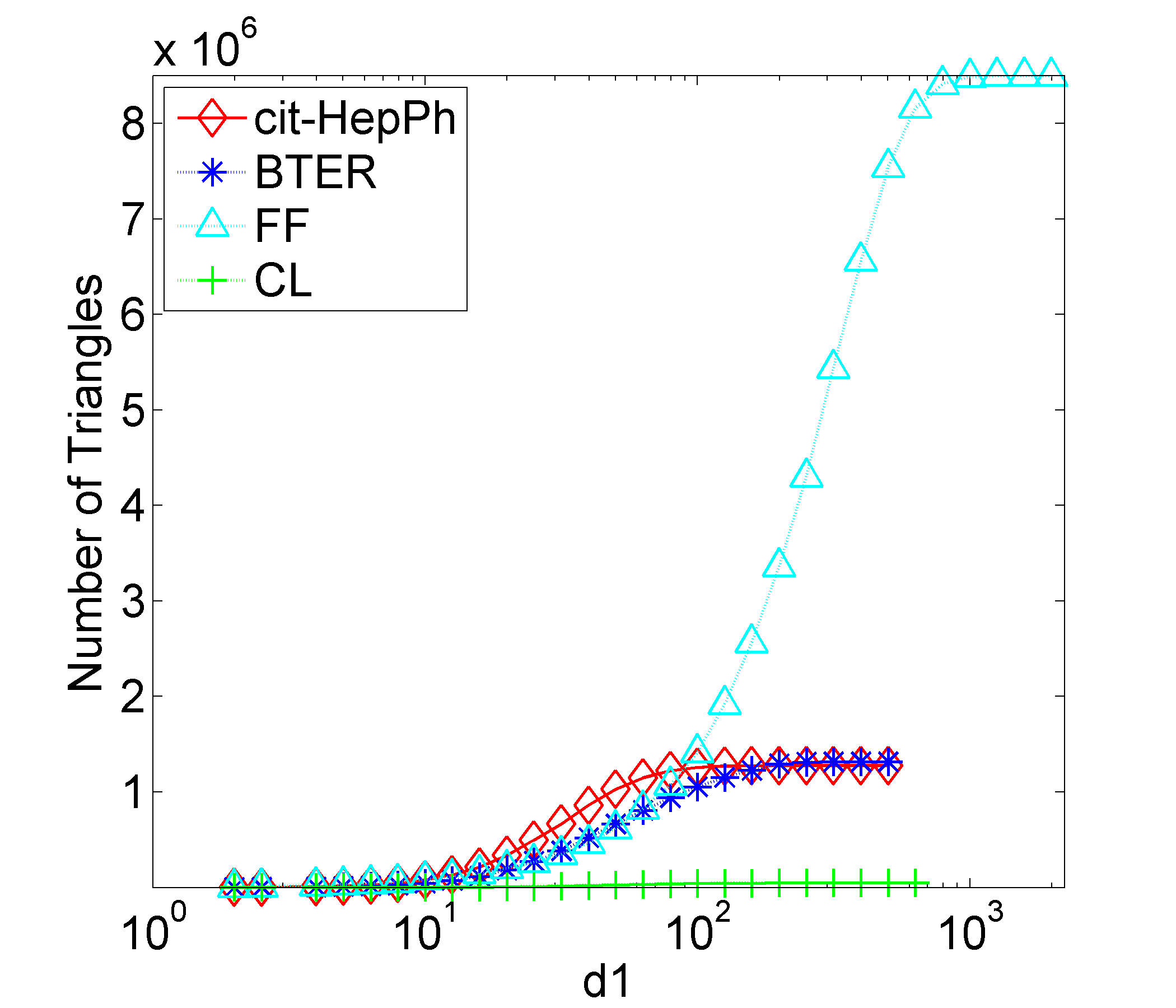} }
  \subfloat[soc-Epinions]{\label{fig:CDF-gm-Soc-Epinions}
    \includegraphics[width=.25\textwidth,trim=0 0 0 0]{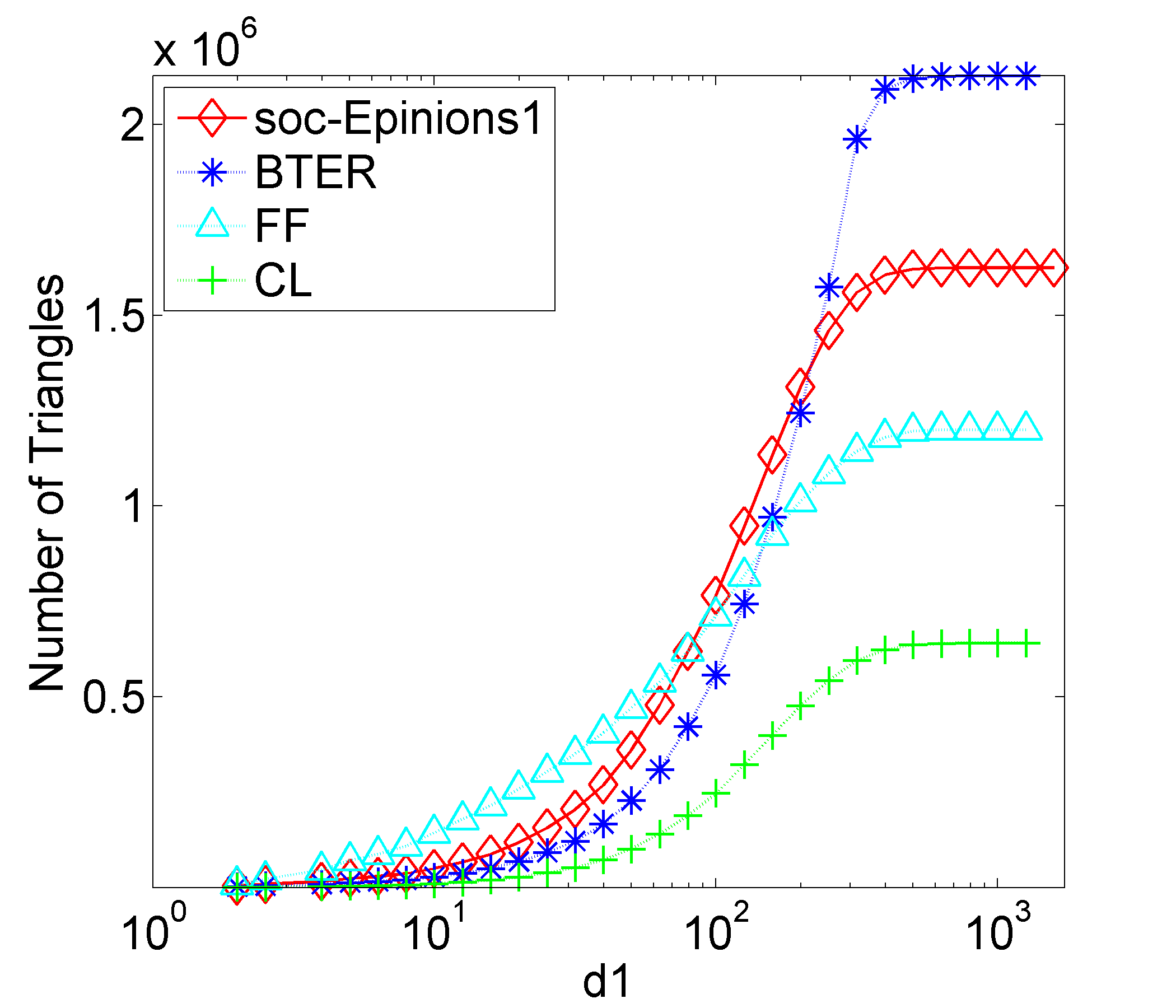}
  }
 \\
  \subfloat[as-caida20071105]{\label{fig:CDF-gm-as-caida}
    \includegraphics[width=.25\textwidth,trim=0 0 0 0]{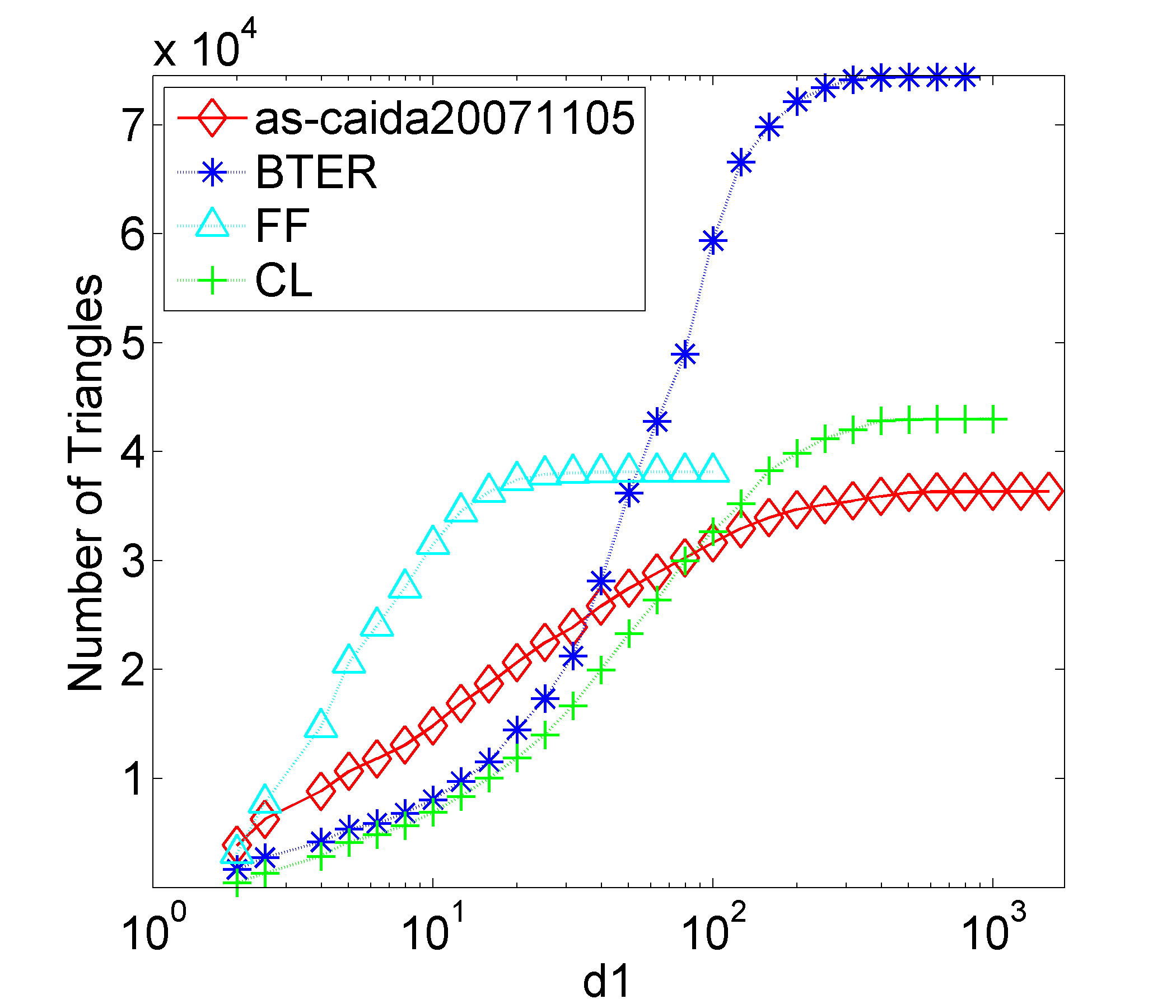}
  }
  \subfloat[oregon1\_010331]{\label{fig:CDF-gm-oregon1}
    \includegraphics[width=.25\textwidth,trim=0 0 0 0]{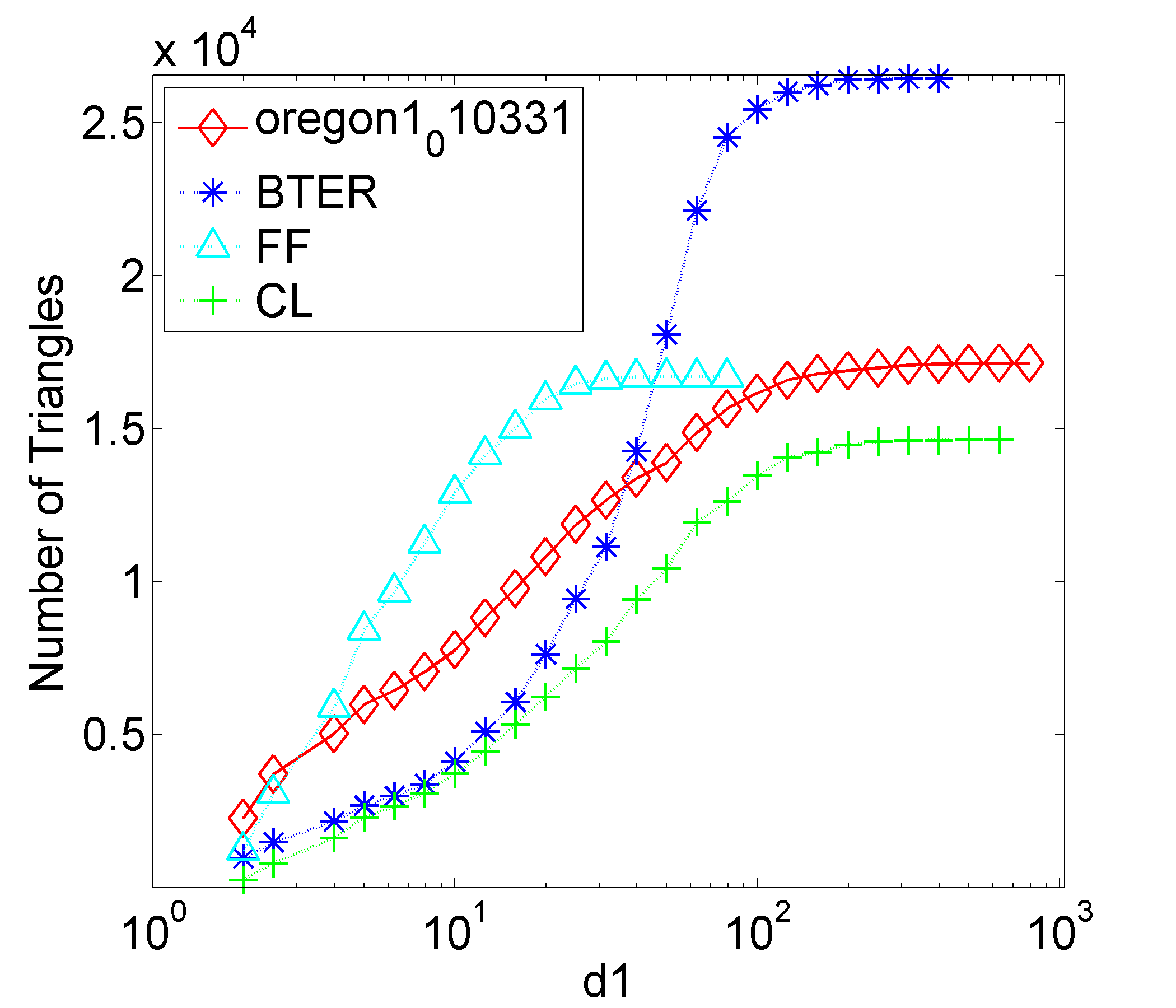}
  }
  \subfloat[web-Stanford]{\label{fig:CDF-gm-web-Stanford}
    \includegraphics[width=.25\textwidth,trim=0 0 0 0]{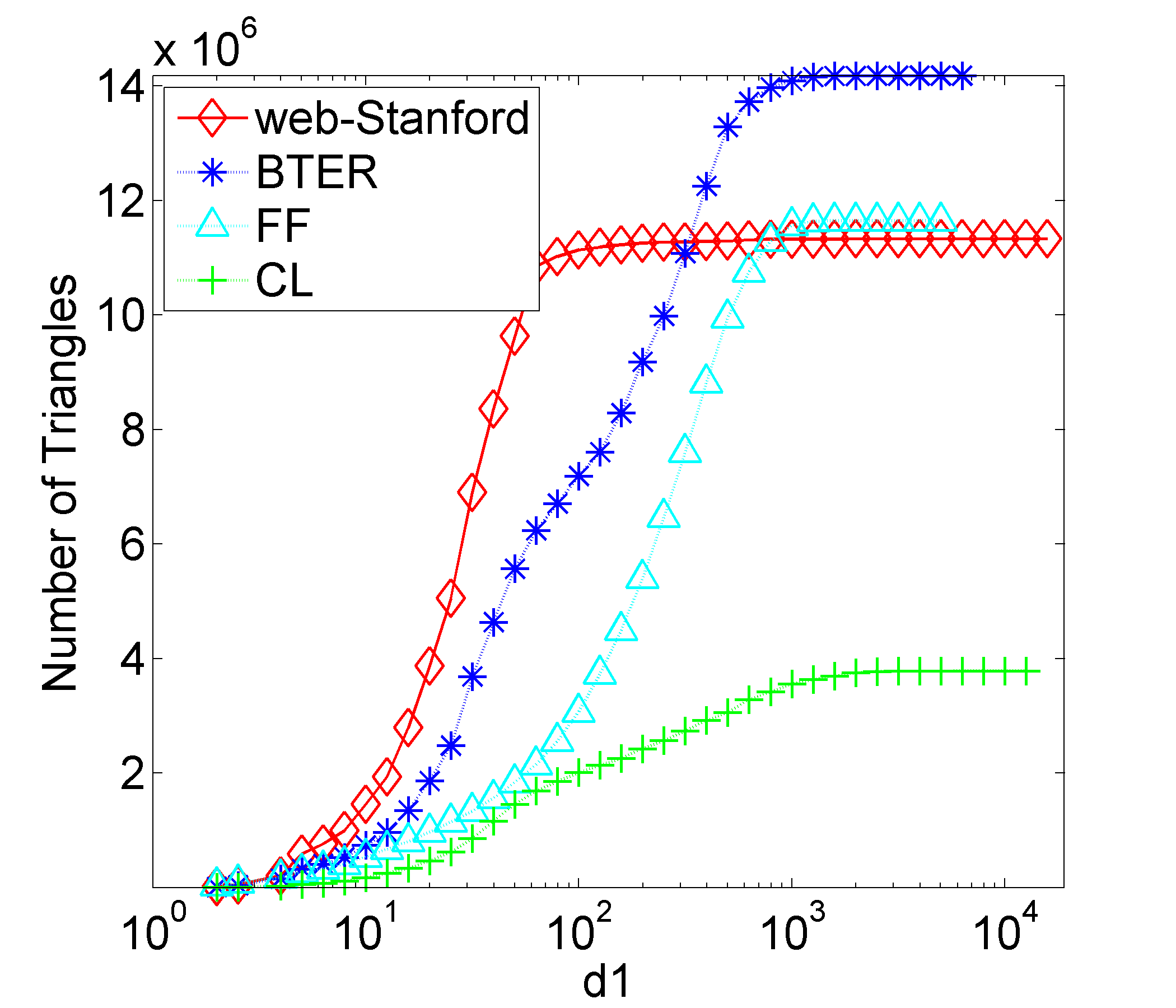}
  }
  \subfloat[wiki-Talk]{\label{fig:CDF-gm-wiki-Talk}
    \includegraphics[width=.25\textwidth,trim=0 0 0 0]{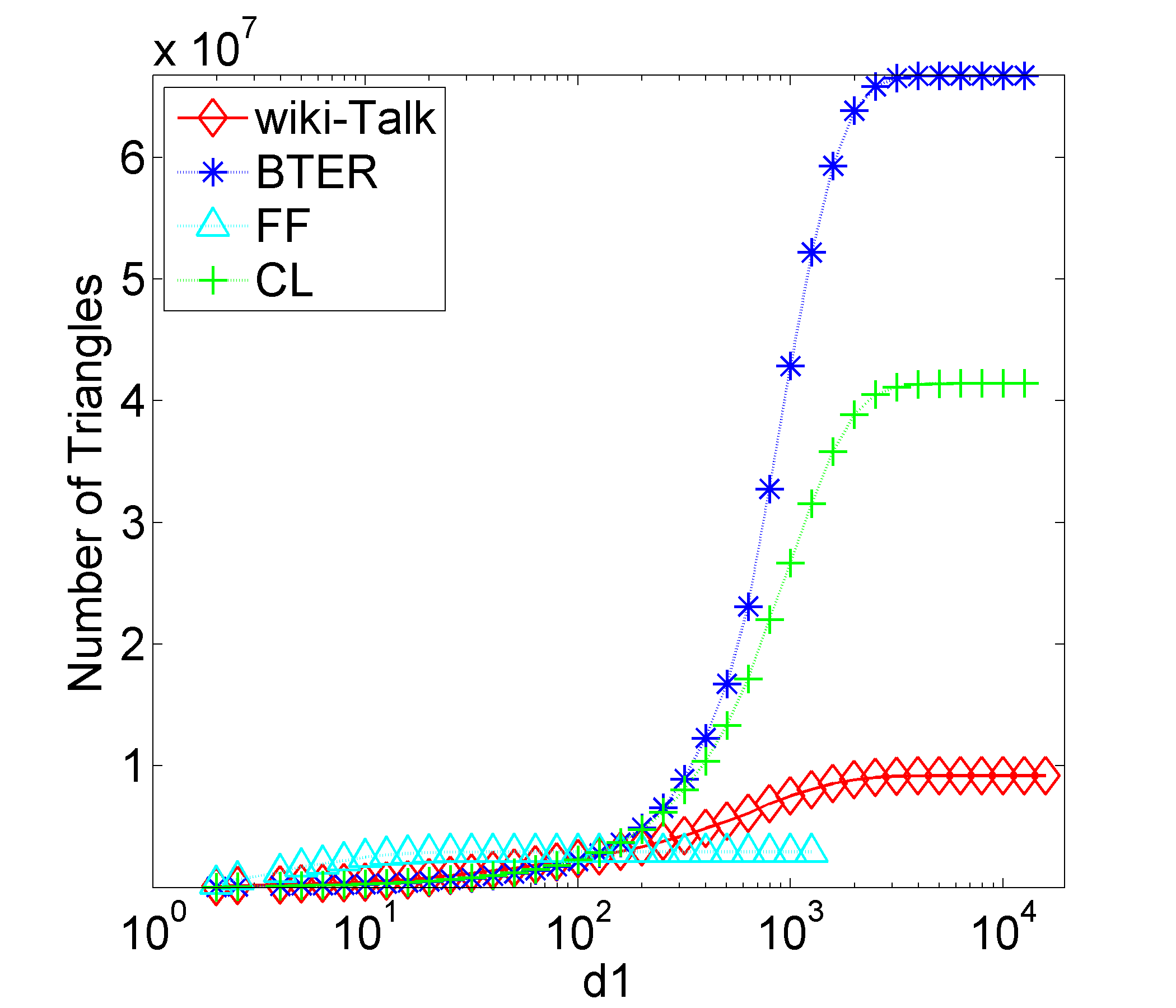}
  }
   \caption{Comparison of cumulative number of triangles for the different models}
    \label{fig:gm-cdf}
\end{figure*}

\smallskip
\textbf{Distance Relations:} \textit{Some models cannot provide a gap between  minimum, medium, and maximum degrees of triangles.}

\smallskip
We provide the degree-comparison plots of generated models in \Fig{gm-degs} for one \emph{high-$C$} and one \emph{low-$C$} network as representative. The other networks in \emph{high-$C$} and \emph{low-$C$} networks have the similar degree-comparison plots.

For \emph{high-$C$} networks, BTER cannot provide the gap among minimum, middle, and maximum degrees. Having the almost same degree for all $\da{i}$, $\db{i}$, and $\dc{i}$ conflicts our \emph{Observation 2}. FF provides a distance between minimum, middle, and maximum degrees, however, $\db{i}$ and $\dc{i}$ strongly deviate from $\da{i}$ in the middle (it generates a bump). CL also provides the gap among degrees but both $\db{i}$ and $\dc{i}$ strongly deviates from $\da{i}$ at the beginning.

For \emph{low-$C$} networks, BTER does not generate distance at the beginning again but in the middle it jumps to high $\dc{i}$ values. FF provides the gap but it has a bump in the middle again. CL provides much wider gap between $\db{i}$ and $\dc{i}$.

It is very clear in \Fig{gm-degs} that CL generates heterogonous triangles in any network, and FF and BTER generate more homogenous triangles and none of the graph models exactly match the triangle behavior in the original network.

\smallskip
\textbf{The number of triangles per $\da{i}$:} \textit{None of the models obtain the number of triangles per $\da{i}$ for \emph{both} \emph{high-$C$} and \emph{low-$C$} networks.}

\smallskip
To understand, how triangles are distributed among degrees, we analyze the number of triangles generated per minimum degree $\da{i}$ for different graph models in \Fig{gm-cdf}. Models behave differently for \emph{high-$C$} and \emph{low-$C$} networks again.

For \emph{high-$C$} networks, BTER model matches the original network behavior very well except Soc-Epinions (which is in the border of \emph{low-$C$}). FF generates way too many triangles for high degree nodes except Soc-Epinions (it produces less triangles for this graph). CL is generating consistently lower number of triangles than the original number per minimum degree $\da{i}$. For Soc-Epinions, CL generates slightly higher triangles but still less than the original. It is observable that when $C$ decreases, CL starts to generate comparably more triangles.
In \emph{high-$C$} networks, FF reaches much higher $\da{i}$ values (except Soc-Epinions). BTER and CL are reaching the similar $\da{i}$  values to the real $\da{i}$ .

For \emph{low-$C$} networks, none of the graph model matches to the original network behavior. BTER is generating more triangles for high degree nodes. CL is catching the similar behavior in \Fig{CDF-gm-as-caida} and in \Fig{CDF-gm-oregon1} but for the others it misses as well. FF tends to only generate triangles with relatively low degree vertices since the graphs are very sparse in \emph{low-$C$} networks. Hence, the plot for FF ends very early.  In \emph{low-$C$} networks, FF reaches only one-tenth of the maximum $\da{i}$ value of the original networks (except web-Stanford).

\smallskip
\textbf{Clustering coefficient plots:} \textit{The clustering coefficient behavior of models differ for \emph{high-$C$} and \emph{low-$C$} networks.}

\begin{figure*}[htb]
    \centering
    \subfloat[amazon0312]{\label{fig:CC-Amazon0312}
    \includegraphics[width=.25\textwidth,trim=0 0 0 0]{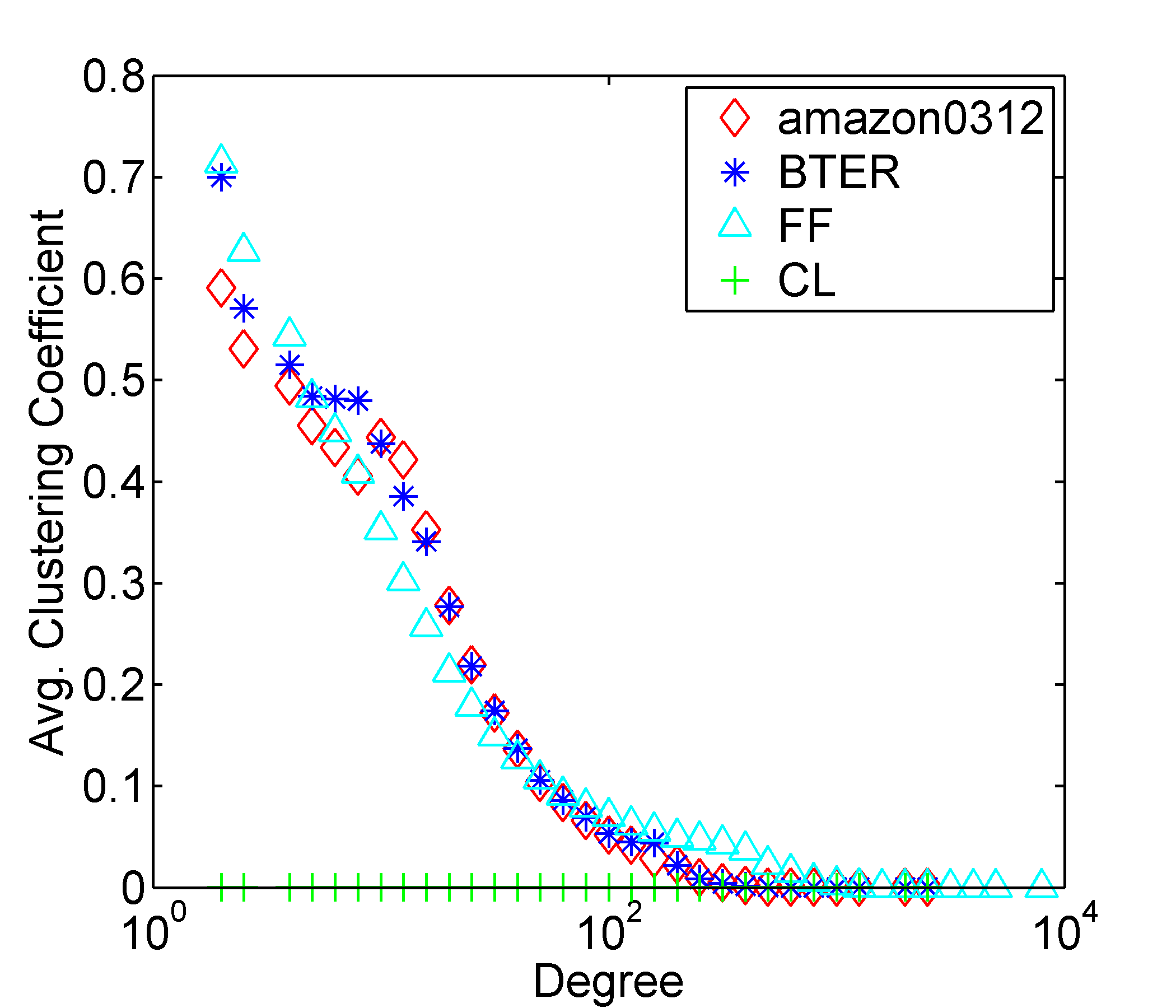}
  }
    \subfloat[ca-AstroPh]{\label{fig:CC-ca-Astro}
	\includegraphics[width=.25\textwidth,trim=0 0 0 0]{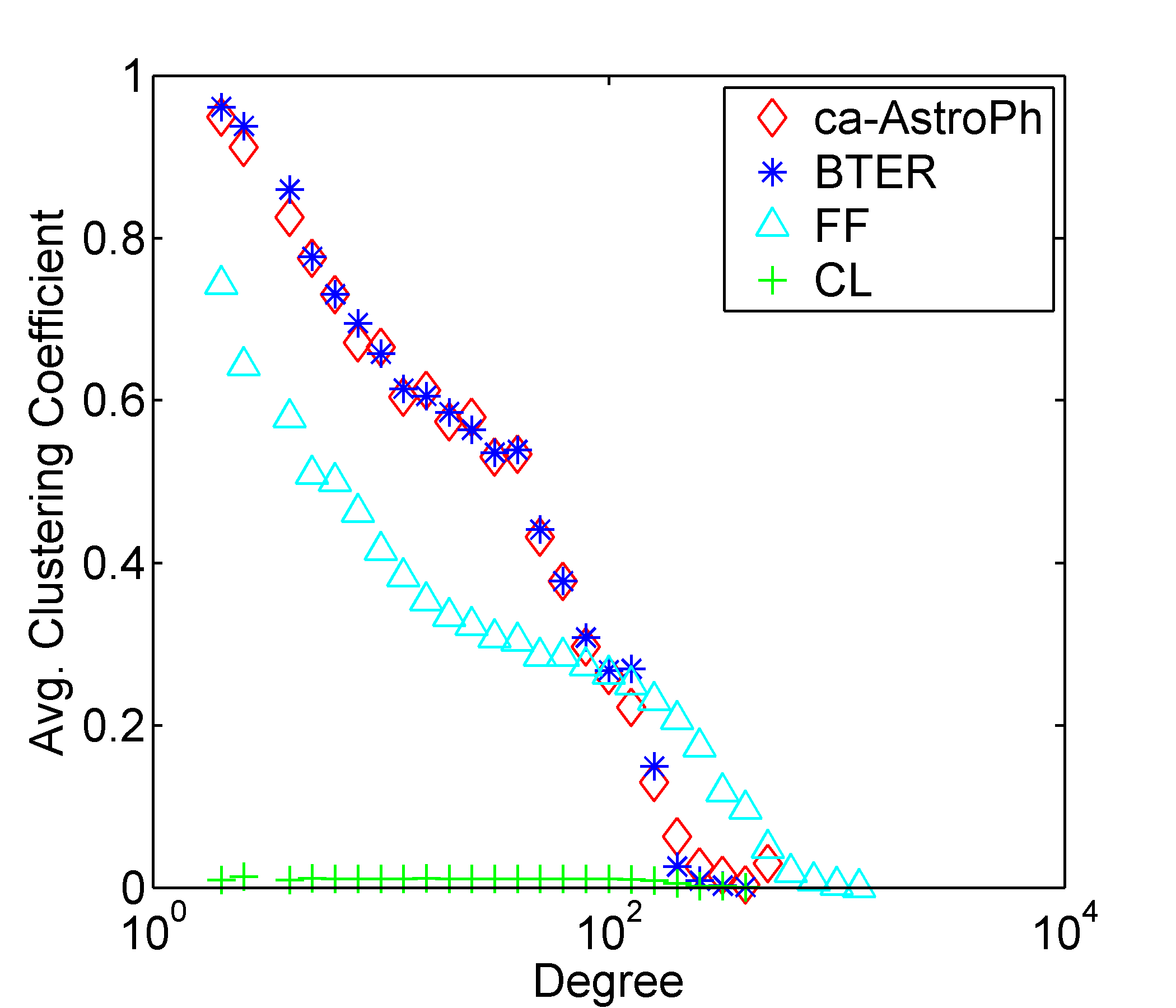}
 }
   \subfloat[cit-HepPh]{\label{fig:CC-cit-HepPh}
    \includegraphics[width=.25\textwidth,trim=0 0 0 0]{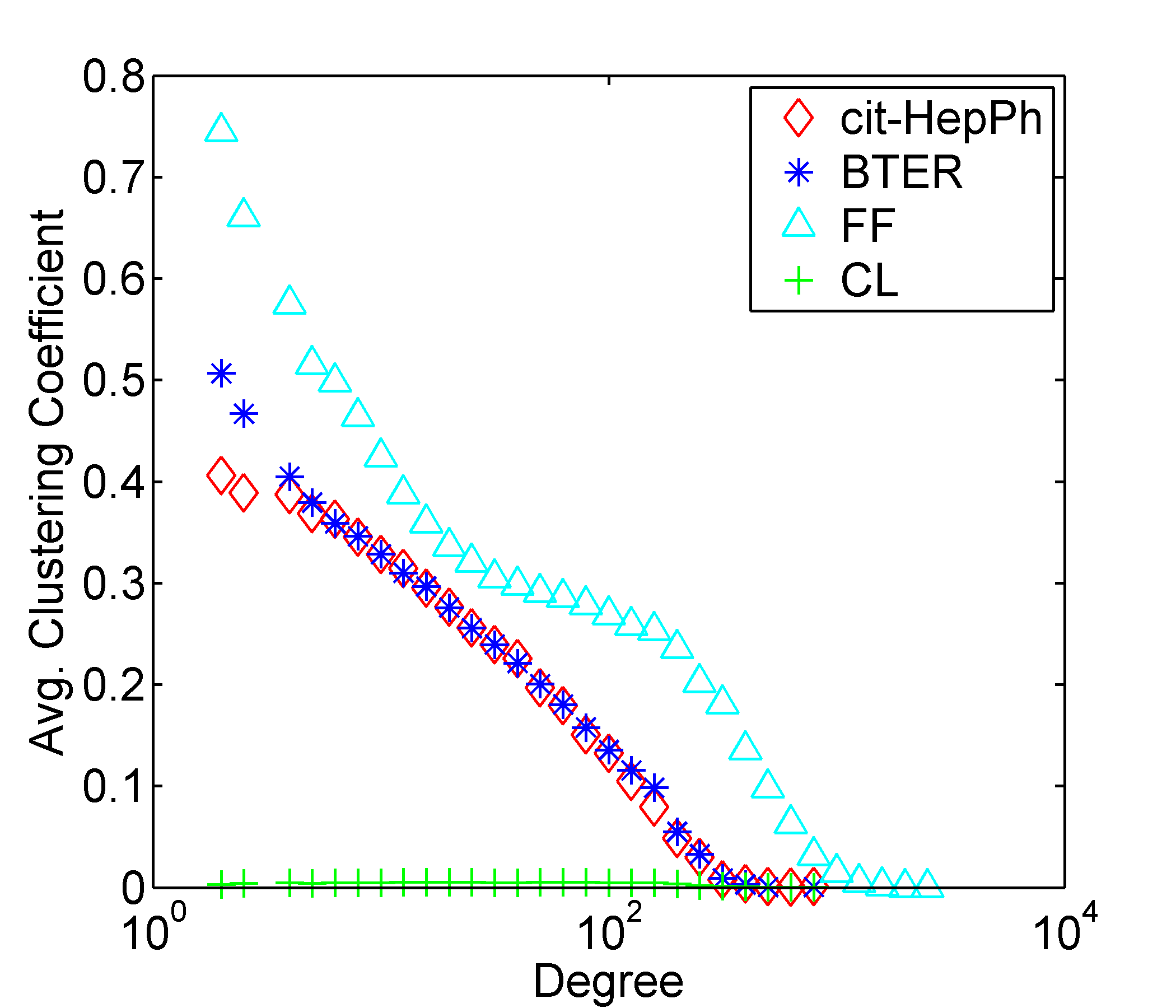}
  }
  \subfloat[soc-Epinions]{\label{fig:CC-Soc-Epinions}
    \includegraphics[width=.25\textwidth,trim=0 0 0 0]{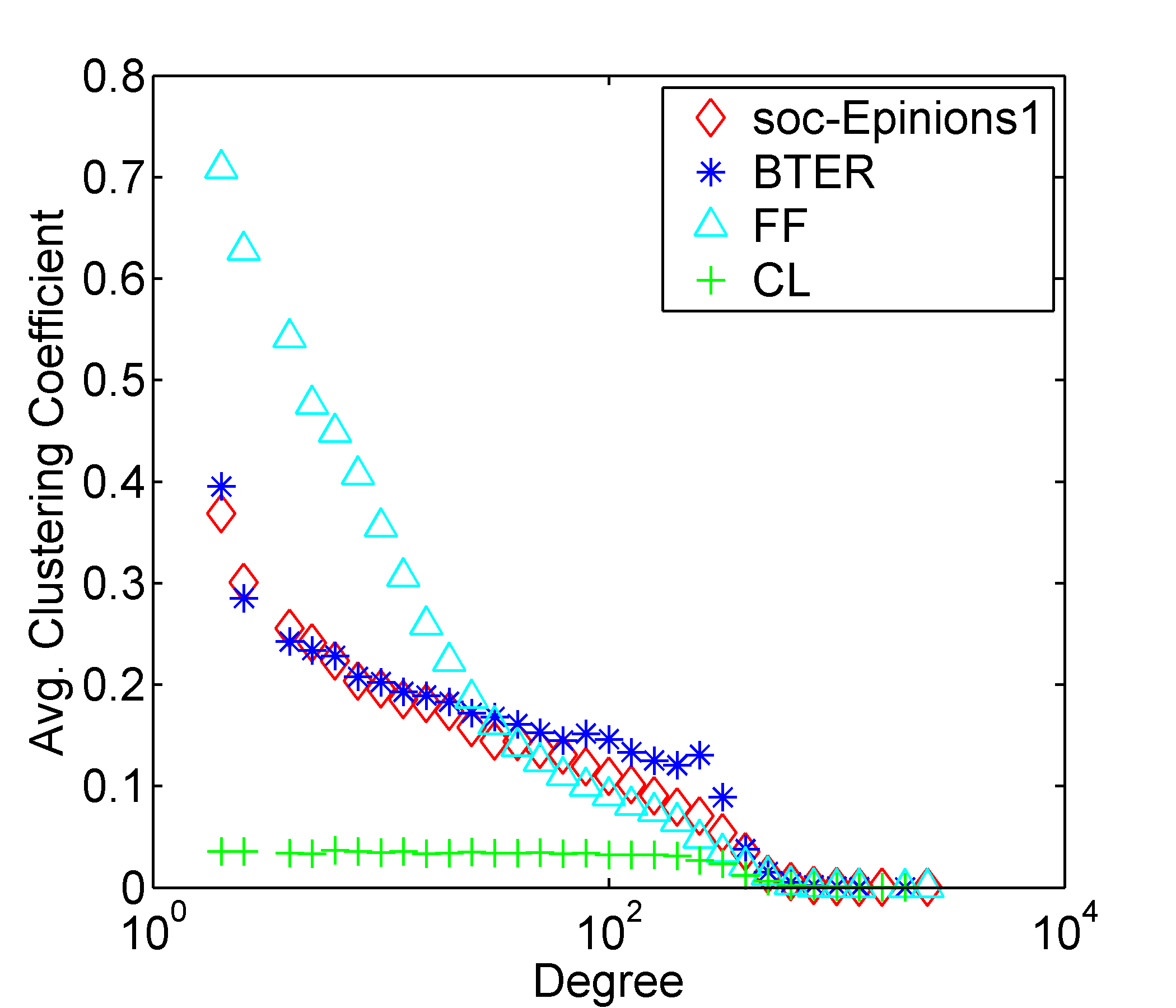}
  }
  \\
  \subfloat[as-caida20071105]{\label{fig:CC-as-caida}
    \includegraphics[width=.25\textwidth,trim=0 0 0 0]{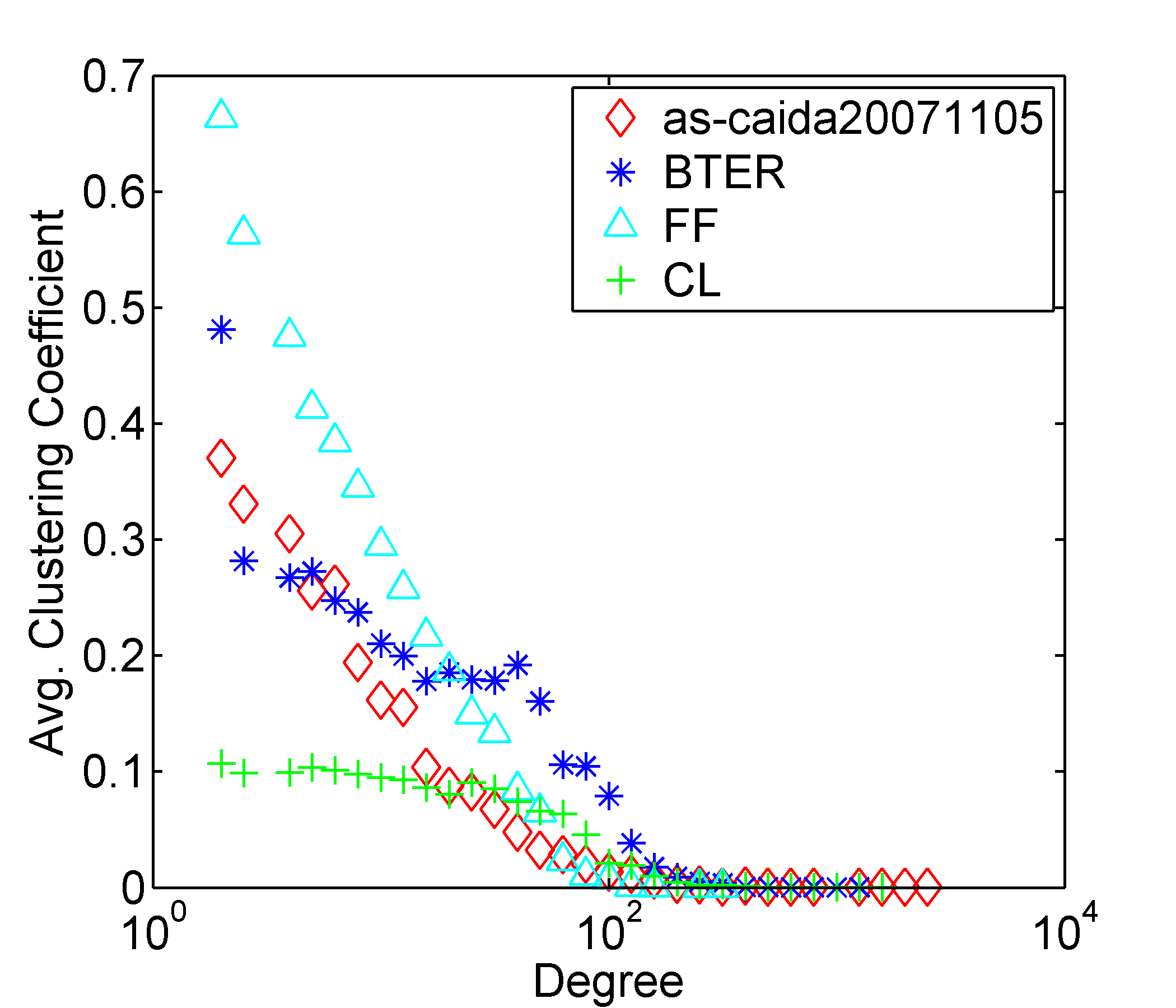}
  }
  \subfloat[oregon1\_010331]{\label{fig:CC-oregon1}
    \includegraphics[width=.25\textwidth,trim=0 0 0 0]{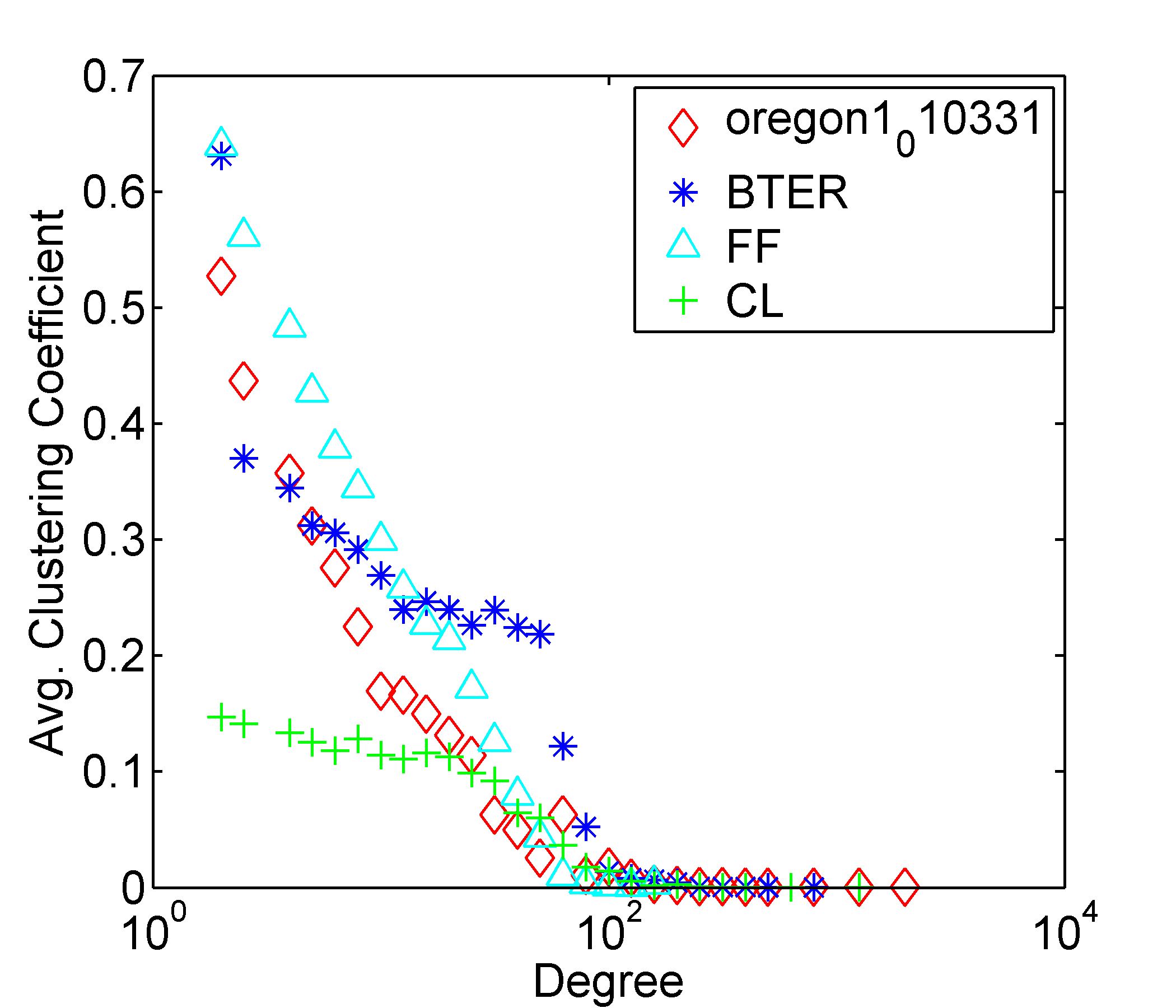}
  }
  \subfloat[web-Stanford]{\label{fig:CC-web-Stanford}
    \includegraphics[width=.25\textwidth,trim=0 0 0 0]{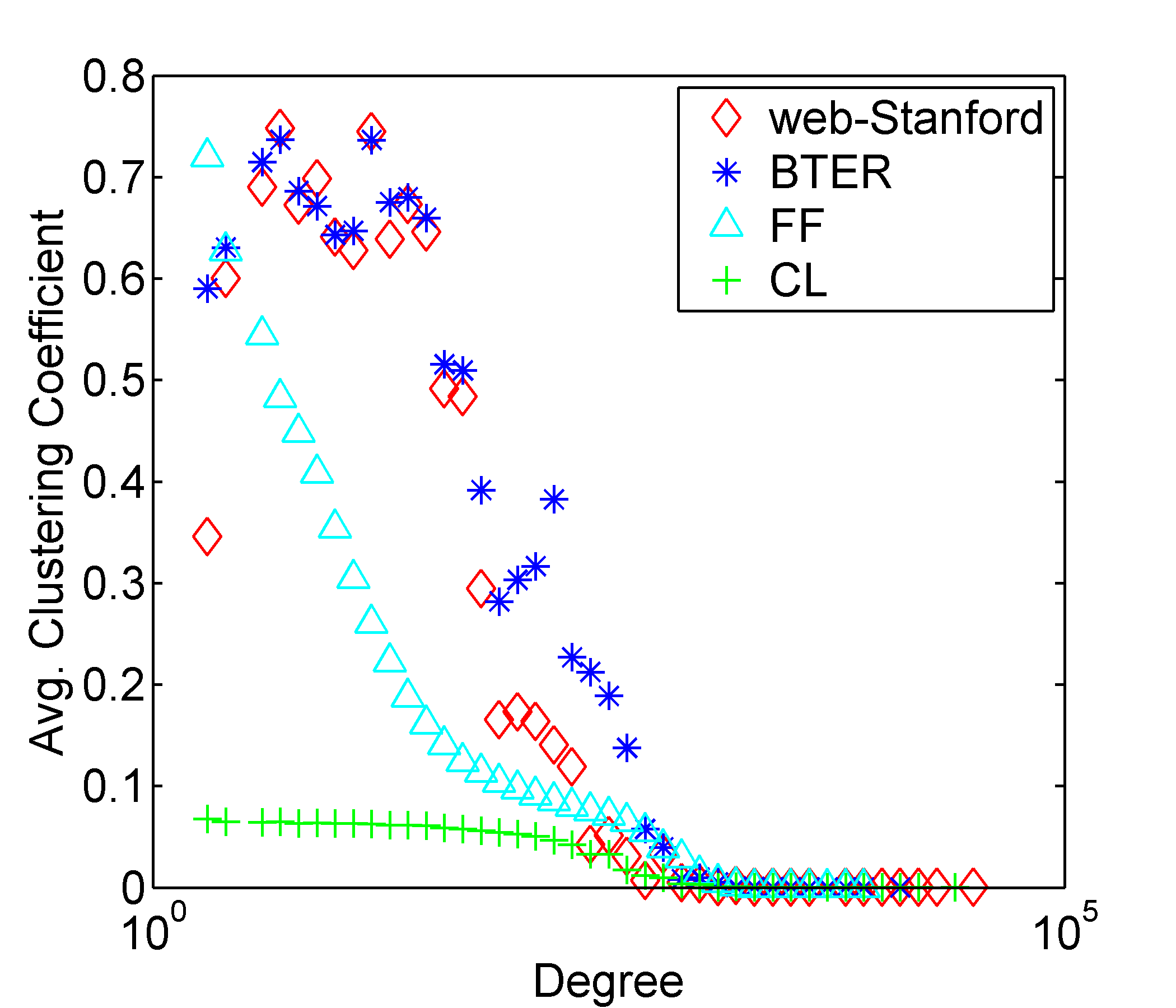}
  }
   \subfloat[wiki-Talk]{\label{fig:CC-wiki-Talk}
    \includegraphics[width=.25\textwidth,trim=0 0 0 0]{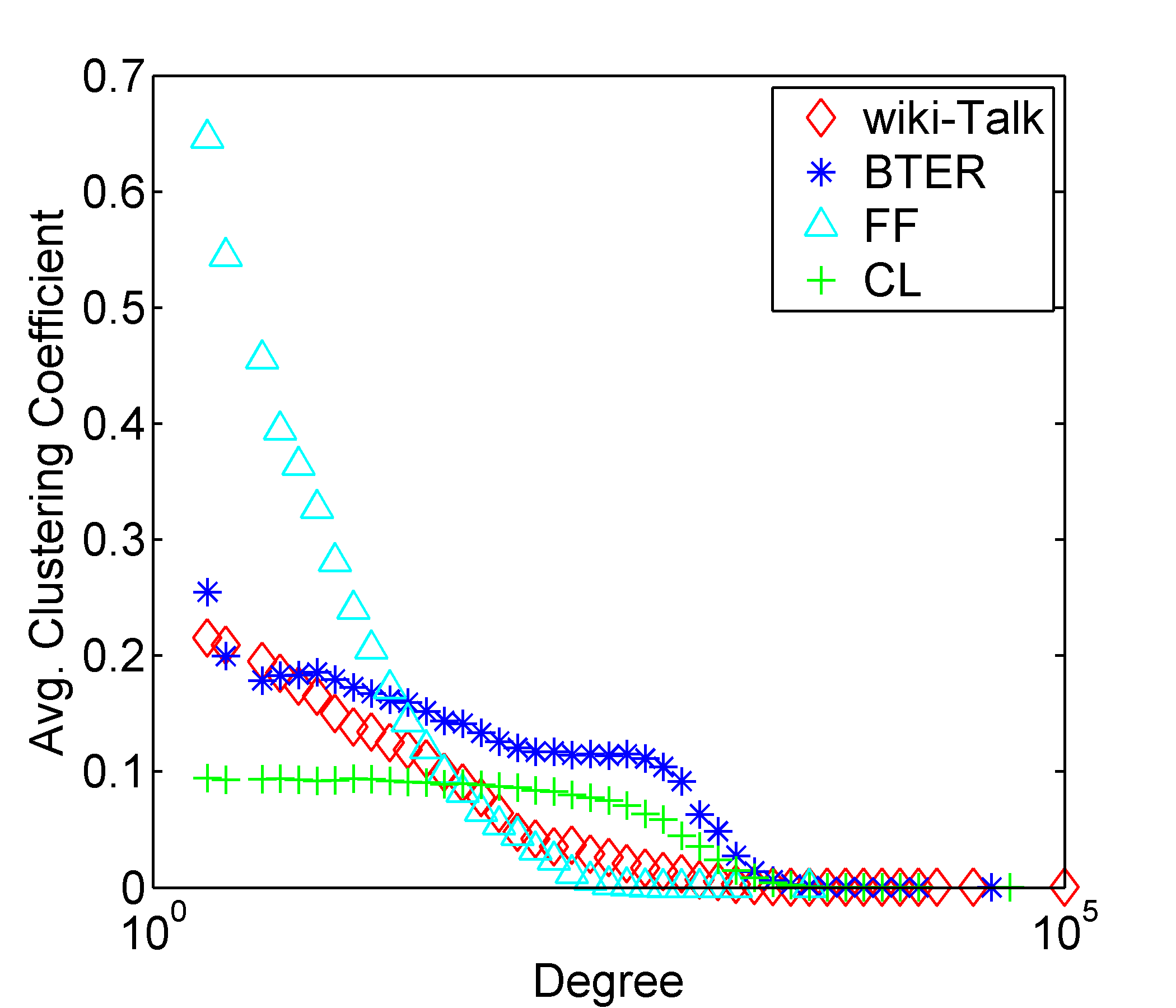}
  }
  \caption{Clustering Coefficient vs Degree Correlation for the different models}
  \label{fig:gm-cc}
\end{figure*}

\smallskip
When we analyze average local clustering coefficient ($C_d$) per degree in graphs models as plotted in \Fig{gm-cc}, we can see the similar trends as in \Fig{gm-cdf}. Clustering coefficient plots confirm that low degree vertices have high local clustering coefficient.

 For \emph{high-$C$} networks, BTER matchs the local clustering coefficient behavior of the real graphs fairly well (in fact, the $C_d$ values are used as input to the model). FF is not matching perfectly but generates high $C_d$ values. CL is failing to reach high clustering coefficient values per degrees for \emph{high-$C$} networks but it has slightly better averages for \emph{low-$C$} networks.

For \emph{low-$C$} networks, BTER deviates from the original $C_d$ values in the middle. FF is not matching perfectly again but getting close. CL is generating higher $C_d$ averages than it generated for \emph{high-$C$} networks.

\section{Conclusions}
\label{sec:conc}

We believe that the dichotomy between homogeneous triangles and heterogeneous triangles is quite useful in characterizing graphs. This also gives a quantifiable method to distinguish the underlying processes creating these graphs.
Social scientists and physicists have long tried to explain the behavior of humans (or appropriate agents) based
on topological structure. The degree-labeled triangles appears to give us a window into this behavior.

The degree-labeled triangle behavior yields fascinating insight into the structure of real-world networks.  As expected the triangles come in various types in different networks. However, our studies showed that the global clustering coefficient is a good indicator of  what kind of  triangles the graph contains. High clustering coefficients ($>0.01$) imply homogenous triangles  (i.e., degrees of the vertices are close), while low clustering coefficients is a sign of heterogenous triangles (i.e., significant variance among the degrees of the vertices).

We have also investigated whether the current graph models can regenerate  the types of triangles  in the real data.  The results showed that  while some models are good at matching the total number of triangles in the real graph, no model can  match the types of triangles for both the high and low clustering coefficient graphs together. Our paper shows that community structures in graph models are not able to capture the community behaviors in the real networks. Therefore, there is a room to improve the existent models and design more realistic graph models which support triangle degree behaviors.

\section*{Acknowledgments}
This work was funded by Defense Advanced Research Projects Agency (DARPA), the applied mathematics program at the United
States Department of Energy
and by an Early Career Award from the Laboratory
Directed Research \& Development (LDRD) program at Sandia National
Laboratories. Sandia National Laboratories is
a multiprogram laboratory operated by Sandia
Corporation, a wholly owned subsidiary of Lockheed Martin Corporation,
for the United States Department of Energy's National Nuclear Security
Administration under contract DE-AC04-94AL85000.

\bibliographystyle{acm}
\bibliography{triangles}

\begin{thebibliography}{10}

\bibitem{BaAl99}
{\sc Barab\'{a}si, A.-L., and Albert, R.}
\newblock Emergence of scaling in random networks.
\newblock {\em Science 286}, 5349 (1999), 509--512.

\bibitem{BeMoSt04}
{\sc Bearman, P.~S., Moody, J., and Stovel, K.}
\newblock {Chains of Affection: The Structure of Adolescent Romantic and Sexual
  Networks}.
\newblock {\em American Journal of Sociology 110}, 1 (July 2004), 44--91.

\bibitem{BeBoCaGi08}
{\sc Becchetti, L., Boldi, P., Castillo, C., and Gionis, A.}
\newblock Efficient semi-streaming algorithms for local triangle counting in
  massive graphs.
\newblock In {\em KDD'08\/} (2008), pp.~16--24.

\bibitem{BeHe11}
{\sc Berry, J.~W., Hendrickson, B., LaViolette, R.~A., and Phillips, C.~A.}
\newblock Tolerating the community detection resolution limit with edge
  weighting.
\newblock {\em Phys. Rev. E 83\/} (May 2011), 056119.

\bibitem{BoLaMoCh06}
{\sc Boccaletti, S., Latora, V., Moreno, Y., Chavez, M., and Hwang, D.-U.}
\newblock Complex networks: Structure and dynamics.
\newblock {\em Physics Reports 424\/} (2006), 175--308.

\bibitem{Burt04}
{\sc Burt, R.~S.}
\newblock Structural holes and good ideas.
\newblock {\em American Journal of Sociology 110}, 2 (2004), 349--399.

\bibitem{Burt07}
{\sc Burt, R.~S.}
\newblock Secondhand brokerage: Evidence on the importance of local structure
  for managers, bankers, and analysts.
\newblock {\em Academy of Management Journal 50\/} (2007).

\bibitem{ChLu02}
{\sc Chung, F., and Lu, L.}
\newblock The average distances in random graphs with given expected degrees.
\newblock {\em PNAS 99\/} (2002), 15879--15882.

\bibitem{ClShNe09}
{\sc Clauset, A., Shalizi, C.~R., and Newman, M. E.~J.}
\newblock Power-law distributions in empirical data.
\newblock {\em SIAM Review 51}, 4 (2009), 661--703.

\bibitem{Co09}
{\sc Cohen, J.}
\newblock Graph twiddling in a {MapReduce} world.
\newblock {\em Computing in Science \& Engineering 11\/} (2009), 29--41.

\bibitem{Co88}
{\sc Coleman, J.~S.}
\newblock Social capital in the creation of human capital.
\newblock {\em American Journal of Sociology 94\/} (1988), S95--S120.

\bibitem{CoWaFa06}
{\sc Contractor, N.~S., Wasserman, S., and Faust, K.}
\newblock Testing multitheoretical organizational networks: {An} analytic
  framework and empirical example.
\newblock {\em Academy of Management Review 31}, 3 (2006), 681--703.

\bibitem{EcMo02}
{\sc Eckmann, J.-P., and Moses, E.}
\newblock Curvature of co-links uncovers hidden thematic layers in the {World
  Wide Web}.
\newblock {\em PNAS 99}, 9 (2002), 5825--5829.

\bibitem{GlSe11}
{\sc Gleich, D.~F., and Seshadhri, C.}
\newblock Neighborhoods are good communities.
\newblock arXiv:1112.0031v1, 2011.

\bibitem{HoZh07}
{\sc Holme, P., and Zhao, J.}
\newblock Exploring the assortativity-clustering space of a network's degree
  sequence.
\newblock {\em Phys. Rev. E 75\/} (Apr 2007), 046111.

\bibitem{KlKu99}
{\sc Kleinberg, J.~M., Kumar, R., Raghavan, P., Rajagopalan, S., and Tomkins,
  A.~S.}
\newblock The web as a graph: measurements, models, and methods.
\newblock In {\em Proceedings of the 5th annual international conference on
  Computing and combinatorics\/} (Berlin, Heidelberg, 1999), COCOON'99,
  Springer-Verlag, pp.~1--17.

\bibitem{La06}
{\sc Lawrence, B.~S.}
\newblock Organizational reference groups: A missing perspective on social
  context.
\newblock {\em Organization Science 17}, 1 (2006), 80--100.

\bibitem{LeChKlFa10}
{\sc Leskovec, J., Chakrabarti, D., Kleinberg, J., Faloutsos, C., and
  Ghahramani, Z.}
\newblock {K}ronecker graphs: An approach to modeling networks.
\newblock {\em J. Machine Learning Research 11\/} (Feb. 2010), 985--1042.

\bibitem{LeFa07}
{\sc Leskovec, J., and Faloutsos, C.}
\newblock Scalable modeling of real graphs using {Kronecker} multiplication.
\newblock In {\em ICML '07\/} (2007), ACM, pp.~497--504.

\bibitem{LeKlFa05}
{\sc Leskovec, J., Kleinberg, J., and Faloutsos, C.}
\newblock Graphs over time: densification laws, shrinking diameters and
  possible explanations.
\newblock In {\em Proceedings of the eleventh ACM SIGKDD international
  conference on Knowledge discovery in data mining\/} (New York, NY, USA,
  2005), KDD '05, ACM, pp.~177--187.

\bibitem{LiHo12}
{\sc Litvak, N., and van~der Hofstad, R.}
\newblock Large scale-free networks are not disassortative.
\newblock arXiv:1204.0266v1.

\bibitem{MaHuKrHu07}
{\sc Mahadevan, P., Hubble, C., Krioukov, D.~V., Huffaker, B., and Vahdat, A.}
\newblock Orbis: {Rescaling} degree correlations to generate annotated internet
  topologies.
\newblock {\em SIGCOMM'07\/} (2007), 325--336.

\bibitem{MaKrFaVa06}
{\sc Mahadevan, P., Krioukov, D., Fall, K., and Vahdat, A.}
\newblock Systematic topology analysis and generation using degree
  correlations.
\newblock {\em SIGCOMM'06\/} (2006), 135--146.

\bibitem{Milo2002}
{\sc Milo, R., {Shen-Orr}, S., Itzkovitz, S., Kashtan, N., Chklovskii, D., and
  Alon, U.}
\newblock Network motifs: Simple building blocks of complex networks.
\newblock {\em Science 298}, 5594 (2002), 824--827.

\bibitem{Ne02}
{\sc Newman, M. E.~J.}
\newblock Assortative mixing in networks.
\newblock {\em Phys. Rev. Letter 89\/} (May~20 2002), 208701.

\bibitem{PiSeKo12}
{\sc Pinar, A., Seshadhri, C., and Kolda, T.~G.}
\newblock The similarity between {S}tochastic {K}ronecker and {C}hung-{L}u
  graph models.
\newblock In {\em Proceedings of SIAM Conference on Data Mining\/} (2012),
  SIAM.
\newblock arXiv:1110.4925, to appear in Proc. SDM12.

\bibitem{Po98}
{\sc Portes, A.}
\newblock Social capital: Its origins and applications in modern sociology.
\newblock {\em Annual Review of Sociology 24}, 1 (1998), 1--24.

\bibitem{Pr07}
{\sc Pržulj, N.}
\newblock Biological network comparison using graphlet degree distribution.
\newblock {\em Bioinformatics 23}, 2 (2007), e177--e183.

\bibitem{SaCaWiZa10}
{\sc Sala, A., Cao, L., Wilson, C., Zablit, R., Zheng, H., and Zhao, B.~Y.}
\newblock Measurement-calibrated graph models for social network experiments.
\newblock In {\em WWW '10\/} (2010), ACM, pp.~861--870.

\bibitem{SeKoPi11}
{\sc Seshadhri, C., Kolda, T.~G., and Pinar, A.}
\newblock Community structure and scale-free collections of {Erd\H{o}s-R\'enyi}
  graphs.
\newblock {\em Physical Review E 85\/} (May 2012).

\bibitem{CPK11}
{\sc Seshadhri, C., Pinar, A., and Kolda, T.}
\newblock Choosing parameters for stochastic kronecker graphs.
\newblock In {\em ICDM 2011\/} (2011).

\bibitem{ShWaGi02}
{\sc Shasha, D., Wang, J. T.-L., and Giugno, R.}
\newblock Algorithmics and applications of tree and graph searching.
\newblock In {\em PODS'02\/} (2002), pp.~39--52.

\bibitem{Ts08}
{\sc Tsourakakis, C.}
\newblock Fast counting of triangles in large real networks without counting:
  Algorithms and laws.
\newblock In {\em ICDM'08\/} (2008), pp.~608--617.

\bibitem{WaSt98}
{\sc Watts, D., and Strogatz, S.}
\newblock Collective dynamics of `small-world' networks.
\newblock {\em Nature 393\/} (1998), 440--442.

\bibitem{FoDeCo10}
{\sc Welles, B.~F., Van~Devender, A., and Contractor, N.}
\newblock Is a friend a friend?: {Investigating} the structure of friendship
  networks in virtual worlds.
\newblock In {\em CHI-EA'10\/} (2010), pp.~4027--4032.

\bibitem{WhAl06}
{\sc Whitney, D.~E., and Alderson, D.}
\newblock {\em Are technological and social networks really different?}
\newblock 2006, pp.~25--30.

\bibitem{YaYuHa04}
{\sc Yan, X., Yu, P.~S., and Han, J.}
\newblock Graph indexing: A frequent structure-based approach.
\newblock In {\em SIGMOD'04\/} (2004), pp.~335--346.

\bibitem{Snap}
{Stanford Network Analysis Project (SNAP)}.
\newblock Available at \url{http://snap.stanford.edu/}.

\end{thebibliography}

 \end{document}